\documentclass[10pt,journal]{IEEEtran}

\usepackage{filecontents}
\usepackage{stmaryrd}
\usepackage{multirow}
\usepackage[noadjust]{cite}
\usepackage{graphicx}
\usepackage{psfrag}

\usepackage{url}
\usepackage{stfloats}

\usepackage{amsmath,epsfig,amsfonts,amssymb,graphics,psfrag,theorem,calc,url,bm,color,nicefrac,epstopdf,algorithmic}
\usepackage{subcaption}

\usepackage{array}
\usepackage{caption}
\usepackage{cases}
\usepackage{verbatim}
\usepackage[algoruled,linesnumbered]{algorithm2e}
\usepackage{soul}
\usepackage{cancel}

\newtheorem{Lemma}{Lemma}
\newtheorem{Prop}{Proposition}
\newtheorem{Theorem}{Theorem}

\theorembodyfont{\rmfamily}

\newtheorem{Remark}{Remark}

\newcommand{\W}{\bm{W}}
\newcommand{\G}{\bm{G}}
\newcommand{\Q}{\bm{Q}}
\newcommand{\X}{\bm{X}}
\newcommand{\Y}{\bm{Y}}
\newcommand{\Z}{\bm{Z}}
\newcommand{\C}{\bm{C}}

\newcommand{\A}{\bm{A}}
\newcommand{\B}{\bm{B}}
\newcommand{\x}{\bm{x}}

\DeclareMathOperator*{\minimize}{\rm minimize}

\usepackage{pifont}
\usepackage{etoolbox}

\begin{document}
	
	\title{Tensor Completion from Regular Sub-Nyquist Samples}

	\author{Charilaos~I.~Kanatsoulis,~\IEEEmembership{Student Member,~IEEE,}
		        Xiao~Fu,~\IEEEmembership{Member,~IEEE,}\\
		        Nicholas~D.~Sidiropoulos,~\IEEEmembership{Fellow,~IEEE,}
		        and~Mehmet~Ak\c{c}akaya,~\IEEEmembership{Member,~IEEE}
		\thanks{
			C.I. Kanatsoulis and M. Ak\c{c}akaya are with the Department of ECE, University of Minnesota, Minneapolis, MN 55455, USA email: kanat003@umn.edu; akcakaya@umn.edu.
			X. Fu is with the School of EECS, Oregon State University, Corvallis, OR 97331, USA email: xiao.fu@oregonstate.edu.
			N. D. Sidiropoulos is with the Department of ECE, University of Virginia, Charlottesville, VA 22904, USA email: nikos@virginia.edu.

		}
	}

	\maketitle

\begin{abstract}
Signal sampling and reconstruction is a fundamental engineering task at the heart of signal processing. The celebrated Shannon-Nyquist theorem guarantees perfect signal reconstruction from uniform samples, obtained at a rate twice the maximum frequency present in the signal. Unfortunately a large number of signals of interest are far from being band-limited. This motivated  research on reconstruction from sub-Nyquist samples, which mainly hinges on the use of random / incoherent sampling procedures. However, uniform or regular sampling is more appealing in practice and from the system design point of view, as it is far simpler to implement, and often necessary due to system constraints. In this work, we study regular sampling and reconstruction of three- or higher-dimensional signals (tensors). We show that reconstructing a tensor signal from regular samples is feasible. Under the proposed framework, the sample complexity is determined by the tensor rank---rather than the signal bandwidth. This result offers new perspectives for designing practical {\it regular} sampling patterns and systems for signals that are naturally tensors, e.g., images and video. For a concrete application, we show that functional magnetic resonance imaging (fMRI) acceleration is a tensor sampling problem, and design practical sampling schemes and an algorithmic framework to handle it. Numerical results show that our tensor sampling strategy accelerates the fMRI sampling process significantly without sacrificing reconstruction accuracy.
	
\end{abstract}


	\IEEEpeerreviewmaketitle

	\section{Introduction}
	\IEEEPARstart{S}{ampling} and reconstruction of signals is a fundamental problem in signal processing. In the first half of the 20th century, Whittaker, Nyquist, Kotelnikov and Shannon \cite{whittaker1915xviii,nyquist1928certain,kotelnikov1933transmission,shannon1949communication} laid the foundation of {\em sampling theory}. It guarantees perfect reconstruction of a signal from uniformly spaced samples, if sampling is performed at a rate at least twice the maximum frequency present in the signal. The Shannon-Nyquist theorem applies to both continuous and discrete signals. 
	It capitalizes on the band-limitedness property and is the first, and one of the very few results, that allow perfect reconstruction of a signal under a uniform, or more generally, regular sampling process. The challenge is that applying Shannon-Nyquist sampling to wideband signals requires very high sampling rates---which can be quite costly in practice.
	
	In the early 2000's {\em compressive sensing} (CS) \cite{candes2006robust,donoho2006compressed,candes2008introduction} emerged as an alternative, enabling reconstruction from a set of measurements, sampled or compressed below the Nyquist rate. CS works under two basic premises: the signal of interest must have a sparse representation in a known transform domain; and the sampling pattern should be `incoherent'. 
	Under these assumptions, tractable algorithms are shown to recover the signal of interest. Compared to the Shannon-Nyquist sampling theorem, CS leverages signal sparsity, rather than bandlimitedness. This result is significant, since some wideband signals of practical interest, are sparse in certain domains \cite{mallat1999wavelet,baraniuk2007compressive}. 
    On the downside, CS entails higher reconstrunction complexity than sinc function interpolation, and relies on incoherent/random sampling -- thus losing the simplicity of regular/uniform sampling. A few exceptions exist, e.g., \cite{delvaux2008rank,haupt2010}, but the results are probabilistic and/or usually quite restrictive in practice.

   Following the ideas of CS, \textit{low-rank matrix completion} (LRMC) techniques were proposed for reconstructing matrix signals from a set of samples \cite{candes2009exact,jamali2010matrix}. This line of research utilizes the \textit{rank} of the matrix as complexity measure for sampling and has attracted significant attention, since it is related to a number of important applications such as recommender systems \cite{koren2009matrix}. However, similar to CS, LRMC is based on incoherent sampling. Furthermore the reconstruction guarantees in both CS and LRMC are probabilistic, contrary to the Shannon-Nyquist theorem which deterministically guarantees signal reconstruction.

Our work is motivated by the following question. \textit{Is there a sub-Nyquist sampling mechanism that works under regular sampling for certain signals of interest}? This research question is very intriguing: regular sampling is efficient, friendly to implementation and often mandatory, and sub-Nyquist sampling is desired since numerous real-world signals are far from being bandlimitted.
	
	In this work, we offer an affirmative to the above research question for a large variety of multidimensional signals. The proposed approach guarantees recoverability under realistic conditions, both generic and deterministic. Specifically, we focus on the problem of sampling tensor signals, i.e., signals whose entries are indexed by three or more coordinates \cite{sidiropoulos2017tensor,kolda2009tensor}. Tensor signals naturally arise in a large number of areas such as machine learning and data analytics \cite{papalexakis2017tensors}, signal processing and communications \cite{fu2015factor}, image processing and remote sensing \cite{kanatsoulishsrtsp,kanatsoulis2018hyperspectralicassp,kanatsoulis2018hyperspectralicip}, medical imaging \cite{chatzichristos2018blind}, genomics \cite{hore2016tensor}, chemometrics \cite{smilde2005multi}, just to name a few. 
	Hence, considering sampling and reconstruction of tensor signals is of broad interest.
	The problem is challenging, since various tensor signals are neither bandlimited, sparse, nor low-rank matrices (via `unfolding')---and thus existing sampling techniques are not always applicable. 

   The reconstruction of sampled tensor signals, known in the literature as {\em tensor completion}, has been studied in machine learning and computer vision \cite{karatzoglou2010multiverse,liu2013tensor}. The majority of existing works \cite{liu2013tensor,gandy2011tensor,vervliet2017canonical,acar2011scalable,yuan2016tensor} focus on the algorithmic aspect of tensor completion. The few that provide recovery guarantees \cite{huang2014provable,zhang65exact} are based on random sampling schemes and/or LRMC ideas, which are not tailored to the tensor specifics. The work that is closest to ours is \cite{sorensen2017fiber}, which offers reconstruction conditions when the tensor `fibers' are sampled. However, the conditions are somehow restrictive, since the rank is constrained to be lower than the fiber dimension, and a variety of other interesting types of regular tensor sampling have not been considered.

		\noindent
		{\textbf{Contributions:}} In this work, we study the task of sampling and reconstruction of signals that are tensors---or {\it tensor sampling} in short.
		We propose a tensor sampling framework that is flexible and easy to implement. Generic as well as deterministic theoretical conditions are derived, under which reconstruction is guaranteed.
		Similar to matrix completion, the sample complexity for tensor signal reconstruction is mainly affected by the canonical polyadic rank and the tensor size---instead of signal bandwidth or sparsity.
		Unlike CS and LRMC, the proposed approach does not require incoherent sampling. Therefore, regular, equispaced and highly structured sampling strategies can be adopted---which has a much broader spectrum of applications in practice. 
		
		Our second major contribution lies in designing accelerated acquisition schemes for \textit{functional magnetic resonance imaging} (fMRI) utilizing the proposed tensor sampling principles.
	    Note that traditional fMRI acquisition is considered an ``agonizingly slow'' scanning process, which strongly motivates exploring appropriate sampling techniques for acceleration. 
	    However, due to hardware limitations, random or incoherent sampling strategies are considered impractical for this task \cite{feinberg2012rapid}.
	    Nevertheless, the proposed tensor sampling framework fits this task very well as fMRI signals are naturally tensors.
	    Extensive simulations using synthetically generated data show that the proposed tensor sampling schemes are promising.
	    More importantly, experiments
	    using real fMRI data demonstrate remarkable acceleration compared to traditional fMRI scanning approaches, without sacrificing reconstruction accuracy.
	    
	    \bigskip
	    
	    An early version of part of this work appears in conference form in {\it Proc. ICASSP 2019} \cite{kanatsoulis2018icassp}. In this journal version, we consider additional sampling schemes, include deterministic reconstruction conditions along with thorough model analysis and proofs. We also design multi-slice fMRI acceleration schemes and conduct detailed experiments.

	\section{Tensor Algebra Preliminaries}
	In this work we heavily use tensor algebra. To facilitate the upcoming discussion we briefly present some preliminary tensor algebra concepts. The reader is referred to \cite{sidiropoulos2017tensor,kolda2009tensor} for further details.
	
\begin{figure}
		\centering
		\includegraphics[width=0.7\linewidth]{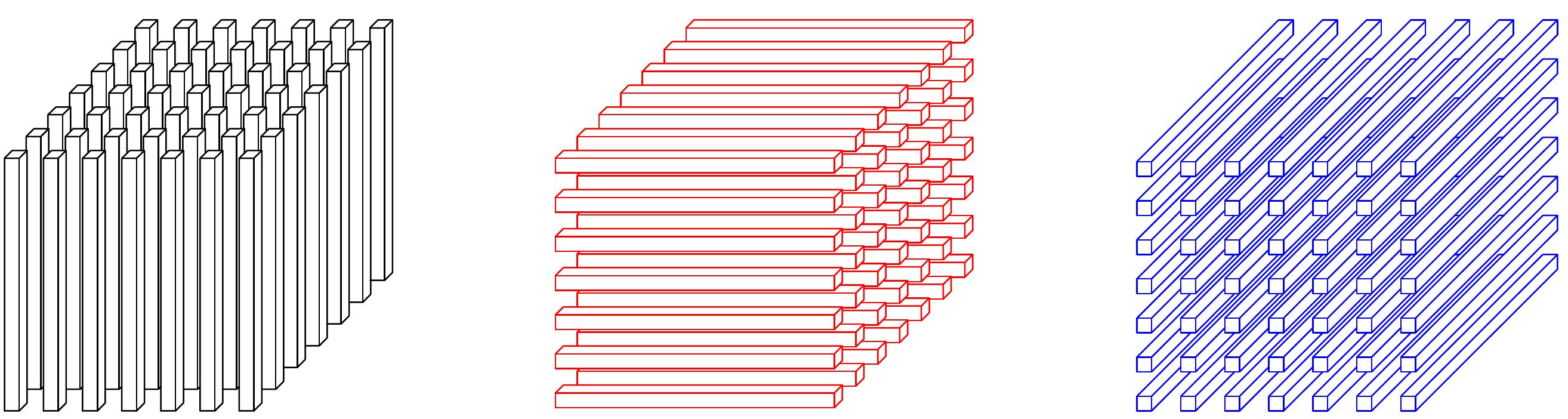}
		\caption{The columns ($\underline{\X}(i,:,j)$), rows ($\underline{\X}(:,j,k)$), and fibers ($\underline{\X}(i,j,:)$) of a third-order tensor, respectively. {Figure taken from \cite{kanatsoulishsrtsp}, \copyright ~ IEEE, 2019. Permission will be sought for reuse.}}
		\label{fig:modes}
	\end{figure}

\begin{figure}
		\centering
		\includegraphics[width=0.7\linewidth]{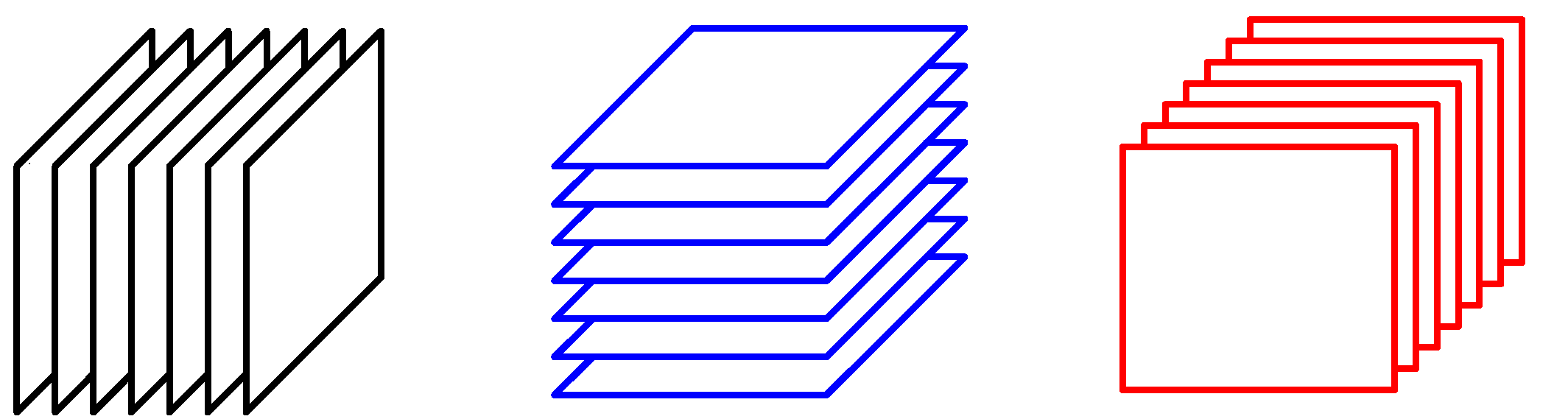}
		\caption{The vertical ($\underline{\X}(:,j,:)$), horizontal ($\underline{\X}(i,:,:)$), and frontal slabs ($\underline{\X}(:,:,k)$) of a third-order tensor, respectively.}
		\label{fig:modes2}
	\end{figure}

	A third-order tensor $\underline{\bm X}\in\mathbb{C}^{I\times J\times K}$ is a three-way array indexed by $i,j,k$ with elements $\underline{\bm X}(i,j,k)$. It consists of three modes: columns $\underline{\bm X}(i,:,k)$, rows $\underline{\bm X}(:,j,k)$, fibers $\underline{\bm X}(i,j,:)$; and three types of slabs: horizontal $\underline{\bm X}(i,:,:)$, vertical $\underline{\bm X}(:,j,:)$ and frontal $\underline{\bm X}(:,:,k)$  -- see Fig. \ref{fig:modes}, \ref{fig:modes2}, respectively. A rank-one tensor $\underline{\bm Z}\in\mathbb{C}^{I\times J\times K}$ is the outer product of three vectors, $\bm a\in\mathbb{C}^{I},~\bm b \in\mathbb{C}^{J},~\bm c \in\mathbb{C}^{K}$, denoted as $\underline{\bm Z}=\bm a\circ \bm b\circ \bm c$, where $\circ$ is the outer product operator. Any tensor can be realized as a sum of three way outer products (rank one tensors), i.e. 
	\begin{equation}\label{PD}
	\underline{\bm X}=\sum_{f=1}^F{\bm a}_f\circ{\bm b}_f\circ{\bm c}_f.
	\end{equation}
	The above expression is known as the polyadic decomposition (PD) of a third-order tensor. If $F$ denotes the minimum number of outer products needed to synthesize $\underline{\X}$, then $F$ is called \textit{tensor rank} or \textit{CP rank} and the decomposition is known as \textit{canonical polyadic decomposition} (CPD) or \textit{parallel factor analysis} (PARAFAC) \cite{harshman1994parafac}. The CPD elementwise representation can be written as $\underline{\bm X}(i,j,k)=\sum_{f=1}^F{\bm A}(i,f){\bm B}(j,f){\bm C}(k,f),$ where $\bm{A}=[\bm a_1,\dots,\bm a_F]\in\mathbb{C}^{I\times F},~\bm{B}=[\bm b_1,\dots,\bm b_F]\in\mathbb{C}^{J\times F},~\bm{C}=[\bm c_1,\dots,\bm c_F]\in\mathbb{C}^{K\times F}$ are called the low rank factors of the tensor. A third-order tensor can be fully characterized by its latent factors, thus we adopt the notation $\underline{\bm X}=\left\llbracket{\bm A},{\bm B},{\bm C}\right\rrbracket$ to represent the tensor.  
	A striking property of tensors is that the CPD is essentially unique under mild conditions. A generic result on the uniqueness of the CPD is as follows.
	
	\begin{Theorem}\label{thm:CPD_generic}
		\cite{chiantini2012generic} 
		Let $\underline{\bm X}=\left\llbracket{\bm A},{\bm B},{\bm C}\right\rrbracket$ with $\bm A : I\times F$, $\bm B : J\times F$, and $\bm C : K\times F$. Assume that ${\bm A}$, ${\bm B}$ and ${\bm C}$ are drawn from some joint absolutely continuous distribution. Also assume $I\geq J\geq K$ without loss of generality. If $F \leq 2^{\lfloor\log_2 J\rfloor+\lfloor\log_2 K\rfloor-2}$, then the decomposition of $\underline{\bm X}$ in terms of $\bm A, \bm B$, and $\bm C$ is essentially unique, almost surely.
	\end{Theorem}

	Here, essential uniqueness means that if $\tilde{\bm A},\tilde{\bm B},\tilde{\bm C}$ also satisfy $\underline{\bm X}=\llbracket \tilde{\bm A},\tilde{\bm B},\tilde{\bm C}\rrbracket$, then ${\bm A}=\tilde{\bm A}{\bm \Pi}{\bm \Lambda}_1$, ${\bm B}=\tilde{\bm B}{\bm \Pi}{\bm \Lambda}_2$, and ${\bm C}=\tilde{\bm C}{\bm \Pi}{\bm \Lambda}_3$, where ${\bm \Pi}$ is a permutation matrix and ${\bm \Lambda}_i$ is a full rank diagonal matrix such that ${\bm \Lambda}_1{\bm \Lambda}_2{\bm \Lambda}_3={\bm I}$.
    
As far as deterministic identifiability is concerned, we have:
\begin{Theorem}\label{theorem:deterministic_CP}
	\cite{kruskal1977three}
	Let $\underline{\bm X}=\left\llbracket{\bm A},{\bm B},{\bm C}\right\rrbracket$ with $\bm A : I\times F$, $\bm B : J\times F$, and $\bm C : K\times F$. The decomposition $\underline{\bm X}=\left\llbracket{\bm A},{\bm B},{\bm C}\right\rrbracket$ is essentially unique with CP rank $F$ if $k_{\A}+k_{\B}+k_{\C}\geq 2F+2$.
\end{Theorem}    
 Here $k_{\A}$ denotes the Kruskal rank of a matrix, i.e., the largest integer $k_{\A}$ such that \textit{any} $k_{\A}$ columns of $\A$ are linearly independent.    
  
	A tensor can be represented in a matrix form using the \textit{matricization} operation. There are three common ways to matricize (or unfold) a third-order tensor, by stacking columns, rows, or fibers of the tensor to form a matrix. For example the following operation stacks the fibers of tensor $\underline{\X}$ as rows of matrix $\X^{(3)}$:
    \begin{align}\label{X3}
	\X^{(3)}:=[\text{vec}(\underline{\bm X}(:,:,1)), \text{vec}(\underline{\bm X}(:,:,2)), \dots, \text{vec}(\underline{\bm X}(:,:,K))],\nonumber
	\end{align}
	where `${\rm vec}(\cdot)$' is the vectorization operator. One can see that: 
	\begin{equation}\label{X3v2}
	\X^{(3)}=({\bm B}\odot{\bm A}){\bm C}^T\in\mathbb{C}^{IJ\times K},
	\end{equation}
	where $\odot$ denotes the Khatri-Rao (column-wise Kronecker) product. The superscript $(3)$ denotes that the unfolding is performed on the third mode of the tensor, i.e. fibers are stacked together.
	
	 Finally we will need the \textit{mode product} operation in tensor analytics. The mode product operator multiplies a matrix to a tensor in a single mode. A third order tensor has three modes (rows, columns, fibers), thus three different mode products are defined. A joint mode-1, mode-2, and mode-3 product of a third-order tensor is represented by the following notation:
	\begin{equation}\label{modeprod}
	\tilde{\underline{\bm X}}=\underline{\bm X}\times_1 {\bm P}_1\times_2 {\bm P}_2\times_3 {\bm P}_3
	\end{equation}
	where ``$\times_1$'' denotes the operation that multiplies each column of $\underline{\bm X}$ with ${\bm P}_1$, ``$\times_2$'' denotes multiplying each row of $\underline{\bm X}$ with ${\bm P}_2$, and ``$\times_3$'' denotes multiplying each fiber of $\underline{\bm X}$ with ${\bm P}_3$. The mode product is reflected in the polyadic decomposition of the tensor, i.e., the outcome of \eqref{modeprod} results in a tensor $\underline{\tilde{\bm X}}$ with polyadic decomposition:
	\[\underline{\tilde{\bm X}}=\llbracket \bm P_1{\bm A},\bm P_2{\bm B},\bm P_3{\bm C}\rrbracket,\]
	The above decomposition is essentially unique under some conditions---this point will turn out to be crucial in the upcoming discussion.

	\section{Tensor Sampling Mechanisms}
	The core of this work discusses the sampling and reconstruction of third-order tensors. The main claim is fundamental: roughly speaking, any third-order tensor that does not have very high rank can always be recovered from a sufficient number of regular samples. The sampling is not constrained to follow a randomized or incoherent process. On the contrary, we focus on {\em regular} and highly structured schemes. Various regular sampling strategies are considered. They involve sampling whole slabs in different modes (slab sampling), certain fibers in a single or multiple modes (fiber sampling) and entries in a systematic manner (entry sampling). Exposition and development use \textit{third-order} tensors, but all the techniques can be naturally extended to higher-order tensors in a conceptually straightforward way. Similar to the case of matrices, even if a tensor is high-rank in the strict mathematical sense, it can often be approximated using low rank, in which case it can be {\em approximately} recovered using the proposed sampling and reconstruction schemes, as we will see. 
	
    \subsection{General Strategy and Insight}
    Let us consider the following general form of tensor sampling:
    \[       \bm y = {\sf Sample}(\underline{\bm X}),    \]	
    where ${\sf Sample}(\cdot):\mathbb{C}^{I\times J\times K}\rightarrow \mathbb{C}^{L}$ is a down-sampling operator with $L\ll IJK$. Our goal is to study under what conditions and sampling strategies, recovering $\underline{\bm X}$ from $\bm y$ is possible. This is an inverse problem like in CS \cite{candes2006robust,donoho2006compressed,candes2008introduction} and LRMC \cite{candes2009exact,jamali2010matrix}. However, unlike in \cite{candes2006robust,donoho2006compressed,candes2008introduction,candes2009exact,jamali2010matrix,gandy2011tensor}, we do not consider random/incoherent down-sampling operators but highly structured ones---which model a plethora of engineering applications, are easier for practical system implementation and computationally more efficient.
    
    Our work rests upon two basic ideas. The first utilizes the uniqueness property of the CPD. Recall that every tensor admits a CPD, and the CPD is essentially unique if the CP rank is not very large (in many cases: not full-rank). The second exploits the relation between a sampled sub-tensor and the original tensor $\underline{\bm X}=\llbracket {\bm A},{\bm B},{\bm C}\rrbracket$. If we sample ${\cal S}_r\subseteq\{1,\ldots,I\}$ rows, ${\cal S}_c\subseteq\{1,\ldots,J\}$ columns and ${\cal S}_f\subseteq\{1,\ldots,K\}$ fibers and form a sub-tensor $\underline{\bm X}({\cal S}_r,{\cal S}_c,{\cal S}_f)$, then:
    \[  \underline{\bm X}({\cal S}_r,{\cal S}_c,{\cal S}_f)= \llbracket {\bm A}({\cal S}_r,:),{\bm B}({\cal S}_c,:),{\bm C}({\cal S}_f,:)\rrbracket.\]
    One key observation is that the above sub-tensor can be decomposed to a sum of rank-one terms of number equal to the rank of the original tensor. Furthermore, the latent factors share certain rows with the original latent factors.
    Intuitively, if ${\rm rank}(\underline{\bm X})$ is not huge, there is a good chance that the sub-tensor admits a unique CPD, and part of the information of $\bm A$, $\bm B$, and $\bm C$ can be extracted from the sub-tensor. Hence, by judiciously sampling and constructing sub-tensors, it seems viable to recover the entire $\bm A$, $\bm B$, and $\bm C$, and thus reconstruct $\underline{\bm X}$. This is the main idea. 
    
 Despite this conceptual simplicity, however, fleshing out this task is nontrivial. First, when factoring the sub-tensors, there are always permutation and scaling ambiguities---even if every sub-tensor admits unique CPD, reconstruction of the whole tensor is not guaranteed. Thus the sampling mechanisms need to be carefully designed to address this issue. Second, balancing the sampling ratio with the ability to identify the original tensor is a key consideration and needs attentive thinking and design. In the remaining section, we propose a series of sampling mechanisms that take into consideration both design challenges. The considered sampling schemes are practical and motivated by real engineering applications, particularly in the field of medical imaging.

	\subsection{Slab sampling}
	First, we study the task of reconstructing a third-order tensor from slab samples, taken from two different modes. Recovering tensor signals from sampled slabs finds applications in fMRI acceleration \cite{moeller2010multiband,smith2013resting,chiew2018recovering} and image/video inpainting \cite{bertalmio2000image,patwardhan2007video}. However, there is no unified characterization for recoverability under regular sampling patterns, to our best knowledge.
	\begin{figure}[h!]
		\centering
		\begin{subfigure}[b]{0.39\linewidth}
			\includegraphics[width=\linewidth]{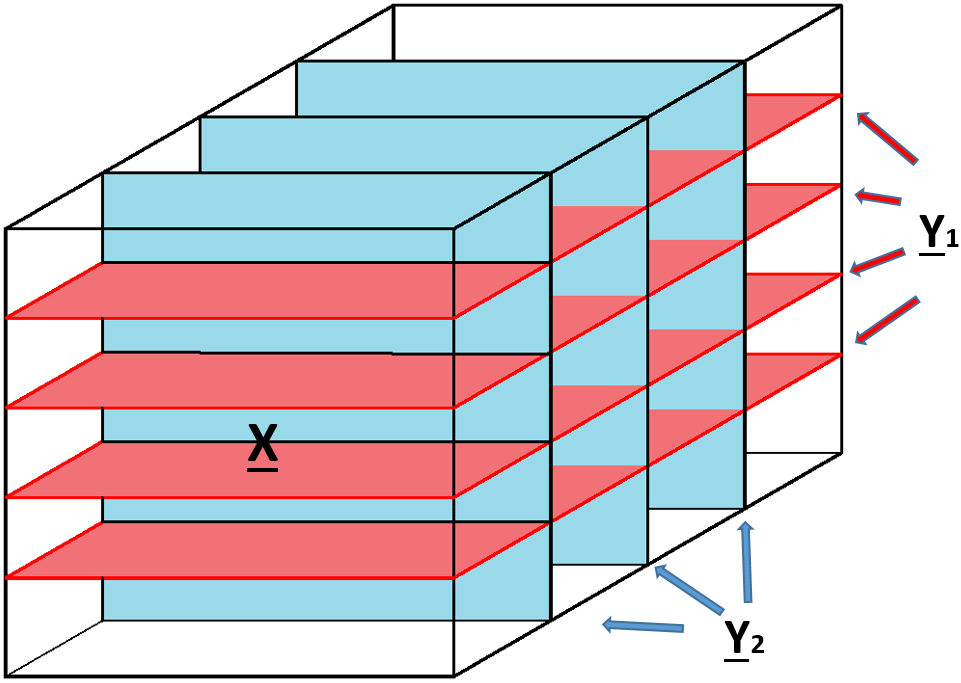}
		\end{subfigure}
		\begin{subfigure}[b]{0.39\linewidth}
			\includegraphics[width=\linewidth]{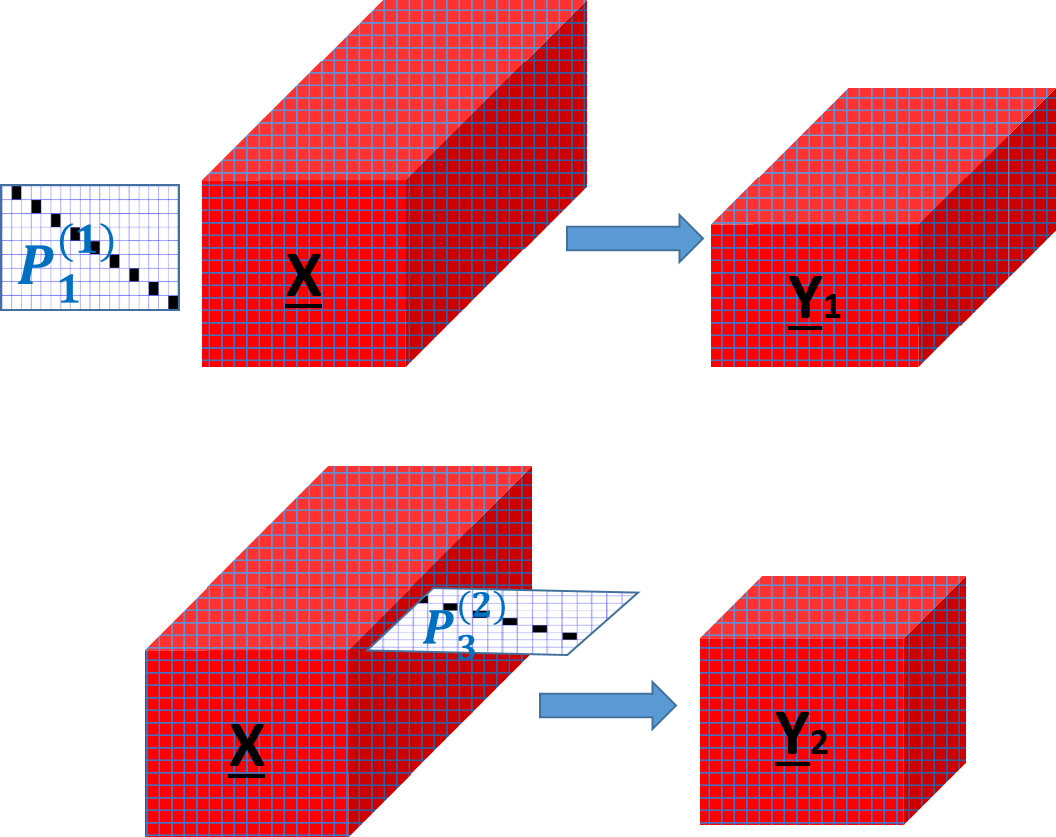}
		\end{subfigure}
		\caption{Tensor slab sampling paradigm.}
		\label{fig:paradigm1}
	\end{figure}
	Let $\underline{\bm X}\in\mathbb{C}^{I\times J \times K}$ be the original full tensor, which is not fully accessible or is subject to sampling. Instead we sample/observe a subset of slabs in one mode, e.g., horizontal slabs, $\mathcal{S}_r\subseteq \{1, \ldots, I\},$ and a subset of slabs in a different mode, e.g. frontal slabs, $\mathcal{S}_f\subseteq \{1, \ldots, K\}$. If $|\mathcal{S}_r|=I_1\geq 2$ and $|\mathcal{S}_f|=K_2\geq 2$, two separate sampled tensors are formed, i.e.,
 $\underline{\bm Y}_{1}\in\mathbb{C}^{I_1\times J \times K}$ and $\underline{\bm Y}_{2}\in\mathbb{C}^{I\times J \times K_2}$, which represent the subset of observable horizontal and frontal slabs of $\underline{\bm X}$ respectively. Apparently, {$\underline{\bm Y}_{1}$ can be written as the mode $1$ multiplication of tensor $\underline{\bm X}$ with selection matrix $\bm{P}_1^{(1)}\in\mathbb{R}^{I_1\times I}$, i.e. 
	\begin{equation}\label{compressed1}
	{\underline{\bm Y}}_{1}=\underline{\bm X}\left(\mathcal{S}_h,:,:\right)=\underline{\bm X}\times_1 {\bm P}_1^{(1)}
	\end{equation}
	and $\underline{\bm Y}_{2}$ as a mode $3$ multiplication with matrix $\bm{P}_3^{(2)}\in\mathbb{R}^{K_2\times K}$, i.e.
	\begin{equation}\label{compressed2}
	{\underline{\bm Y}}_{2}=\underline{\bm X}\left(:,:,\mathcal{S}_f\right)=\underline{\bm X}\times_3 {\bm P}_3^{(2)}
	\end{equation}
	The sampling matrices $\bm{P}_1^{(1)},~\bm{P}_3^{(2)}$ perform slab selection in a single mode of $\underline{\bm X}$, thus $I_1<I,~K_2<K$ (they are `fat') and also have full row rank.
	A schematic illustration of the tensor slab sampling model is given in Fig. \ref{fig:paradigm1}. 
	Note that, $\bm{P}_1^{(1)},~\bm{P}_3^{(2)}$ are not constrained to be randomly drawn in our framework. On the contrary, the sampling process is allowed to be regular or highly structured, see Fig. \ref{fig:paradigm1}. Assuming 
	$\underline{\bm X} = \left\llbracket{\bm A},{\bm B},{\bm C}\right\rrbracket$, following \eqref{compressed1}, \eqref{compressed2}, the sub-tensors $\underline{\bm Y}_{1},~\underline{\bm Y}_{2}$ can be expressed in a PD form:
\begin{subequations}\label{eq:para1}
	\begin{align}
	\underline{\bm Y}_{1} &= \left\llbracket\bm P_1^{(1)}\bm A,{\bm B},{\bm C}\right\rrbracket\\
	\underline{\bm Y}_2 &= \left\llbracket{\bm A},{\bm B},\bm{P}_3^{(2)}{\bm C}\right\rrbracket
	\end{align}
\end{subequations}
	Using \eqref{eq:para1} identifiability of $\underline{\X}$ from $({\bm Y}_{1}, {\bm Y}_{2})$  can be established: 
	\begin{Theorem}\label{theorem:generic_slab}
		Let $\underline{\bm X}\in\mathbb{C}^{I\times J \times K}$ be the original tensor signal to recover, with CPD $\underline{\bm X} = \left\llbracket {\bm A}, {\bm B},{\bm C}\right\rrbracket$ of rank $F$. Assume that ${\bm A}$, ${\bm B}$ and ${\bm C}$ are drawn from a joint absolutely continuous distribution over $\mathbb{C}^{(I+J+K)F}$, and that ${\bm A}^\star,{\bm B}^\star,{\bm C}^\star$ satisfy the equations in~\eqref{eq:para1}. Then, $\underline{\hat{\bm X}}(i,j,k)=\sum_{f=1}^F{\bm A}^\star(i,f){\bm B}^\star(j,f){\bm C}^\star(k,f)$ recovers the ground-truth $\underline{\bm X}$ almost surely if {$\min\left\lbrace{ I_1J}, { JK}, { I_1K}, 4JK_2\right\rbrace \geq 4\lceil F\rceil$} or
		{$\min\left\lbrace{ IJ}, { JK_2}, {IK_2}, 4I_1J\right\rbrace \geq 4\lceil F\rceil$}; where $\lceil x\rceil$ denotes the closest power of $2$ which is greater than $x$.
	\end{Theorem}
The proof is presented in Appendix \ref{appendix:proof_slab}.
 The intuition is that if $\underline{\Y}_1$ or $\underline{\Y}_2$ admit a unique CPD, under Theorem \ref{thm:CPD_generic}, the factors $\B,~\C$ or $\A,~\B$ respectively can be identified. Then $\A$ or $\C$ are recovered from the other tensor, where $\A$ or $\C$ have been left uncompressed. Note that in slab sampling only one sub-tensor is required to admit a unique CPD. The reason is that identifying the latent factors of one sub-tensor, directly estimates two original latent factors. Then, the remaining factor can be obtained via solving a linear system of equations. Furthermore, permutation and scaling ambiguities are automatically resolved, since $\underline{\Y}_1,~\underline{\Y}_2$ sample common rows of $\underline{\X}$. Overall reconstruction is performed as $\underline{\hat{\bm X}}(i,j,k)=\sum_{f=1}^F{\bm A}^\star(i,f){\bm B}^\star(j,f){\bm C}^\star(k,f)$. Deterministic conditions can also be derived, and this discussion is postponed to the following section.
	
The previous analysis can be easily extended to the case where slab sampling is performed in all 3 modes of the tensor.

	\subsection{Fiber sampling}\label{model:fiber}
	 Next, we consider the reconstruction of tensor $\underline{\X}$ from a subset of fibers, sampled along a single mode of the tensor. Fiber sampling is also of interest to a number of applications in chemometrics \cite{acar2011scalable,tomasi2005parafac} nuclear magnetic resonance spectroscopy \cite{orekhov2003optimizing} and fMRI acceleration (see Sec. \ref{fMRI}). To make the discussion concrete, consider the scenario illustrated in Fig. \ref{fig:fiber_new}, where $D=3$ fiber patterns appear in the tensor sampling scheme.
	\begin{figure}[h!]
		\centering
		\includegraphics[width=0.6\linewidth]{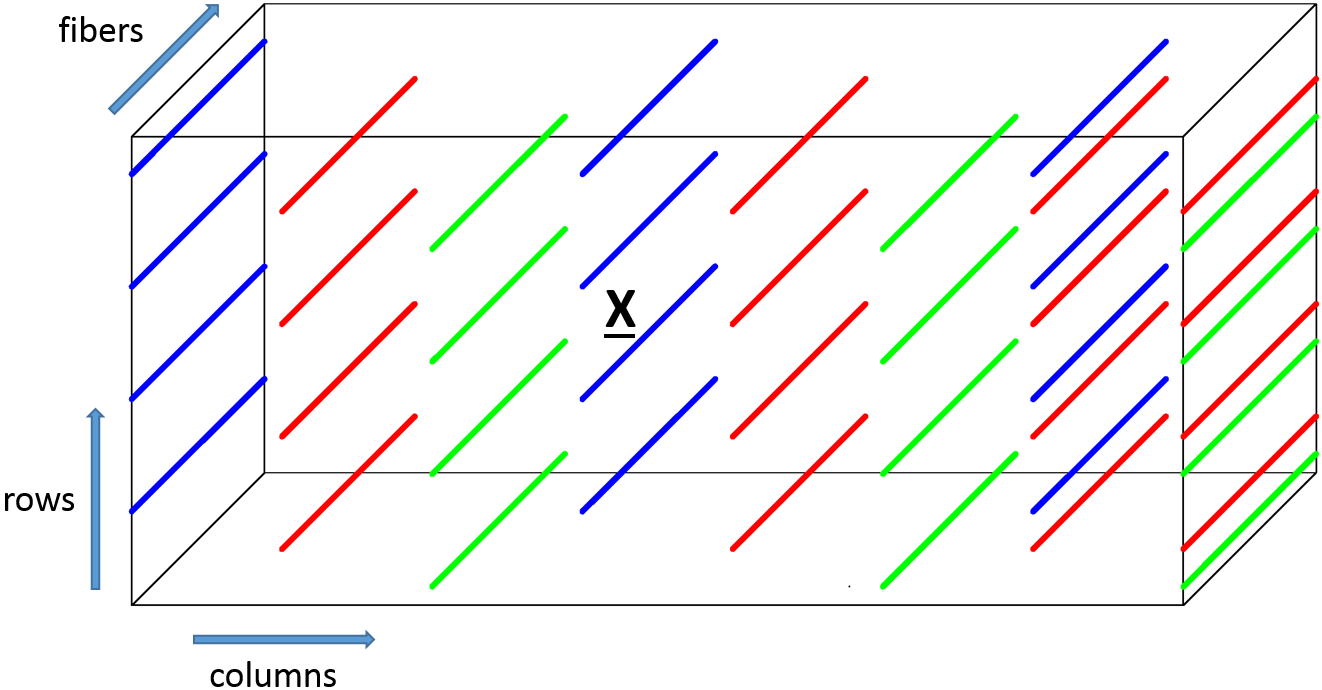}
		
		\caption{Tensor fiber sampling paradigm.}
		\label{fig:fiber_new}
	\end{figure}
	A pattern will be defined as a subset of rows $\mathcal{S}_r^{(d)}\subseteq \{1, \ldots, I\}$ and columns $\mathcal{S}_c^{(d)}\subseteq \{1, \ldots, J\}$ for which every point $\underline{\X}(i,j,k),~i\in\mathcal{S}_r,~j\in\mathcal{S}_c,~ k\in\left\lbrace1,\dots,K\right\rbrace$ belongs to the pattern. 
	In the illustrated scenario, each pattern (blue, d=1; red, d=2; green, d=3) samples fibers defined by the following subset of rows and columns: $\mathcal{S}_r^{(1)}= \{1,4,7,10\},~\mathcal{S}_c^{(1)}= \{1,4,7\},~\mathcal{S}_r^{(2)}= \{2,5,8,11\},\mathcal{S}_c^{(2)}= \{2,5,7,8\},~\mathcal{S}_r^{(3)}= \{3,6,9,12\},\mathcal{S}_c^{(3)}= \{3,6,8\}$. Rearranging the order of the columns results in the model shown in Fig. \ref{fig:paradigm2}. 
	\begin{figure}[h!]
		\centering
		\begin{subfigure}[t]{0.54\linewidth}
			\includegraphics[width=\linewidth]{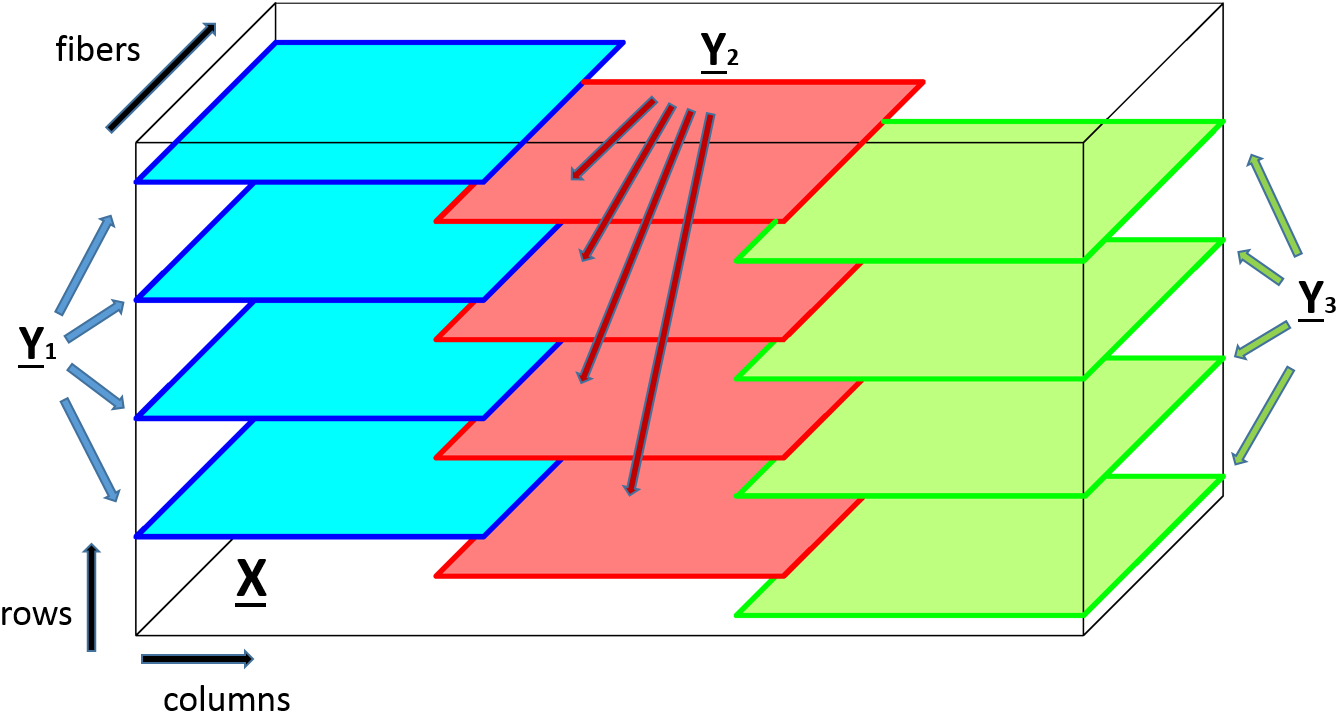}
		\end{subfigure}
		\begin{subfigure}[t]{0.44\linewidth}
			\includegraphics[width=\linewidth]{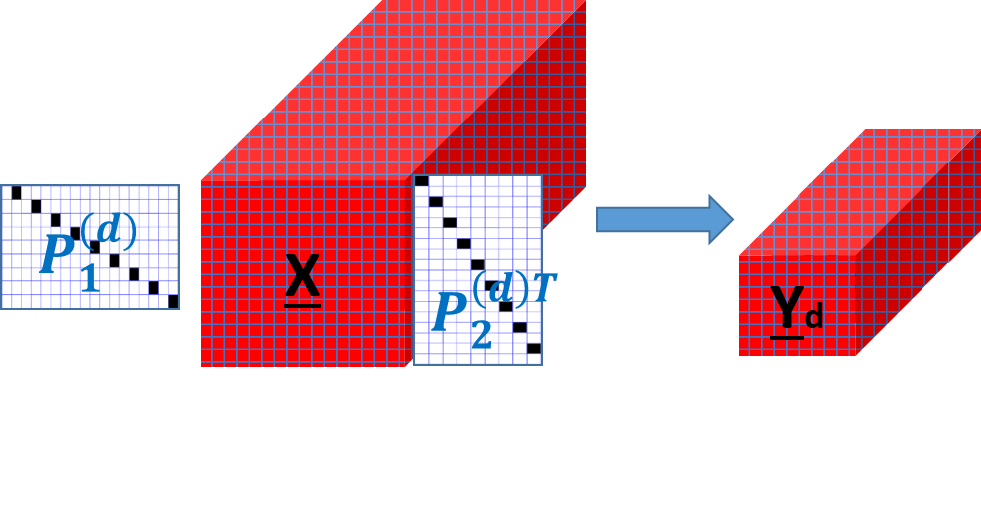}
		\end{subfigure}
		\caption{Fiber sampling model in a single mode}
		\label{fig:paradigm2}
		\vspace{-0.25cm}
	\end{figure}
	
	In the general case, the proposed fiber sampling framework entails each pattern forming a third-order tensor, i.e., $|\mathcal{S}_r^{(d)}|,~|\mathcal{S}_c^{(d)}|\geq 2$ and that samples are taken from every row and column of the tensor. The latter is a necessary condition for every factorization based completion approach, since completely unobserved slabs are impossible to recover. Furthermore, each pattern is required to sample from a common row or column with at least one more, thus creating an overlapping chain between patterns. The reason is that for pairwise mutually exclusive patterns, there exists a non-trivial scaling ambiguity, which cannot be determined. 
	Formally the necessary sampling rules, for the proposed fiber sampling framework, are expressed as:
	\begin{subequations}\label{constraints_fiber1}
		{\small\begin{align}
		&|\mathcal{S}_r^{(d)}|,~|\mathcal{S}_c^{(d)}|\geq 2
		\label{rule:patterns}\\
		&\bigcup\limits_{d=1}^{D} \mathcal{S}_r^{(d)} = \{1, \ldots, I\},~ \bigcup\limits_{d=1}^{D} \mathcal{S}_c^{(d)} = \{1, \ldots, J\}\label{rule:mandatory}\\
		&\bigcap\limits_{d=1}^{D}\bigcup\limits_{d^{\prime}\neq d}\mathcal{S}_c^{(d)}\cap\mathcal{S}_c^{(d^{\prime})}\bigcup\mathcal{S}_{r}^{(d)}\cap\mathcal{S}_{r}^{(d^{\prime})}\neq\emptyset,\label{rule:graph}
		\end{align}}\noindent
	\end{subequations}\noindent
where $d,d^{\prime}\in\left\lbrace 1,\dots D\right\rbrace$. The rules in \ref{constraints_fiber1} handle a plethora of sampling schemes. Specifically, each pattern is allowed to be equispaced, regular, random etc. This shows that reconstruction from regular samples is indeed doable. The sampling in Fig. \ref{fig:fiber_new}, \ref{fig:paradigm2}, for example, is regular and each pattern consists of equispaced rows and deterministically spaced columns.

	Following similar analysis as in slab sampling, let $\underline{\bm Y}_{d}\in\mathbb{C}^{I_d\times J_d\times K}$ be the sampled subtensor, formed by pattern $d$. Also let ${\bm P}_1^{(d)}\in\mathbb{R}^{I_d\times I},~{\bm P}_2^{(d)}\in\mathbb{R}^{J_d\times J}$ be the row and column selection matrices determining the $d$ pattern. Then $\underline{\bm Y}_{d}$ is written as follows. 

	\begin{align}\label{eq:para2}
			\underline{\bm Y}_{d} =&\underline{\bm X}\left(\mathcal{S}_r^{(d)},\mathcal{S}_c^{(d)},:\right)= \underline{\bm X}\times_1 {\bm P}_1^{(d)}\times_2 {\bm P}_2^{(d)}\nonumber\\ =& \left\llbracket\bm P_1^{(d)}\bm A,\bm P_2^{(d)}{\bm B},{\bm C}\right\rrbracket
		~~~~~~	d=1,\ldots,D
	\end{align}

Using the equation in \eqref{eq:para2} we can establish generic identifiability of fiber sampling as: 
\begin{Theorem}\label{theorem:generic_fiber}
	Let $\underline{\bm X}\in\mathbb{C}^{I\times J \times K}$ be the original tensor signal, fiber sampled according to \eqref{constraints_fiber1}, with CPD $\underline{\bm X} = \left\llbracket {\bm A}, {\bm B},{\bm C}\right\rrbracket$ of rank $F$. Assume that ${\bm A}$, ${\bm B}$ and ${\bm C}$ are drawn from a joint absolutely continuous distribution over $\mathbb{C}^{(I+J+K)F}$, and that ${\bm A}^\star,{\bm B}^\star,{\bm C}^\star$ satisfy the equations in~\eqref{eq:para2}. Then, $\underline{\hat{\bm X}}(i,j,k)=\sum_{f=1}^F{\bm A}^\star(i,f){\bm B}^\star(j,f){\bm C}^\star(k,f)$ recovers the ground-truth $\underline{\bm X}$ almost surely if {$\min_d\left\lbrace{I_dJ_d}, { J_dK}, {I_dK}\right\rbrace \geq 4\lceil F\rceil$}; where $\lceil x\rceil$ denotes the closest power of $2$ which is greater than $x$.
\end{Theorem}
The proof is relegated to Appendix \ref{appendix:proof_generic_fiber_entry}.
In contrast to the previous case of slab sampling, where identifiability of one sampled tensor $\underline{\Y}_i$ is enough, fiber sampling requires all $\underline{\Y}_i$'s to admit a unique CPD model---otherwise certain rows of $\A,~\B$ would be impossible to identify. The claim is simple and intuitive: The number of samples required to recover a fiber sampled tensor are proportional to the rank of the tensor.
\begin{Remark} Theorem \ref{theorem:generic_fiber} studies general tensors where factor $\C$ is not required to have full column rank, and thus $K<F$ can be easily handled. Fiber sampling and recovery of tensors with $\C$ having full column rank, is extensively studied in \cite{sorensen2017fiber}. Compared to our work, the sampling strategy therein has to follow rules \eqref{rule:mandatory}, \eqref{rule:graph}, whereas \eqref{rule:patterns} can be relaxed. On the other hand, the results of this paper are tailored to cases where the sampling process exhibits some regularity and $K<F$ is allowed. Note that $\C$ being full column rank, which is mandatory in \cite{sorensen2017fiber}, is a quite restrictive condition and prohibitive for several applications, e.g., fMRI acceleration as we will see next. 
\end{Remark}
	\subsection{Entry sampling}\label{model:entry}
	So far we have discussed slab, fiber sampling of third-order tensors and provided conditions under which reconstruction is guaranteed. In this subsection, we move a step further and study the more general problem of tensor reconstruction from a subset of entries, sampled in a regular fashion along the tensor. Entry sampling is another important sampling mechanism, which along with fiber sampling will prove very useful in accelerating the fMRI scan acquisition (see Sec. \ref{fMRI}). 
	
	We are interested in cases where the sampling process can be viewed as a series of patterns. A pattern is defined, similarly to fiber sampling, as a subset of rows $\mathcal{S}_r^{(d)}\subseteq \{1, \ldots, I\},$ columns $\mathcal{S}_c^{(d)}\subseteq \{1, \ldots, J\}$ and fibers $\mathcal{S}_f^{(d)}\subseteq \{1, \ldots, K\}$, for which every point $\underline{\X}(i,j,k),~i\in\mathcal{S}_r, j\in\mathcal{S}_c, k\in\mathcal{S}_f$ belongs to the pattern. For example consider the scenario illustrated in Fig. \ref{fig:entry_sampling1}.
	\begin{figure}[h!]
		\centering
		\begin{subfigure}[b]{0.6\linewidth}
			\includegraphics[width=\textwidth]{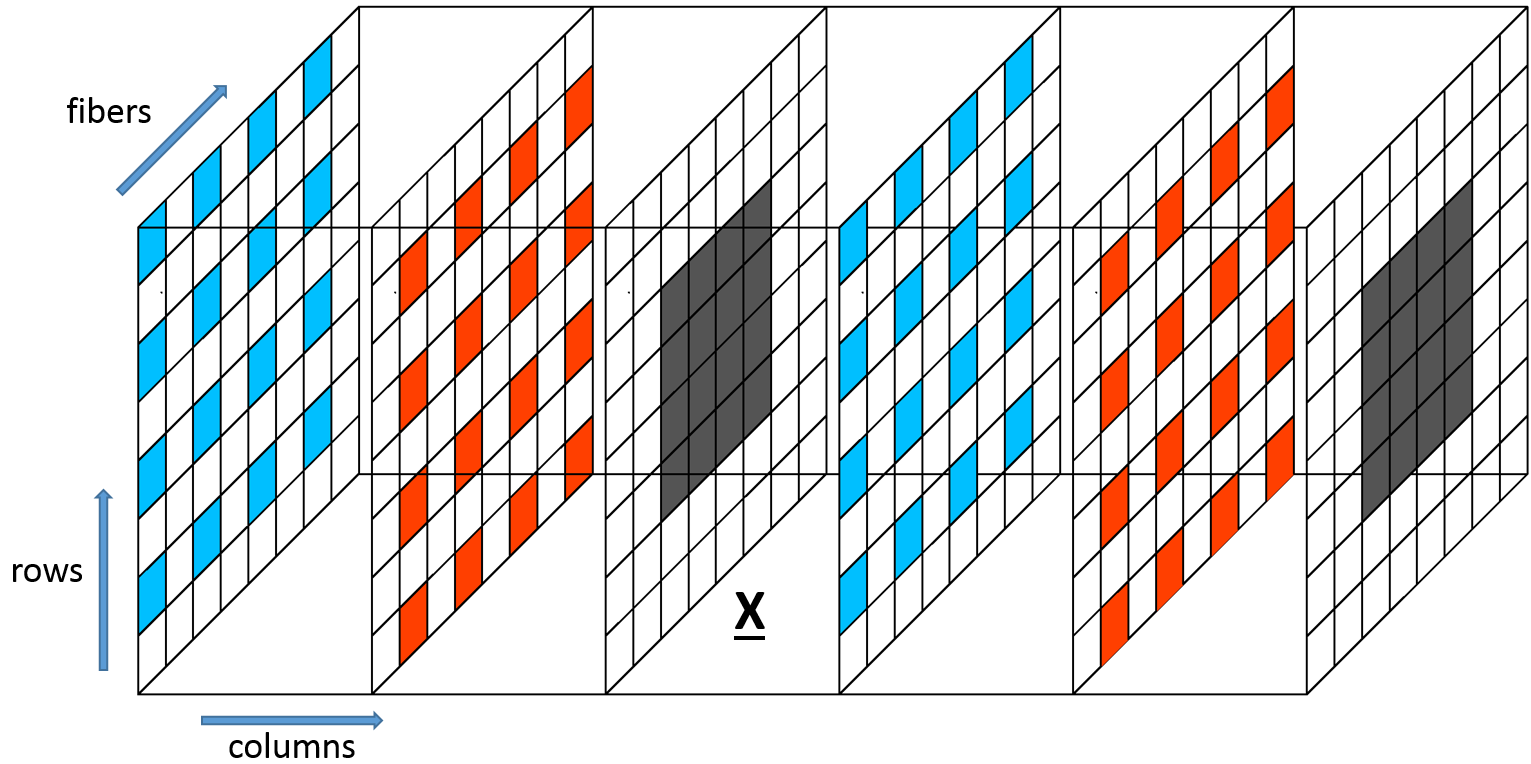}
		\end{subfigure}
		\begin{subfigure}[b]{0.38\linewidth}
			\includegraphics[width=\textwidth]{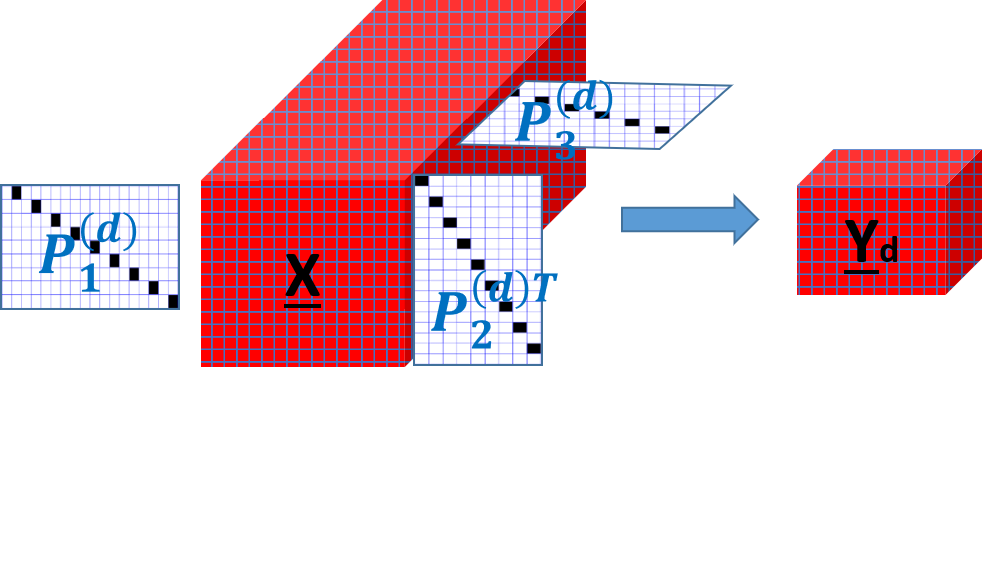}
		\end{subfigure}
		\caption{Tensor entry sampling paradigm.}
		\label{fig:entry_sampling1}
	\end{figure}
	The number of patterns is $D=3$ and $\mathcal{S}_r^{(1)}=\mathcal{S}_c^{(1)}= \{1,3,5,7\},~\mathcal{S}_f^{(1)}= \{1,4\},~\mathcal{S}_r^{(2)}=\mathcal{S}_c^{(2)}= \{2,4,6,8\},\mathcal{S}_f^{(2)}= \{2,5\},~\mathcal{S}_r^{(3)}=\mathcal{S}_c^{(3)}= \{3,4,5,6\},\mathcal{S}_f^{(3)}= \{3,6\}$. In general, the proposed framework requires samples to be taken from all rows, columns and fibers of the tensor, thus $D\geq 2$ and each pattern should include $2$ rows, $2$ columns and $2$ fibers at minimum. Furthermore, the patterns need to overlap like a domino, i.e. any pattern should sample from some common rows, columns or fibers with another. Furthermore, the overlap between two patterns is required to involve 2 elements in one mode and 1 element in a different mode. For example, a pair of patterns should sample 2 common rows and 1 common column. The latter is a necessary condition resulting from the inherent permutation and scaling ambiguity of the CPD. Formally the rules of entry sampling are:
	\begin{subequations}\label{constraints_entry1}
		{\small\begin{align}
			&|\mathcal{S}_r^{(d)}|,~|\mathcal{S}_c^{(d)}|,~|\mathcal{S}_f^{(d)}|\geq 2
			\label{rule:patterns2}\\
			&\bigcup\limits_{d=1}^{D} \mathcal{S}_r^{(d)} = \{1, \ldots, I\},~ \bigcup\limits_{d=1}^{D} \mathcal{S}_c^{(d)} = \{1, \ldots, J\},\nonumber\\ &\bigcup\limits_{d=1}^{D} \mathcal{S}_f^{(d)} = \{1, \ldots, K\}\label{rule:mandatory2}\\
			&\bigcup\limits_{m\neq m^{\prime}}\bigcap\limits_{d=1}^{D}\bigcup\limits_{d^{\prime}\neq d}\left\lbrace\mathcal{S}_m^{(d)}\cap\mathcal{S}_m^{(d^{\prime})}\bigcap\mathcal{S}_{m^{\prime}}^{(d)}\cap\mathcal{S}_{m^{\prime}}^{(d^{\prime})}\right\rbrace\neq\emptyset
			\label{rule:graph2}
			\\
			&\bigcup\limits_{m}\left\lbrace|\mathcal{S}_m^{(d)}\cap\mathcal{S}_{m}^{(d^{\prime})}|\geq 2\right\rbrace\neq\emptyset,
			\label{rule:perm}
			\end{align}}
	\end{subequations}
\noindent where $d,d^{\prime}\in\left\lbrace 1,\dots D\right\rbrace,~m,m^{\prime}\in\left\lbrace c,r,f\right\rbrace$.
Following similar analysis as in single mode fiber sampling, let $\underline{\bm Y}_{d}\in\mathbb{C}^{I_d\times J_d\times K_d}$ be the sampled subtensor representation of pattern $d$. Also let ${\bm P}_1^{(d)}\in\mathbb{R}^{I_d\times I},~{\bm P}_2^{(d)}\in\mathbb{R}^{J_d\times J},~{\bm P}_3^{(d)}\in\mathbb{R}^{J_d\times J}$ be the row, column and fiber selection matrices determining pattern $d$. Then $\underline{\bm Y}_{d}$ is written as:

\begin{align}\label{model:entry_sampling}
\underline{\bm Y}_{d} =&\underline{\bm X}\left(\mathcal{S}_r^{(d)},\mathcal{S}_c^{(d)},\mathcal{S}_f^{(d)}\right)= \underline{\bm X}\times_1 {\bm P}_1^{(d)}\times_2 {\bm P}_2^{(d)}\times_3 {\bm P}_3^{(d)}\nonumber\\ =& \left\llbracket\bm P_1^{(d)}\bm A,\bm P_2^{(d)}{\bm B},\bm P_3^{(d)}{\bm C}\right\rrbracket
~~~~~~	d=1,\ldots,D
\end{align}

The model in \eqref{model:entry_sampling} is identifiable, under generic conditions presented in the following theorem.
\begin{Theorem}\label{theorem:generic_entry}
	Let $\underline{\bm X}\in\mathbb{C}^{I\times J \times K}$ be the original tensor signal, sampled according to \eqref{constraints_entry1}, with CPD $\underline{\bm X} = \left\llbracket {\bm A}, {\bm B},{\bm C}\right\rrbracket$ of rank $F$. Assume that ${\bm A}$, ${\bm B}$ and ${\bm C}$ are drawn from a joint absolutely continuous distribution over $\mathbb{C}^{(I+J+K)F}$, and that ${\bm A}^\star,{\bm B}^\star,{\bm C}^\star$ satisfy the equations in~\eqref{model:entry_sampling}. Then, $\underline{\hat{\bm X}}(i,j,k)=\sum_{f=1}^F{\bm A}^\star(i,f){\bm B}^\star(j,f){\bm C}^\star(k,f)$ recovers the ground-truth $\underline{\bm X}$ almost surely if {$\min_d\left\lbrace{ I_dJ_d}, { J_dK_d}, { I_dK_d}\right\rbrace \geq 4\lceil F\rceil$}; where $\lceil x\rceil$ denotes the closest power of $2$ which is greater than $x$.
\end{Theorem}
The proof is presented in Appendix \ref{appendix:proof_generic_fiber_entry}.
Similar to fiber sampling, reconstruction of a tensor from entries, sampled as described in \eqref{constraints_entry1}, is guaranteed, if all the sub-sampled tensors formed by the emerging patterns admit a unique CPD.

\section{Deterministic Identifiability and insights}
The sampling mechanisms, discussed so far, can be realized as separate, yet coupled, sub-sampled versions of the original third-order tensor $\underline{\X}$. Recovery of $\underline{\X}$, under various sampling mechanisms, was established by applying generic identifiability results on the CPD of the sub-tensors. However, reconstruction of the original tensor is also guaranteed under purely deterministic conditions. In the case of slab sampling we have the following theorem.
	\begin{Theorem}\label{theorem:deterministic_general}
		Let $\underline{\bm X}\in\mathbb{C}^{I\times J \times K}$ be the original tensor signal to recover, with CPD $\underline{\bm X} = \left\llbracket {\bm A}, {\bm B},{\bm C}\right\rrbracket$ of rank $F$. Assume that ${\bm A}^\star,{\bm B}^\star,{\bm C}^\star$ satisfy the equations in~\eqref{eq:para1}. Then, $\underline{\hat{\bm X}}(i,j,k)=\sum_{f=1}^F{\bm A}^\star(i,f){\bm B}^\star(j,f){\bm C}^\star(k,f)$ recovers the ground-truth $\underline{\bm X}$ if {$2F+2 \leq k_{{\bm P_1^{(1)}\A^{\star}}}+k_{\B^{\star}}+k_{\C^{\star}}$} and $\B^{\star}\odot{\bm P}_3^{(2)}\C^{\star}$ has full column rank, or if
		$2F+2 \leq k_{{\A^{\star}}}+k_{{\B^{\star}}}+k_{\bm P_3^{(2)}\C^{\star}}$ and $\B^{\star}\odot{\bm P}_1^1\A^{\star}$ has full column rank.
	\end{Theorem}
When fiber or entry sampling is employed, we have:
\begin{Theorem}\label{theorem:deterministic_general2}
			\vspace{-0.33cm}
	Let $\underline{\bm X}\in\mathbb{C}^{I\times J \times K}$ be the original tensor signal, fiber or entry sampled according to \eqref{constraints_fiber1} or \eqref{constraints_entry1} respectively. Also let $\underline{\bm X} = \left\llbracket {\bm A}, {\bm B},{\bm C}\right\rrbracket$ denote the rank-$F$ CPD of $\underline{\bm X}$. Assume that ${\bm A}^\star,{\bm B}^\star,{\bm C}^\star$ have no repeated entries and satisfy the equations in~ \eqref{eq:para2}, \eqref{model:entry_sampling}, according to the sampling mechanism. Then, $\underline{\hat{\bm X}}(i,j,k)=\sum_{f=1}^F{\bm A}^\star(i,f){\bm B}^\star(j,f){\bm C}^\star(k,f)$ recovers the ground-truth $\underline{\bm X}$ if {$~2F+2 \leq \min_d\left\lbrace k_{{\bm P_1^{(d)}\A^{\star}}}+k_{{\bm P_2^{(d)}\B^{\star}}}+k_{{\bm P_3^{(d)}\C^{\star}}}\right\rbrace$}, where ${\bm P}_3^{(d)}=\bm I$ for fiber sampling.
\end{Theorem}
Proof of both theorems is presented in Appendix \ref{appendix:proof_deterministic}.

The implication of Theorems \ref{theorem:generic_slab}- \ref{theorem:deterministic_general2} is significant and intuitive.
Recovery of $\underline{\X}$ is based on two basic principles: identifiability of the factors of the sub-sampled tensors and ability to reconcile for the permutation and scaling ambiguities. 
The first is a property of the both the signal of interest and the sampling mechanism. 
In particular the rank of the tensor signal, along with Kruskal or generic conditions on the factors determine the number of samples required for full tensor recovery. Note that the number and structure of samples varies according to the applied sampling mechanism. Hence, there is a clear correlation between the rank of the tensor and the number of samples needed---higher ranks require higher number of samples. The second principle is a necessary property of the sampling mechanism. Although slab sampling automatically handles permutation and scaling ambiguities, fiber and entry sampling schemes have to be carefully designed to satisfy \eqref{constraints_fiber1} or \eqref{constraints_entry1} and eliminate permutation and scaling mismatches.

In a nutshell,  one can learn the rows of factors $\A,~\B,~\C$ from the sub-tensors, up to column permutation and scaling and resolve the mismatches using common information between the sub-tensors. Then reconstruction of the original tensor is attained as $\underline{\bm X}=\left\llbracket{\bm A},{\bm B},{\bm C}\right\rrbracket$.

To build some further intuition on the theoretical conditions, consider the following example.
Let $\underline{\bm X}\in\mathbb{C}^{512\times512\times 513}$ be the tensor with CP rank $F=1000$ which is subject to sampling. First we sample the $I_1$ equispaced horizontal slabs and $K_2$ equispaced frontal slabs. Following Theorem \ref{theorem:generic_slab}, reconstruction of $\underline{\bm X}$ is guaranteed if we sample at least $I_1=8$ horizontal slabs and $K_2=2$ frontal slabs and vise versa. This results in sampling ratio $r=\frac{\#\text{observed entries}}{\# \text{total entries}}=0.019$. Next we sample fibers of the tensor in a regular fashion, similarly to Fig. \ref{fig:fiber_new}. According to theorem \ref{theorem:generic_fiber}, $33,280$ fibers are sufficient to recover the original tensor, which gives sampling ratio $r=0.13$. Finally, entries are sampled in a regular fashion, as shown in Fig. \ref{fig:entry_sampling1}. The total number of entries required to reconstruct the original tensor is $2,097,664$, according to theorem \ref{theorem:generic_entry} , which results in sampling ratio $r=0.016$. Note that for smaller rank, e.g, $F=250$ the total number of samples can be significantly reduced, giving sampling ratio $r_{slab}=0.008$, $r_{fiber}=0.063$, $r_{entry}=0.004$.

	\section{Application to parallel fMRI acceleration}\label{fMRI}
	 Interestingly, the previously described sampling mechanisms find application in accelerating fMRI scan acquisition. fMRI is used to measure brain activity associated with changes in blood oxygen levels. MRI acquisitions typically use a set of coils (sensors), that in parallel collect a series of frames focusing on different parts of the brain. In fMRI, the three-dimensional (3D) volume covering the whole brain is typically acquired using multiple two-dimensional (2D) slices. These are discrete-space signals, sampled along a particular trajectory in the k-space, which is a 2-D frequency domain ($k_x,~k_y$), for each brain slice. Therefore an fMRI scan, can be represented by a five-way array with coil, $k_x,~k_y$, slice and time (frame) dimensions. 
	 
	 Acquiring high spatial resolution fMRI is challenging due to time restrictions. On the one hand, the $k$-space sampling has to follow the Shannon-Nyquist theorem to avoid artifacts, when inverse Fourier transform is used for reconstruction. On the other hand sampling at a Nyquist rate leads to prolonged scan acquisition time for each frame, which is prohibitive for high temporal resolution, required in fMRI and neuroscience research. The objective is therefore two-fold: accelerate the scanning process and capture fast brain activity changes. Since the scan acquisition time is proportional to the number of $k$-space samples, ongoing efforts focus on sampling part of the $k$-space ($k_y$ frequencies) of each slice and/or measuring the $k$-space of only a subset of slices. The majority of work is mainly proposed for MRI scans. Classic methods use learning and calibration type techniques \cite{grappa,lustig2010spirit,akccakaya2018scan}, while others employ the CS framework \cite{lustig2010spirit,sparse_sense,kt_focuss} or LRMC \cite{kt_focuss,sake,kt_slr,chiew2015k} to perform the reconstruction. 
	
	While MRI offers significant freedom in designing the $k$-space sampling trajectories for each frame, fMRI acquisition is more restrictive. Specifically, fMRI is performed using a special fast imaging acquisition, called echo planar imaging, that is practically only used with equispaced sub-sampling patterns due to restrictions associated with magnetic field inhomogenities and Eddy currents \cite{feinberg2012rapid}. In simple words, the $k_y$ frequencies sampled for each frame have to be equispaced and all coils need to measure the same frequencies. For example the sampling scheme illustrated in Fig. \ref{fig:fMRI} is typical in fMRI and performs 3-fold acceleration.
	\begin{figure}[h!]
		\centering
		\includegraphics[width=0.6\linewidth]{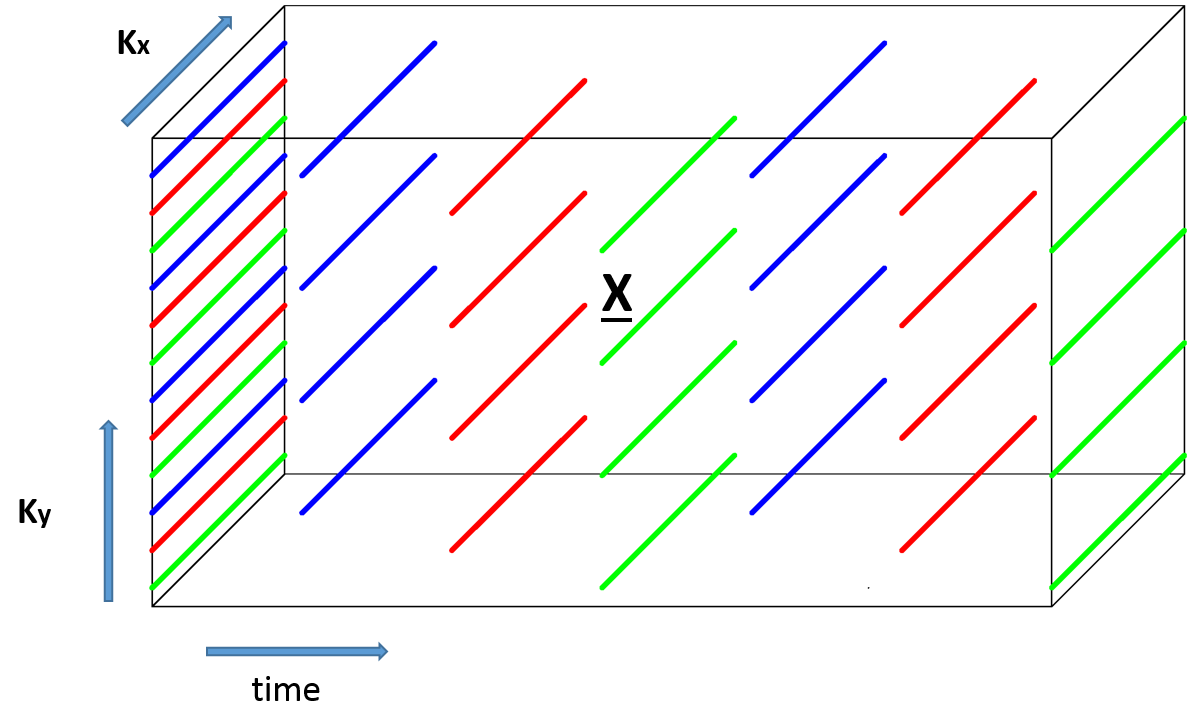}
		
		\caption{Single-slice fMRI sampling at each coil.}
		\label{fig:fMRI}
	\end{figure}
In general, a fully sampled scan is acquired first, which is beneficial for calibration purposes. Then $n$-fold acceleration is achieved by sampling $1/n$ of equispaced $k_y$ frequencies. The frequencies to sample for each frame can be the same for the whole procedure or can be circularly shifted as in Fig. \ref{fig:fMRI}. For the benefit of our method we propose to circularly shift between $n$ equispaced set of frequencies in order to capture the temporal behavior of the brain accurately.

 Sampling the $k_y$ dimension is one way to accelerate the fMRI scanning process. Another idea that is being used is to observe the $k$-space of only a subset of slices at each time slot. In this work we propose to combine these two ideas to further reduce scanning time. Specifically at each time instance sub-sampled $k$-space measurements are acquired for only a subset of slices, instead of the complete set. To design a sampling mechanism that fits our tensor models and fMRI constraints, we first need to acquire a fully sampled scan for every slice. Then for each frame $1/r$ of equispaced $k_y$ frequencies is observed for $1/s$ of the brain slices, so that at the $(rs)$-th frame we have measured every frequency for every slice. This results in $(rs)$-fold acceleration. Fig. \ref{fig:multifMRI} illustrates an fMRI acceleration technique, where $r,s=2$.
\begin{figure}[h!]
	\centering
	\includegraphics[width=0.95\linewidth]{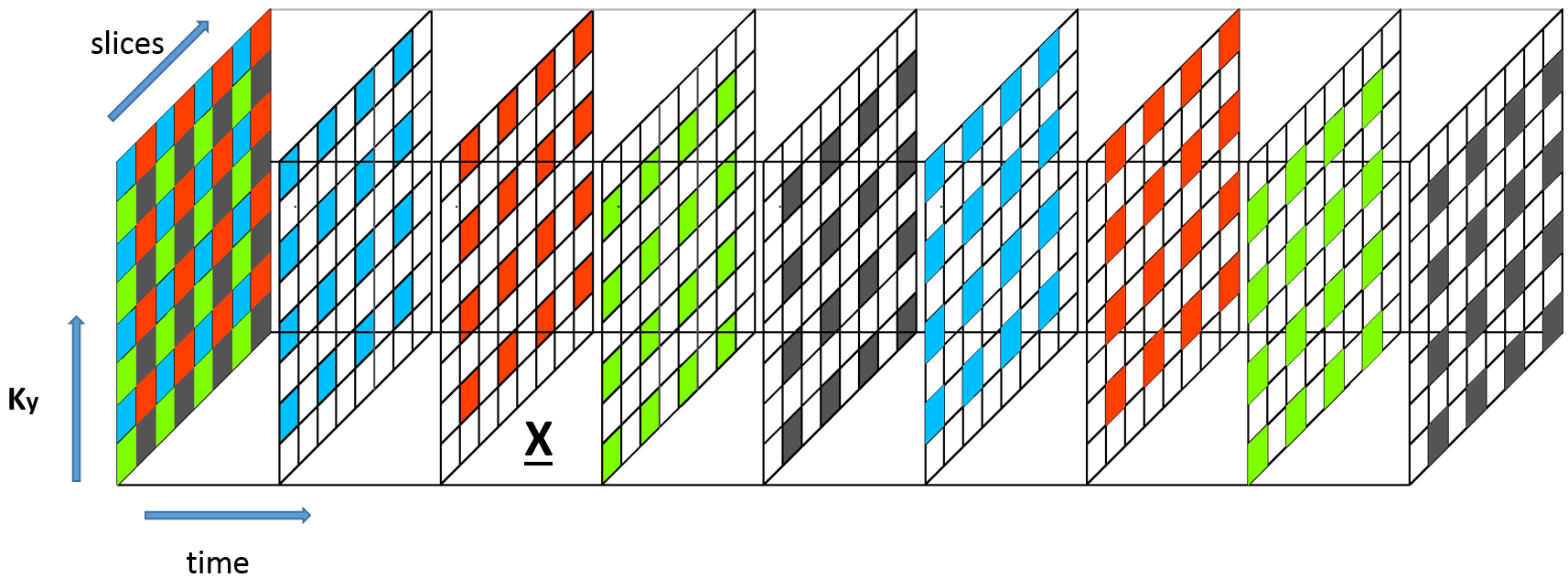}
	
	\caption{Multi-slice fMRI sampling at each coil.}
	\label{fig:multifMRI}
	\vspace{-0.2cm}
\end{figure} 

	 Note that the two aforementioned sampling procedures can be tricky for classic techniques. {On the one hand, calibration-based techniques such as GRAPPA \cite{grappa} are linear and suffer from noise amplification at high acceleration rates. On the other hand, CS and LRMC schemes have difficulties in operating with regular samples, since their success rests upon incoherent sampling.}
	 
	 On the contrary, the proposed tensor sampling and reconstruction framework is exactly designed to handle these highly structured and constrained sampling schemes used in fMRI acquisitions. In particular, the single-slice fMRI acceleration task, as illustrated in Fig. \ref{fig:fMRI}, can be cast as a tensor fiber sampling mechanism, analyzed in subsection \ref{model:fiber}. As mentioned earlier the raw fMRI scan is originally a five-way array and thus each slice is a four-way array. Although the previous analysis could be easily extended to tensors of order higher than three, we choose to work with third-order ones. Specifically the $k$-space is processed in a single dimension (mode) by concatenating $k_x$ and $k_y$. The reason is that the relation between $k_x$ and $k_y$ is often hard to be captured by a multilinear tensor model. As a result one fMRI slice is modeled as a third-order tensor $\underline{\X}\in\mathbb{C}^{I\times J\times K}$, where $I=m_xm_y$ with $m_x,~m_y$ representing the number of frequencies in $k_x,~k_y$ space respectively, $J$ represents the total number of frames (time slots) and $K$ the number of coils. Following the analysis of subsection \ref{model:fiber} we have the following result. 
	 
	 \begin{Prop}
Let $\underline{\X}\in\mathbb{C}^{I\times J\times K}$ be the a single-slice fMRI tensor with rank $F$, modeled as previously explained. Under the assumptions of Theorem \ref{theorem:generic_fiber}, $n$-fold acceleration can be achieved if $n\leq\min\left\lbrace\sqrt{\frac{IJ}{4\lceil F\rceil}}, \frac{JK}{4\lceil F\rceil}, \frac{IK}{4\lceil F\rceil}\right\rbrace$.
	 \end{Prop}

	 Similarly, the proposed  multi-slice fMRI acceleration scheme, which performs joint $k$-space and slice sampling, is cast as an entry tensor sampling procedure, introduced in \ref{model:entry}. To do so, the $k$-space is considered as a single mode, as before, and we also concatenate coils and slices in one dimension. The resulting third-order tensor $\underline{\X}\in\mathbb{C}^{I\times J\times K}$ has the $k$-space in the first mode, i.e, $I=m_xm_y$, $J$ represents the total number of frames and the third mode includes the concatenation of coils and slices, i.e., $K=m_sm_c$ with $m_s,~m_c$ being the number of slices and coils respectively. Following the analysis of subsection \ref{model:entry} we have:
	 
	 	 \begin{Prop}
	 	 	Let $\underline{\X}\in\mathbb{C}^{I\times J\times K}$ be the a multi-slice fMRI tensor with rank $F$, modeled as previously explained. Under the assumptions of Theorem \ref{theorem:generic_entry}, ${(rs)}$-fold acceleration can be achieved if $rs\leq\frac{1}{4\lceil F\rceil}\min\left\lbrace IK, \frac{JK}{s}, \frac{IJ}{r}\right\rbrace$. 
	 	 \end{Prop}
	 
 We should mention that tensor approaches have also been proposed in medical imaging \cite{mardani2016tracking,yaman2017locally,christodoulou2016fast,ma2017high,he2016accelerated}. The work in \cite{mardani2016tracking}, for example, uses a tensor model to approach the MRI sampling and reconstruction problem in an on-line fashion. However, \cite{mardani2016tracking} works under different sampling schemes, which are not regular and appropriate for fMRI, and reconstruction guarantees are not discussed. Moreover, a tensor model is also used in \cite{yaman2017locally}, in the context of MRI denoising, which is different from the fMRI or MRI acceleration problem. Finally the works in \cite{christodoulou2016fast,ma2017high,he2016accelerated} adopt a Tucker model \cite{sidiropoulos2017tensor}, and also require auxiliary acquisition data to estimate the bases in non-spatial modes prior to reconstruction, which are not available in most fMRI acquisitions.

	\section{General Algorithmic framework for Tensor Sampling}
	Previously we studied sampling and reconstruction of third-order tensors under different, identifiable sampling mechanisms. In the current section, the algorithmic component of our approach is discussed. In general, reconstruction of a tensor from a subset of entries falls under the framework of tensor completion. A plethora of algorithms have been proposed, e.g. \cite{acar2011scalable,tomasi2005parafac}. The idea is to use the CPD factors, computed from the incomplete tensor, and reconstruct the original one. Popular methods approach the problem as a system of non-linear equations and handle it using descent direction approaches, such as gradient descent, alternating optimization, or the Gauss-Newton method. 
	
Existing tensor completion works could be employed to approach the recovery task of a regularly sampled tensor. However, the unique characteristics of our models would be ignored. To put it in context, the special structure of {\em regular sampling allows tensor completion by computing the factors of complete tensors}. Note that CPD computation of a complete tensor is a considerably easier task than that of an incomplete one. Several polynomial time algebraic algorithms \cite{sanchez1990tensorial,domanov2014canonical} have been shown to retrieve the original factors,under certain conditions, or effectively initialize optimization approaches with significant success. 

We propose a three step approach to tackle the completion task, which follows the insights of Theorems \ref{theorem:generic_slab}-\ref{theorem:deterministic_general2}. The first step solves the CPD of the sub-sampled tensors independently, the second reconciles for permutation/scaling ambiguities and gets an initial estimate of the factors, and the third solves the coupled CPD problem. Detailed analysis follows.
	\subsection{Step 1: Computing the CPD of sub-tensors}
	First, the CPD of the sub-sampled tensors $\underline{\Y}_i$ is computed. This step is guided from the requirements of each sampling mechanism. To be more precise, CPD of $\underline{\Y}_i$ is computed if the reconstruction conditions require $\underline{\Y}_i$ to admit an identifiable CPD. The slab sampling model, for instance, requires only $\underline{\Y}_1$ or $\underline{\Y}_2$ to admit unique CPD. Therefore the CPD of only one sub-tensor is needed. On the other hand when fiber or entry sampling is considered, the CPD computation of every sub-tensor is performed, following the previous identifiability analysis.
	\subsection{Step 2: Initializing the factors}
	After computing the CPD of the sub-tensors, step 2 computes an initial estimate of the $\A,\B,\C$ factors, after resolving possible permutation and scaling mismatches. We distinguish between 2 different cases:
	
	\noindent
	\textbf{Case 1, slab sampling:} As mentioned earlier slab sampling automatically reconciles for permutation and scaling ambiguities. Furthermore, two of the factors have been already computed from step 1 (e.g., $\A$,$\B$$\leftarrow$CPD($\underline{\Y}_2$)). What remains to be obtained is the third factor (e.g., $\C$), which is revealed by the other sub-tensor ($\underline{\Y}_1$), via solving a linear system of equations: 
	\[\Y_1^{(3)}=({\bm B}\odot{\bm P_1^{(1)}\bm A}){\bm C}^T\]

	\textbf{Case 2, fiber and entry sampling:} Contrary to slab sampling, the permutation and scaling ambiguity is an important issue when fiber or entry sampling is applied. To be more precise, let $\underline{\Y}_d =  \left\llbracket {\bm A}_d, {\bm B}_d,{\bm C}_d\right\rrbracket,~d\in\{1,\dots D\}$, be the sub-tensors formed after fiber or entry sampling. Then:
	
	\begin{subequations}\label{eq1}
		\begin{align}
		&{\bm A_d}=\bm P_1^{(d)}{\bm A}{\bm \Pi}^{(d)}{\bm \Lambda}^{(d)}_1=\A(\mathcal{S}_r^{(d)},:){\bm \Pi}^{(d)}{\bm \Lambda}^{(d)}_1,~\\&{\bm B_d}=\bm P_2^{(d)}{\bm B}{\bm \Pi}^{(d)}{\bm \Lambda}^{(d)}_2={\bm B}(\mathcal{S}_c^{(d)},:){\bm \Pi}^{(d)}{\bm \Lambda}^{(d)}_2,\\&{\bm C_d}=\bm P_3^{(d)}{\bm C}{\bm \Pi}^{(d)}{\bm \Lambda}_3^{(d)}={\bm C}(\mathcal{S}_f^{(d)},:){\bm \Pi}^{(d)}{\bm \Lambda}_3^{(d)},
		\end{align}    
	\end{subequations} 	
	where ${\bm \Pi}^{(d)}\neq{\bm \Pi}^{({d^{\prime}})}$ are permutation matrices and ${\bm \Lambda}_i^{(d)}\neq{\bm \Lambda}_i^{({d^{\prime}})}$ are full rank diagonal matrices such that ${\bm \Lambda}_1^{(d)}{\bm \Lambda}_2^{(d)}{\bm \Lambda}_3^{(d)}={\bm I}$, $d,d^{\prime}\in\{1,\dots,D\}$. Clearly, in order to synthesize $\A,\B,\C$ from $\A_d,\B_d,\C_d$ and reconstruct $\underline{\X}$, the permutation and scaling mismatch should be resolved, i.e., ${\bm \Pi}^{(d)}={\bm \Pi}^{({d^{\prime}})}$, ${\bm \Lambda}_i^{(d)}={\bm \Lambda}_i^{({d^{\prime}})}$ for every $d,~d^{\prime}$.
		
	To overcome this issue, the common information between sub-tensors is utilized. In simple words, \eqref{constraints_fiber1} or \eqref{constraints_entry1} require $\C_d-\C_{d^{\prime}}$ (or $\A_d-\A_{d^{\prime}}$, or $\B_d-\B_{d^{\prime}}$) to share some common rows, i.e., $|\mathcal{S}_f^{(d)}\cap\mathcal{S}_{f}^{(d^{\prime})}:=\mathcal{S}_{f}^{(d-d^{\prime})}|\geq 2$. Now let:
		\begin{subequations}
			\begin{align}\label{eq3}
			{\bm C_d^{(d-d^{\prime})}}={\bm C}(\mathcal{S}_{f}^{(d-d^{\prime})},:){\bm \Pi}^{(d)}{\bm \Lambda}_3^{(d)}\\
			{\bm C_{d^{\prime}}^{(d-d^{\prime})}}=\C(\mathcal{S}_f^{(d-d^{\prime})},:){\bm \Pi}^{({d^{\prime}})}{\bm \Lambda}^{({d^{\prime}})}_3
			\end{align}
		\end{subequations}
	and normalize $\C_{d}^{(d-d^{\prime})},~\C_{d^{\prime}}^{(d-d^{\prime})}$, such that the share a common row, e.g.:
	 \[\bar{\C}_{d}^{(d-d^{\prime})}={\C}_{d}^{(d-d^{\prime})}\G_d^{-1},~\bar{\C}_{d^{\prime}}^{(d-d^{\prime})}={\C}_{d^{\prime}}^{(d-d^{\prime})}\G_{d^{\prime}}^{-1},\]
	  where $\G_d=\text{diag}\left({\C}_{d}^{(d-d^{\prime})}(1,:)\right)$, $\G_{d^{\prime}}=\text{diag}\left({\C}_{{d^{\prime}}}^{(d-d^{\prime})}(1,:)\right)$ and $\text{diag}(\x)$ is the diagonal matrix of row vector $\x$. Then: 
	  \begin{align}\label{eq4}
	  \bar{\C}_{d}^{(d-d^{\prime})}=\bar{\C}_{d^{\prime}}^{(d-d^{\prime})}{\bm \Pi}^{({d^{\prime}})^{-1}}{\bm \Pi}^{({d})}=\bar{\C}_{d^{\prime}}^{(d-d^{\prime})}\bar{\bm\Pi}
	  \end{align}
	  and we can solve for $\bar{\bm\Pi}$ using the Hungarian algorithm \cite{kuhn1955hungarian}. This procedure resolves the permutation mismatch between the factors, i.e., 
	  ${\bm \Pi}^{(d)}={\bm \Pi}^{({d^{\prime}})}$. Note that in case of fiber sampling $\mathcal{S}_f^{(d)}=\mathcal{S}_{f}^{(d^{\prime})}=\mathcal{S}_{f}^{(d-d^{\prime})}=\{1\dots K\}$.

To reconcile for the scaling ambiguity we require extra information coming from the factors that were not involved in permutation match, i.e., $\A_d-\A_{d^{\prime}}$ or $\B-\B_{d^{\prime}}$ in our example. The necessary rules \eqref{rule:graph},~\eqref{rule:graph2}  enforce that there is at least one common row between $\A_d-\A_{d^{\prime}}$ or $\B-\B_{d^{\prime}}$. Then the scaling mismatch can be solved by the following set of equations:
\begin{subequations}
	\begin{align}\label{eq5}
	&{\A}_{d}^{(d-d^{\prime})}={\A}_{d^{\prime}}^{(d-d^{\prime})}{\bm \Lambda}_1^{({d^{\prime}})^{-1}}{\bm \Lambda}_1^{({d})}\\
	&{\B}_{d}^{(d-d^{\prime})}={\B}_{d^{\prime}}^{(d-d^{\prime})}{\bm \Lambda}_3^{({d^{\prime}})^{-1}}{\bm \Lambda}_3^{({d})}\\
	&{\C}_{d}^{(d-d^{\prime})}={\C}_{d^{\prime}}^{(d-d^{\prime})}{\bm \Lambda}_3^{({d^{\prime}})^{-1}}{\bm \Lambda}_3^{({d})}\\
	&{\bm \Lambda}_1^{(d)}{\bm \Lambda}_2^{(d)}{\bm \Lambda}_3^{(d)}={\bm I},~{\bm \Lambda}_1^{(d^{\prime})}{\bm \Lambda}_2^{(d^{\prime})}{\bm \Lambda}_3^{(d^{\prime})}={\bm I}
	\end{align}
\end{subequations}
 ${\A}_{d}^{(d-d^{\prime})},~{\A}_{d^{\prime}}^{(d-d^{\prime})}$ and  ${\B}_{d}^{(d-d^{\prime})},~{\B}_{d^{\prime}}^{(d-d^{\prime})}$ represent the common rows between $\A_d-\A_{d^{\prime}}$ and $\B-\B_{d^{\prime}}$ respectively.
	Next, an initial estimate of the factors is extracted by reading out the appropriate rows from the sub-tensor factors to synthesize $\A,~\B,~\C$ i.e., 
	\begin{equation*}
\A(\mathcal{S}_r^{(d)},:)\leftarrow\A_d,~\B(\mathcal{S}_c^{(d)},:)\leftarrow\B_d,~\C(\mathcal{S}_f^{(d)},:)\leftarrow\C_d,~\forall d.
	\end{equation*}
	\subsection{Step 3: Coupled CPD}
	Finally, $\A,~\B,~\C$ are jointly computed as a classic tensor factorization problem with missing entries, which is equivalent to the following coupled CPD estimator:
	\begin{equation}\label{eq:ctensor1}
	\begin{aligned}
	\minimize_{{\bm A},{\bm B},{\bm C}}&~\sum_{d=1}^{D}\left\|\underline{\bm Y}_{d} - \left\llbracket {\bm P}_1^{(d)}{\bm A}, {\bm P}_2^{(d)}{\bm B},{\bm P}_3^{(d)}{\bm C}\right\rrbracket\right\|_F^2.
	\end{aligned}
	\end{equation}
	
	In slab sampling, ${\bm P}_2^{(1)}, {\bm P}_3^{(1)}, {\bm P}_1^{(2)}, {\bm P}_2^{(2)}$ are identity matrices and in fiber sampling ${\bm P}_3^{(d)}$ is always the identity.
	There are several ways to handle the above non-convex problem. We choose to employ the tensorlab \cite{vervliettensorlab} toolbox, which uses a Gauss Newton approach to solve this nonlinear least squares (NLS) problem. After obtaining the estimates of $\A,\B$ and $\C$,
	$\underline{\X}$ can be reconstructed by:\\
	$\underline{\hat{\bm X}}(i,j,k)=\sum_{f=1}^F\hat{\bm A}(i,f)\hat{\bm B}(j,f)\hat{\bm C}(k,f)$.  
	\vspace{-0.2cm}	
	\subsection{REgular Tensor Sampling and INterpolation Algorithm (RETSINA)}
	As mentioned earlier the accelerated fMRI acquisition can be cast as a tensor sampling and reconstruction task. Therefore it falls under the class of problems that the previously described framework can handle. However we choose to follow a different initialization approach tailored to the specific application. We design two algorithms, one for single-slice fMRI and one for multi-slice fMRI. 
	
	\textbf{Single slice acceleration:}
The REgular Tensor Sampling and INterpolation Algorithm (\texttt{RETSINA}) is presented in Algorithm \ref{algo:Retsina}. We follow a 3 step procedure. In step 1 (initialization), for $n$-fold acceleration we sum every $n$ vertical slabs (frames), where the missing $k$-space measurements are considered zeros, and obtain a tensor $\underline{\X}_n\in\mathbb{C}^{I\times J/n\times K}$ without missing entries. Then we compute the CPD of $\underline{\X}_n$ and get a rough estimate of $\A,~\C$ factors. In step 2 (refinement), we compute the CPD of $\underline{\Y}_1$ (initialized by step 1) and the CPD of $\{\underline{\Y}_i\}_{i\neq 1}$ with known $\C$. Finally, in step 3, we compute the final factors by solving \eqref{eq:ctensor1} with tensorlab's Gauss-Newton algorithm. Compared to the previously presented general framework, \texttt{RETSINA} empirically yields enhanced reconstruction accuracy and reduces the operational time.
\setlength{\textfloatsep}{0pt}
\begin{algorithm}[t!]
	\caption{\texttt{RETSINA}}\label{algo:Retsina}
	\begin{algorithmic}
		{\small
			\STATE \textbf{Input:} $n$, $F$, $\underline{\tilde{\X}}$.
			\STATE \textbf{Initialization:}\vspace{-.15cm}
			\STATE $\underline{\X}_n(:,j,:)=\sum\limits_{l=(j-1)n+2}^{jn+1}\underline{\tilde{\X}}(:,l,:)$
			\vspace{-.08cm}
			\STATE $\A,\C\leftarrow$ CPD$(\underline{\X}_n)$ 
			\STATE Form $\{\underline{\Y}_i\}_{i=1}^n$ from $\underline{\tilde{\X}}$ and set ${\mathcal{S}_i=[1,i+1:n:J]}$.
			\STATE ${\bm B}(\mathcal{S}_i,:)={\bm P}_2^{(i)}{\bm B}=\arg\min_{\bm Z} \lVert\underline{\bm Y}_{i} - \llbracket\bm P_1^{(i)}\bm A,\bm Z,\C\rrbracket\rVert_F^2$
			\STATE \textbf{Refinement:}
			\STATE ${\bm P}_1^{(1)}\A,{\bm P}_2^{(1)}\B,{\bm C}\leftarrow$ CPD$(\underline{\Y}_1)$.
			\STATE ${\bm P}_1^{(i)}\A,{\bm P}_2^{(i)}{\bm B},\sim\leftarrow$ CPD$(\underline{\Y}_i),~~~i\neq 1$ .
			\STATE \textbf{Solve} \eqref{eq:ctensor1} using Gauss-Newton.		
			\STATE	\textbf{Reconstruct~}{\footnotesize the missing entries of $\underline{\bm X}$ using $\underline{\hat{\bm X}}=\left\llbracket{\bm A},{\bm B},{\bm C}\right\rrbracket$.}}
	\end{algorithmic}
\end{algorithm}
\setlength{\textfloatsep}{0pt}
\begin{algorithm}[t!]
	\caption{\texttt{MS-RETSINA}}\label{algo:MS-Retsina}
	\begin{algorithmic}
		{\small
			\STATE \textbf{Input:} $r,~s$, $F$, $\underline{\tilde{\X}}:$ incomplete tensor (zeros are missing entries).
			\STATE \textbf{Initialization:}\vspace{-.15cm}
			\STATE $\underline{\X}_n(:,j,:)=\sum\limits_{l=(j-1)n+2}^{jn+1}\underline{\tilde{\X}}(:,l,:)$, $n=rs$
			\vspace{-.08cm}
			\STATE $\A,\C\leftarrow$ CPD$(\underline{\X}_n)$ 
			\STATE Form $\{\underline{\Y}_i\}_{i=1}^n$ from $\underline{\tilde{\X}}$ and set ${\mathcal{S}_i=[1,i+1:n:J]}$.
			\STATE ${\bm B}(\mathcal{S}_i,:)={\bm P}_2^{(i)}{\bm B}=\arg\min_{\bm Z} \lVert\underline{\bm Y}_{i} - \llbracket\bm P_1^{(i)}\bm A,\bm Z,\C\rrbracket\rVert_F^2$
			\STATE \textbf{Solve} \eqref{eq:ctensor1} using Gauss-Newton.		
			\STATE	\textbf{Reconstruct~}{\footnotesize the missing entries of $\underline{\bm X}$ using $\underline{\hat{\bm X}}=\left\llbracket{\bm A},{\bm B},{\bm C}\right\rrbracket$.}}
	\end{algorithmic}
\end{algorithm}

\textbf{Multi slice acceleration:} 
The Multi-Slice \texttt{RETSINA} (\texttt{MS-RETSINA}) is presented in Algorithm \ref{algo:MS-Retsina}. Compared to \texttt{RETSINA} the initialization step has been modified to this specific case and the refinement step is skipped, since it has been observed that it does not improve the overall performance. 


	
	\section{Simulations}
	In this section, we showcase the effectiveness of the proposed tensor sampling framework using numerical experiments. The experiments involve synthetically generated data as well as fMRI scans in the $k$-space. All simulations are performed in {\sc Matlab} on a Linux server with 3.6GHz cores and 32GB RAM, except part C which is performed on a Linux server with 2GHz cores and 128GB RAM.

\subsection{Synthetic Experiments}
First synthetically generated experiments are conducted to examine the validity of our claims and the performance of the proposed framework.
In particular, a tensor $\underline{\bm X}\in\mathbb{R}^{500\times 500 \times 500}$ is generated as $\underline{\bm X}=\left\llbracket{\bm A},{\bm B},{\bm C}\right\rrbracket$. The elements of the factor matrices are drawn from an independent identically distributed (i.i.d.) zero mean, unit variance Gaussian distribution. We regularly sample $\underline{\bm X}$ according to the previously presented sampling mechanisms, i.e,. slab sampling, fiber sampling, and entry sampling, as shown in Figs. \ref{fig:paradigm1}- \ref{fig:entry_sampling1}. Specifically, for slab sampling we sample equispaced frontal and horizontal slabs. Regarding fiber sampling, each tensor $\underline{\Y}_i$ is a set of fibers, defined by equispaced rows and columns of $\underline{\bm X}$. Note that one vertical slab is fully observed to reconcile for permutation and scaling ambiguities. Equivalently entry sampling is designed to observe different sets of equispaced entries plus a fully sampled vertical slab.

For the experiments, we define the sampling ratio, ${r=\frac{\#\text{of sampled entries}}{IJK}}$,
and vary it from $0.75$ to $0.001$. We also alter the tensor rank $F$ from $5$ to $1000$. To evaluate the performance of tensor reconstruction, we measure the normalized reconstruction error, i.e. 
\begin{equation*}
{\small	\texttt{NRE}=\frac{\sum_{k=1}^{K}\lVert \hat{\underline{\bm X}}(:,:,k)-\underline{\bm X}(:,:,k)\rVert_F}{\sum_{k=1}^{K}\lVert\underline{\bm X}(:,:,k)\rVert_F}}
\end{equation*}
Fig. \ref{fig:Fvsr} presents the results for the three sampling schemes. 
\begin{figure}[h!]
	\centering
	\begin{subfigure}[t]{0.49\linewidth}
		\includegraphics[width=\linewidth]{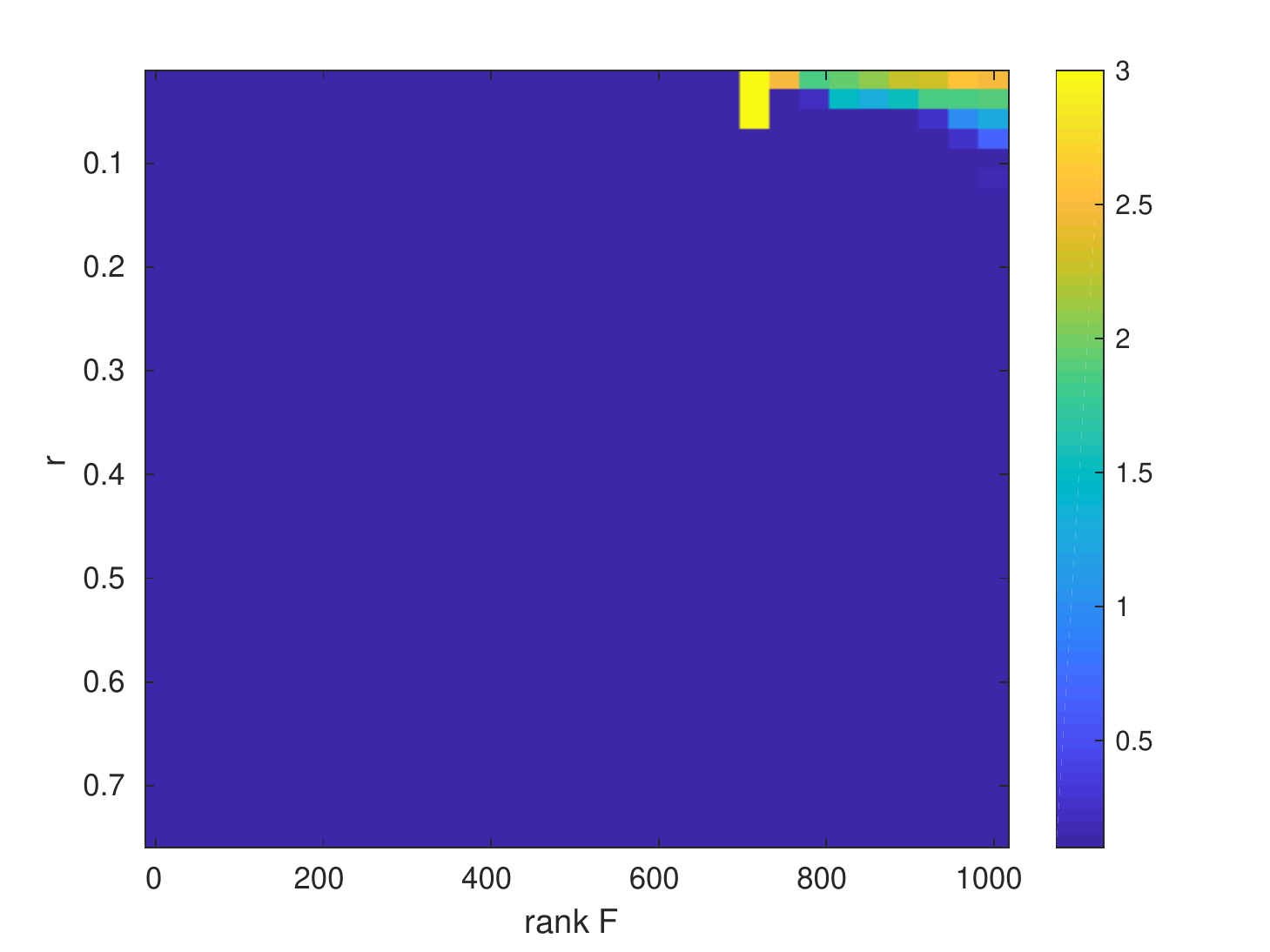}
		\caption{slab sampling}
	\end{subfigure}
	\begin{subfigure}[t]{0.49\linewidth}
		\includegraphics[width=\linewidth]{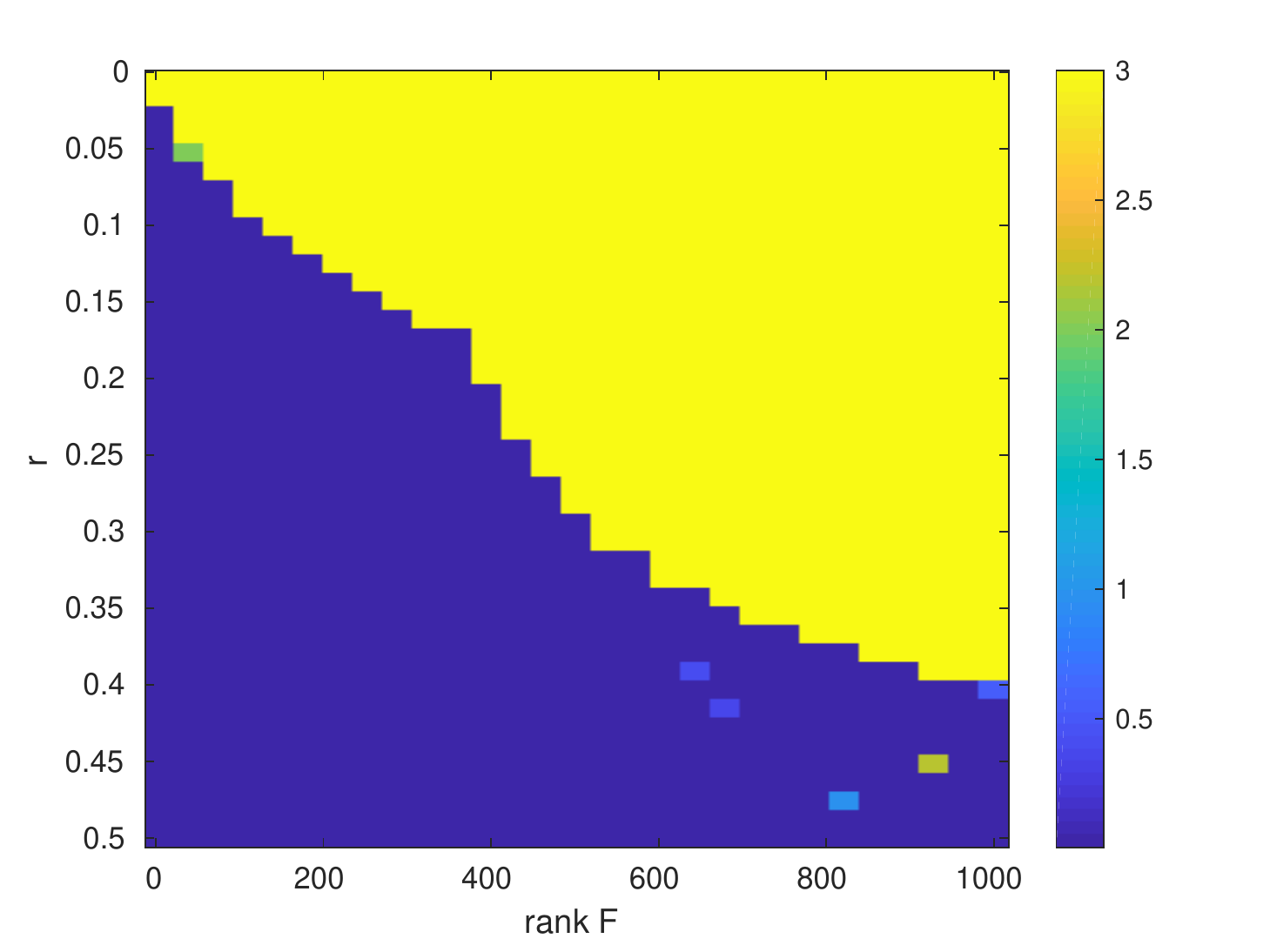}
		\caption{fiber sampling}
	\end{subfigure}
\\
\begin{subfigure}[t]{0.49\linewidth}
	\includegraphics[width=\linewidth]{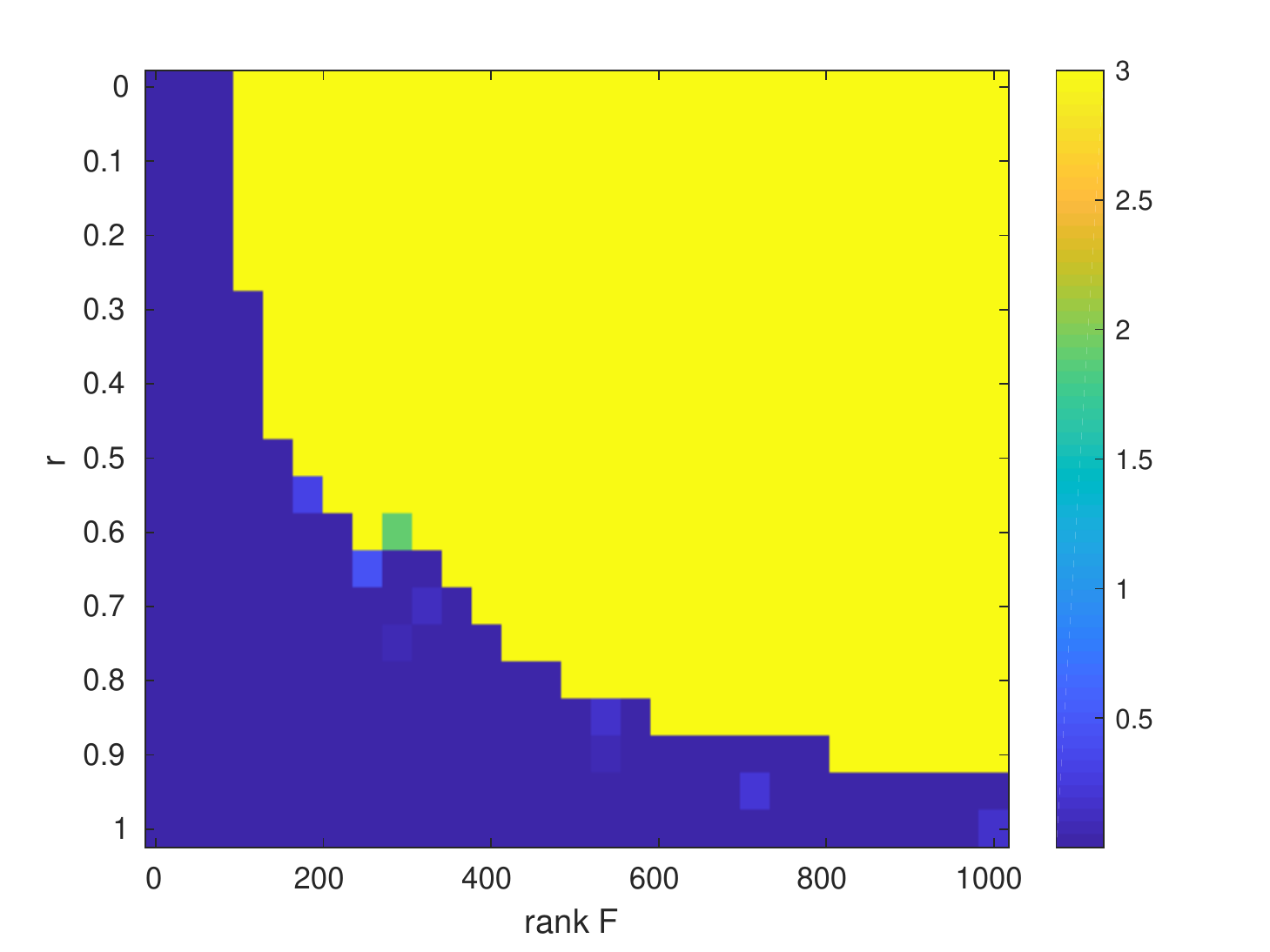}
	\caption{entry sampling}
\end{subfigure}

	\caption{rank $F$ vs sampling ratio $r$.}
	\label{fig:Fvsr}
\end{figure}
As expected the reconstruction accuracy is deteriorating as the rank $F$ increases or the sampling ratio $r$ decreases. For reasonably small ranks and high number of samples the reconstruction is perfect.

\subsection{Accelerated parallel fMRI}
Next, the tensor sampling and reconstruction framework is tested in a real and important problem, that of parallel fMRI acceleration. 
First, we test the performance of the proposed \texttt{RETSINA} with fMRI scans, fully sampled in the $k$-space, obtained from the Center for Magnetic Resonance Research (CMRR) at the University of Minnesota. 

The single slice raw scan is originally a fourth-order tensor of size $104\times104\times32\times 490$ and we unfold it as a third-order tensor $\underline{\X}\in\mathbb{C}^{10816\times 32\times 490}$. We apply 3-fold acceleration by observing $1/3$ of the $k_y$ frequencies, as shown in Fig. \ref{fig:fMRI}.
We choose $F=100$ and run step 1, 2, 3 for 50, 2 and 5 iterations respectively. The baseline algorithms used for comparison are \texttt{k-t Focuss} \cite{kt_focuss}, which is a CS type algorithm, \texttt{k-t SLR} \cite{kt_slr} which combines ideas from both LRMC and CS, and the zero padding inverse discrete Fourier transform (IDFT). The performance of IDFT is an indicator on how difficult the reconstruction is. Note that \texttt{k-t SLR} directly reconstructs the fMRI signal in the absolute $(x-y)$-time-coil space. To be more precise it reconstructs signal $\underline{\W}=|\mathcal{Q}(\underline{\X})|$, where $\mathcal{Q}$ denotes the inverse Fourier transform from the $k_x-k_y$ to the $x-y$ space and $|\cdot|$ is the absolute value. Thus we also measure the NRE of signal $\underline{\W}$, denoted as $\text{NRE}_2$, for fair comparisons. For both \texttt{k-t Focuss} and \texttt{k-t SLR} the publicly available code was used. Note that \texttt{k-t Focuss} and \texttt{k-t SLR} are single coil algorithms in their original implementation, thus we treated each coil separately. \texttt{k-t SLR} requires parameter tuning and so we used a validation step to tune effectively. 

The results are presented in Table \ref{tableG}, which includes the NRE in $k$-space and absolute $x-y$ space as well as runtime. The proposed \texttt{RETSINA} achieves highest reconstruction quality in the $k$-space and works comparably well (but markedly faster) with \texttt{k-t SLR} in reconstructing the signal magnitude in the $x-y$ space. This is expected, since \texttt{RETSINA} reconstructs both the magnitude and phase (which is very important in images) in the $k$-space, whereas \texttt{k-t SLR} reconstructs the magnitude in the $x-y$ space. In terms of runtime \texttt{RETSINA} works faster than \texttt{k-t SLR} and \texttt{k-t Focuss}, but slower than \texttt{IDFT}. However \texttt{IDFT} exhibits very poor reconstruction performance. It is worth noting, that \texttt{k-t Focuss} and \texttt{k-t SLR} qualify for parallel implementation, which could potentially speed up operation time. This would be held, however, at the cost of computational resources.
\begin{table}[h!]
	\centering
	\caption{Reconstruction performance of the competing algorithms.}
	\vspace{-.15cm}
	\label{tableG}
	\resizebox{1\linewidth}{!}{
		\begin{tabular}{ |l| c | c | c |c|c|}
			\hline	
			\textbf{Algorithm}&\texttt{RETSINA} &	\texttt{k-t Focuss} & \texttt{k-t SLR} & \texttt{IDFT} \\ \hline
			$\text{NRE}$ & \textbf{0.124}& \text{0.339} & 1.41& \text{0.8156}  \\
			\hline
			$\text{NRE}_2$ & \text{0.081}& \text{0.286} & \textbf{0.073}& \text{0.7376}  \\
			\hline
			\text{runtime} & 12\text{min}& 25.6\text{min} (48\text{sec}/\text{coil})& 480\text{min} (15\text{min}/\text{coil}) & \textbf{14sec}  \\
			\hline
	\end{tabular}}
	\vspace{-.15cm}
\end{table}
	
	Fig. \ref{fig:rec} shows the reconstructed fMRI scans at different time frames produced by \texttt{RETSINA} along with the fully sampled data. The quality of the reconstruction is significantly high, rendering the proposed \texttt{RETSINA} a good alternative for fMRI acceleration.
	\begin{figure}[h!]
		\centering
		\begin{subfigure}[t]{0.493\linewidth}
			\includegraphics[width=1.0\linewidth]{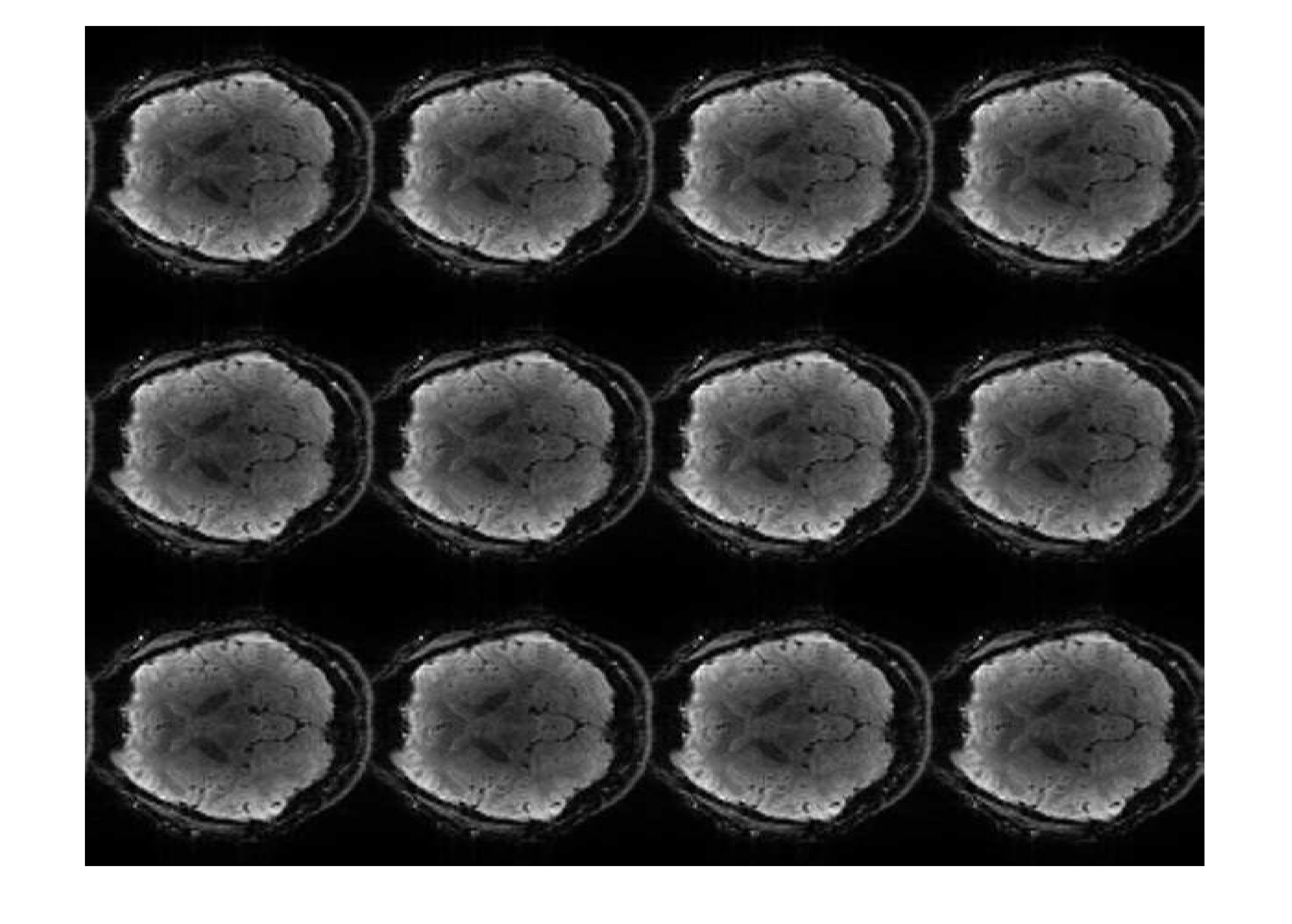}
			\caption{fully sampled scan}
		\end{subfigure}
		\begin{subfigure}[t]{0.493\linewidth}
			\includegraphics[width=1.0\linewidth]{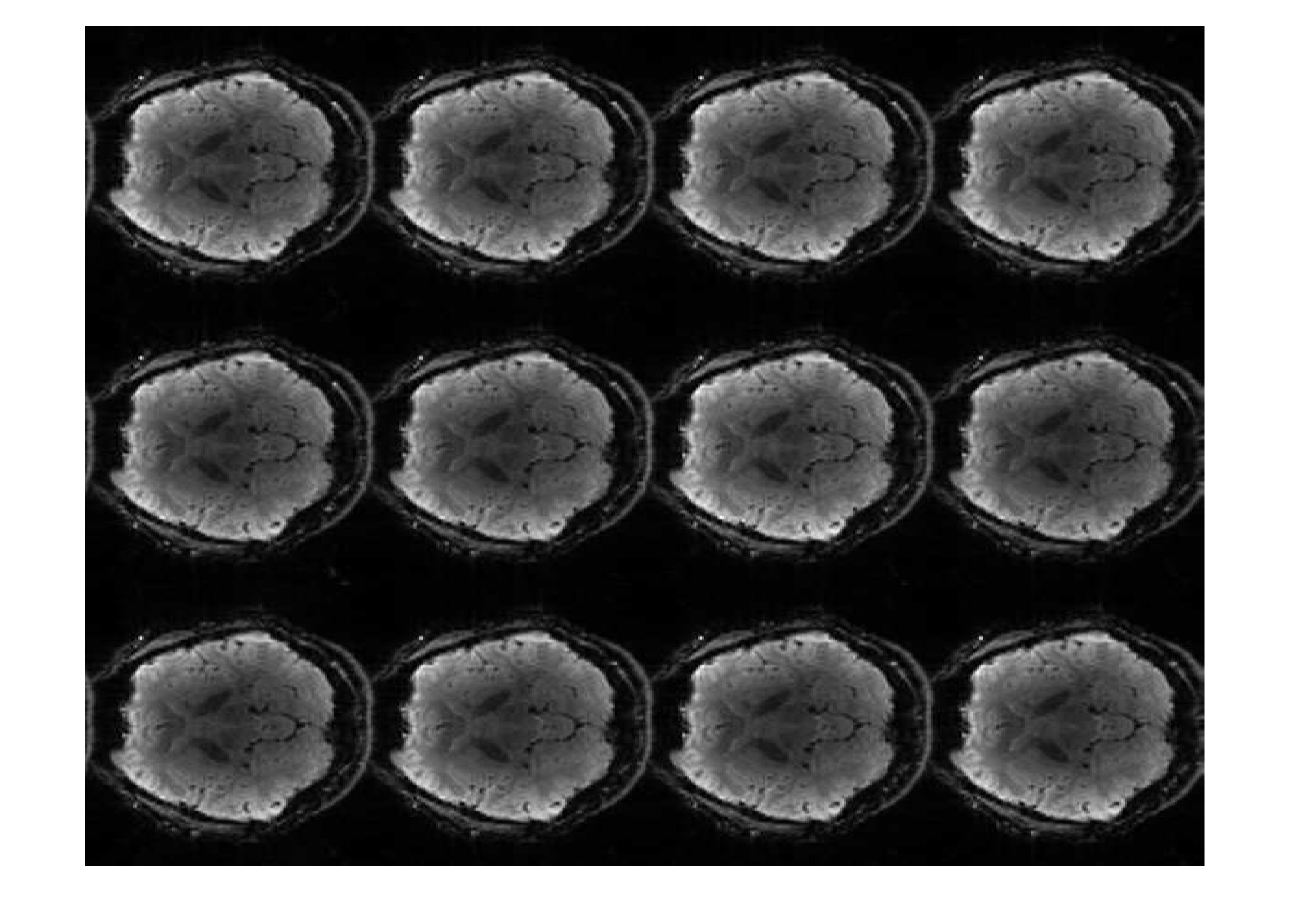}
			\caption{\texttt{RETSINA}}
		\end{subfigure}
		\caption{fMRI reconstruction with 3-fold acceleration}
		\label{fig:rec}
	\end{figure}
	Finally, in Fig. \ref{fig:comparisons} we illustrate the reconstruction performance at a single frame for the competing algorithms. \texttt{IDFT} gives an illustration of the downsampled image, \texttt{RETSINA} works the best and \texttt{k-t SLR} work comparably well, although being slightly off in contrast.		
	\begin{figure}[h!]
		\centering
		\begin{subfigure}[b]{0.092\textwidth}
			\includegraphics[width=1\textwidth]{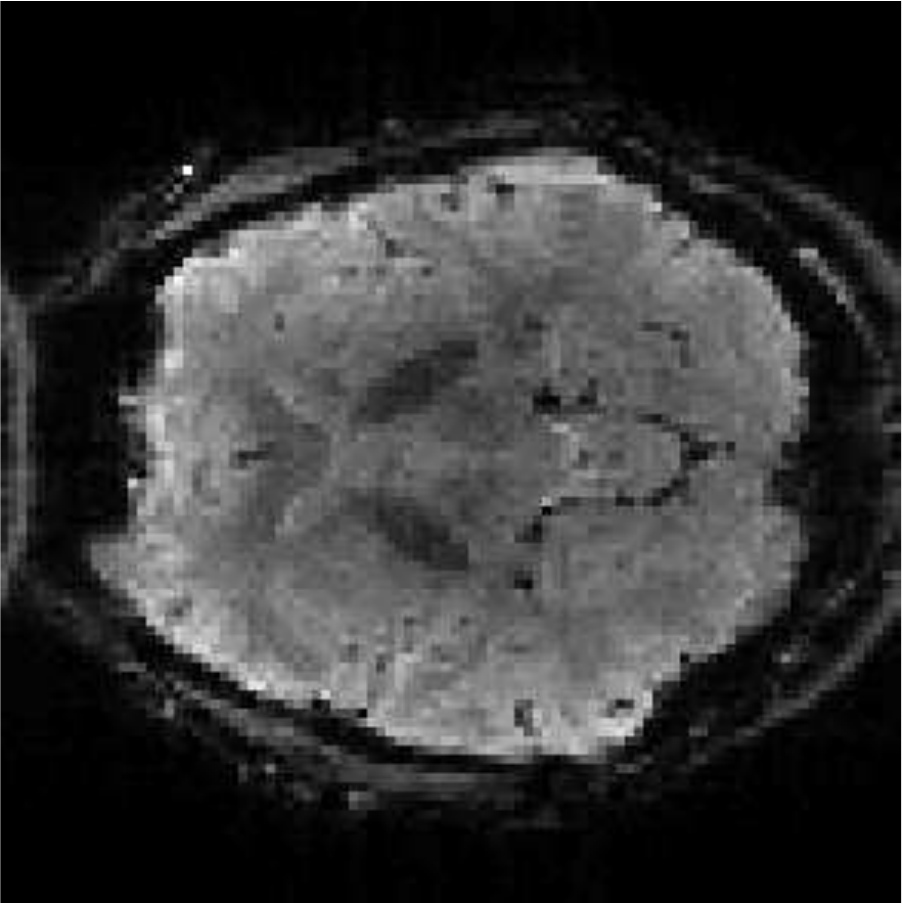}
			\vspace{-2.5pt}
			\caption{\footnotesize original}
		\end{subfigure}
		\hspace{-3.6pt} 
		\begin{subfigure}[b]{0.092\textwidth}
			\includegraphics[width=1\textwidth]{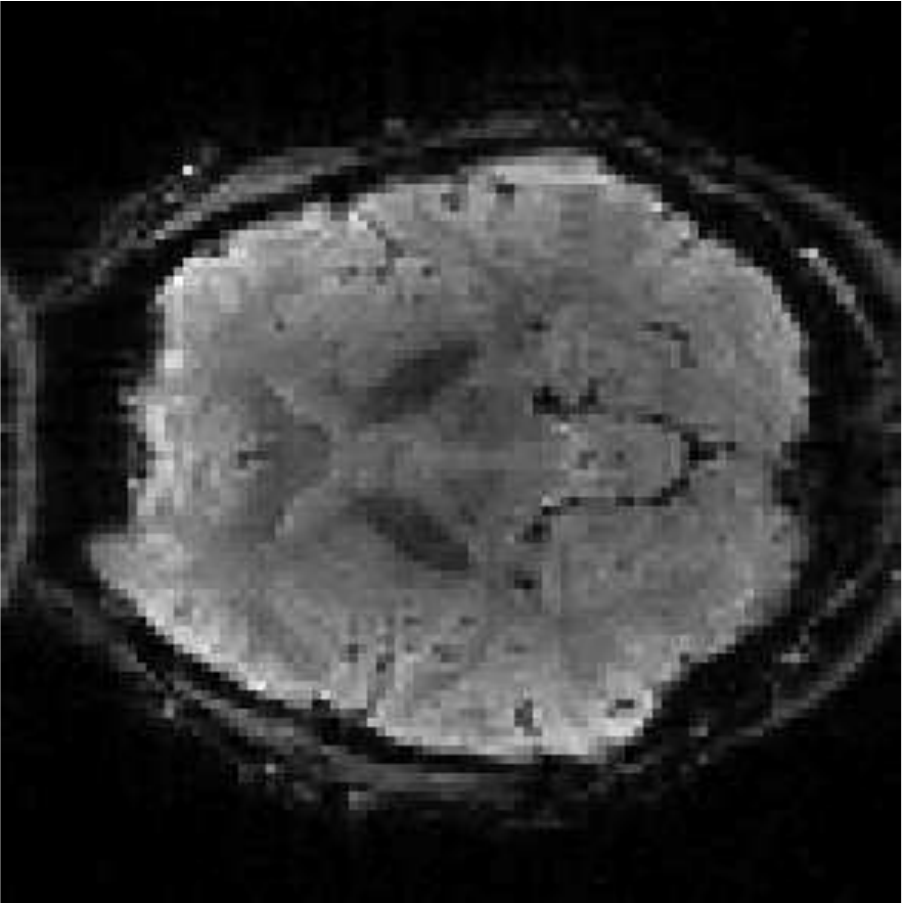}
			\vspace{-2.5pt}
			\caption{\footnotesize \texttt{RETSINA}}
		\end{subfigure}
		\vspace{-2.5pt} 
		\hspace{-3.6pt} 
		\begin{subfigure}[b]{0.092\textwidth}
			\includegraphics[width=1\textwidth]{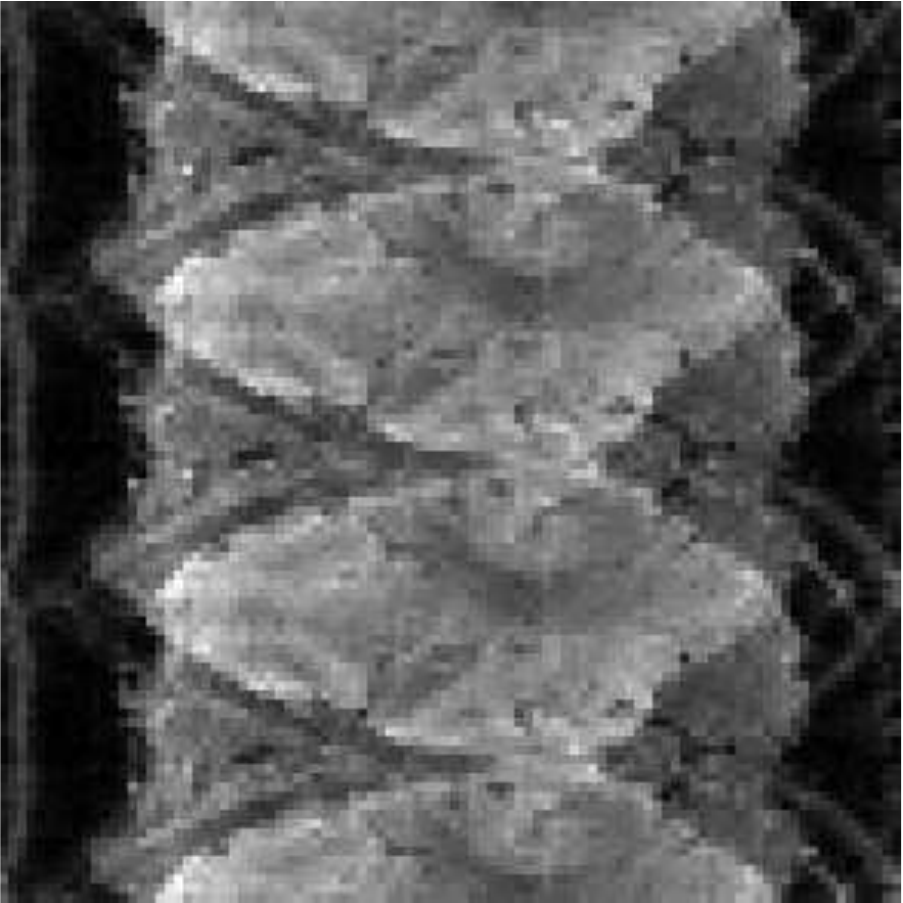}
			\vspace{-2.5pt}
			\caption{\footnotesize IDFT}
		\end{subfigure}
		\hspace{-3.6pt} %
		\begin{subfigure}[b]{0.092\textwidth}
			\includegraphics[width=1\textwidth]{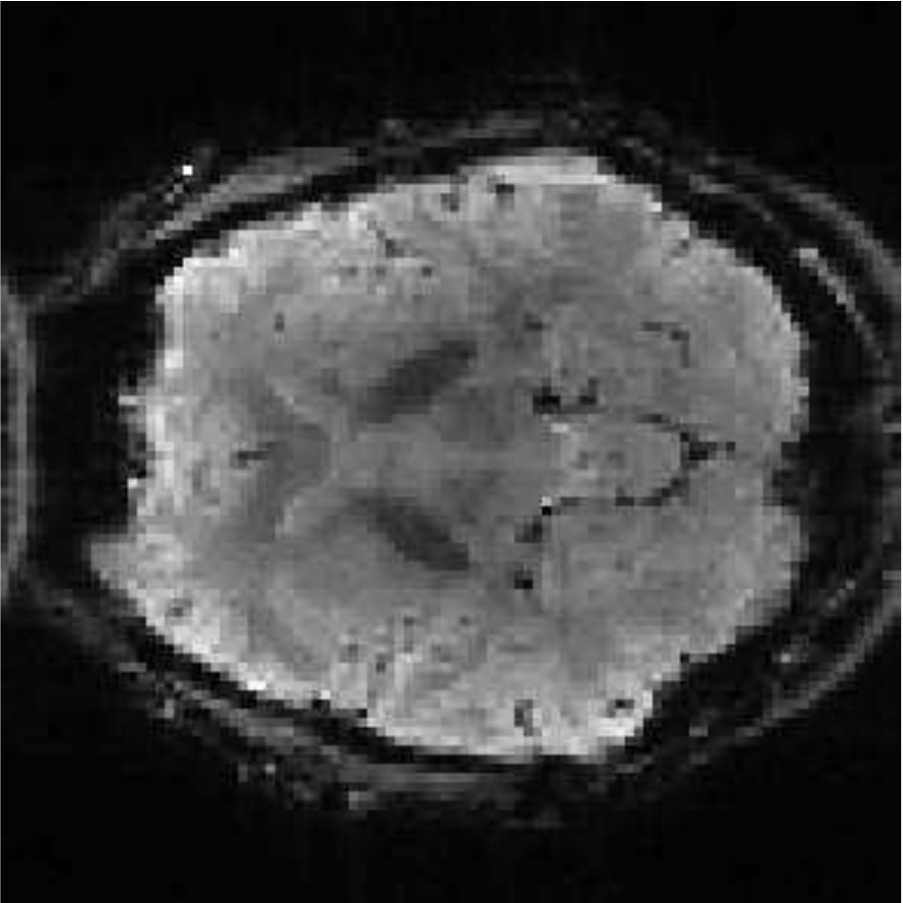}
			\vspace{-2.5pt}
			\caption{\footnotesize  k-t SLR}
		\end{subfigure}
		\hspace{-3.6pt}
		\begin{subfigure}[b]{0.092\textwidth}
			\includegraphics[width=1\textwidth]{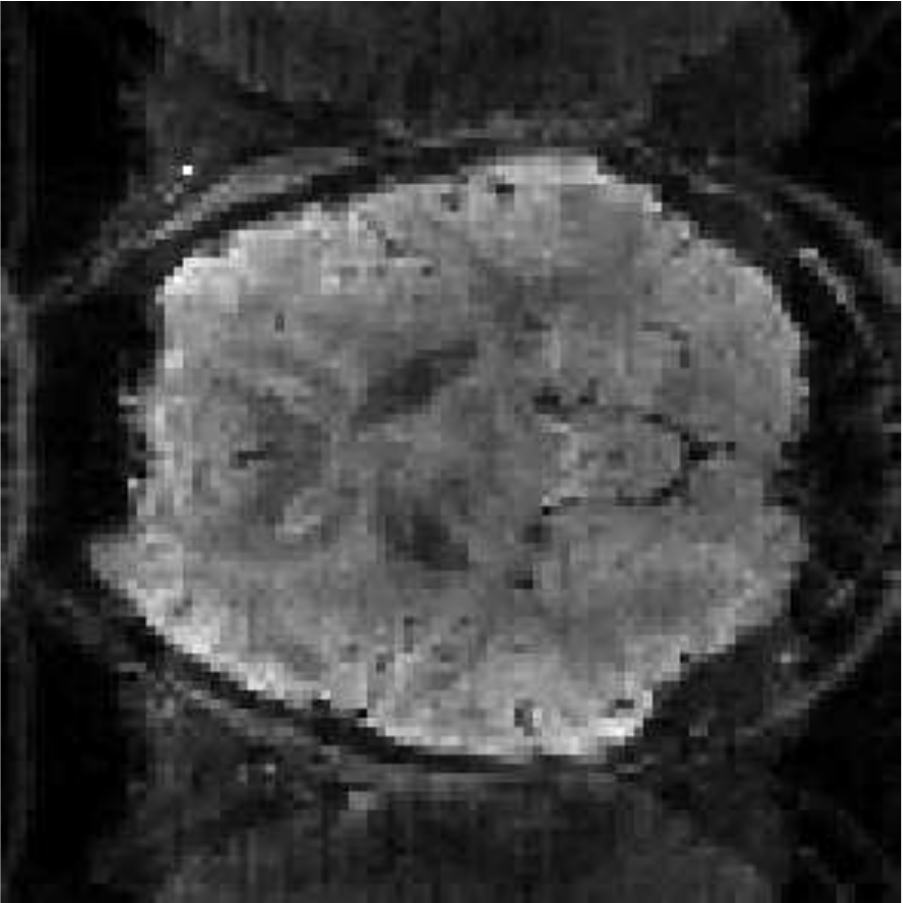}
			\vspace{-2.5pt}
			\caption{\footnotesize  k-t Focuss}
		\end{subfigure}
		\caption{Reconstruction at a single frame}
		\label{fig:comparisons}
		\vspace{-2.5pt}
	\end{figure}
	\vspace{-0.2cm}
\subsection{Accelerated multi-slice parallel fMRI}
Finally, the proposed framework is tested in the task of accelarated multi-slice fMRI acquisition. Recall that acceleration is performed at 2 levels, since at each time slot we measure the sampled $k$-space of only a subset of slices. The multi-slice fMRI raw scan is a fifth-order tensor of size $104\times104\times32\times 490 \times 8$, where the number of slices is $8$. We unfold it as a third-order tensor $\underline{\X}\in\mathbb{C}^{10816\times 490\times 256}$ and observe $1/r$ of the $k_y$ frequencies and $1/s$ of slices, which leads to $rs$-fold acceleration, as illustrated in Fig. \ref{fig:multifMRI}. The tensor rank used in \texttt{MS-RETSINA} is $F=100$. Table \ref{table2} shows the performance of the proposed \texttt{MS-RETSINA} in terms of NRE for various values of $r$ and $s$. An illustrative example of the reconstruction performance is presented in Fig. \ref{fig:rec2}.

\begin{table}[h!]
	\centering
	\caption{\text{NMSE} performance of \texttt{MS-RETSINA}.}
	\label{table2}
	\resizebox{0.95\linewidth}{!}{
		\begin{tabular}{ |l| c | c | c |c|c|c|c|c|c|}
			\hline	
			\textbf{$s\backslash r$}& 2&	3 & 4 &5&6&7&8&9&10 \\ \hline
			$2$ & \textbf{0.16}& 0.17 & 0.18&0.19&0.20&0.19&0.19&0.19&0.21  \\
			\hline
			$3$ & \text{0.18}& 0.19 & 0.20&0.19&0.20&0.22&0.21&0.20&0.22  \\
			\hline
			$4$ & 0.20& 0.20 & 0.20&0.22&0.21&0.21&0.23&0.22&0.23  \\
			\hline
	\end{tabular}}
\end{table}

\begin{figure*}[t!]
	\centering
	\begin{subfigure}[t]{0.24\linewidth}
		\includegraphics[width=1.0\linewidth]{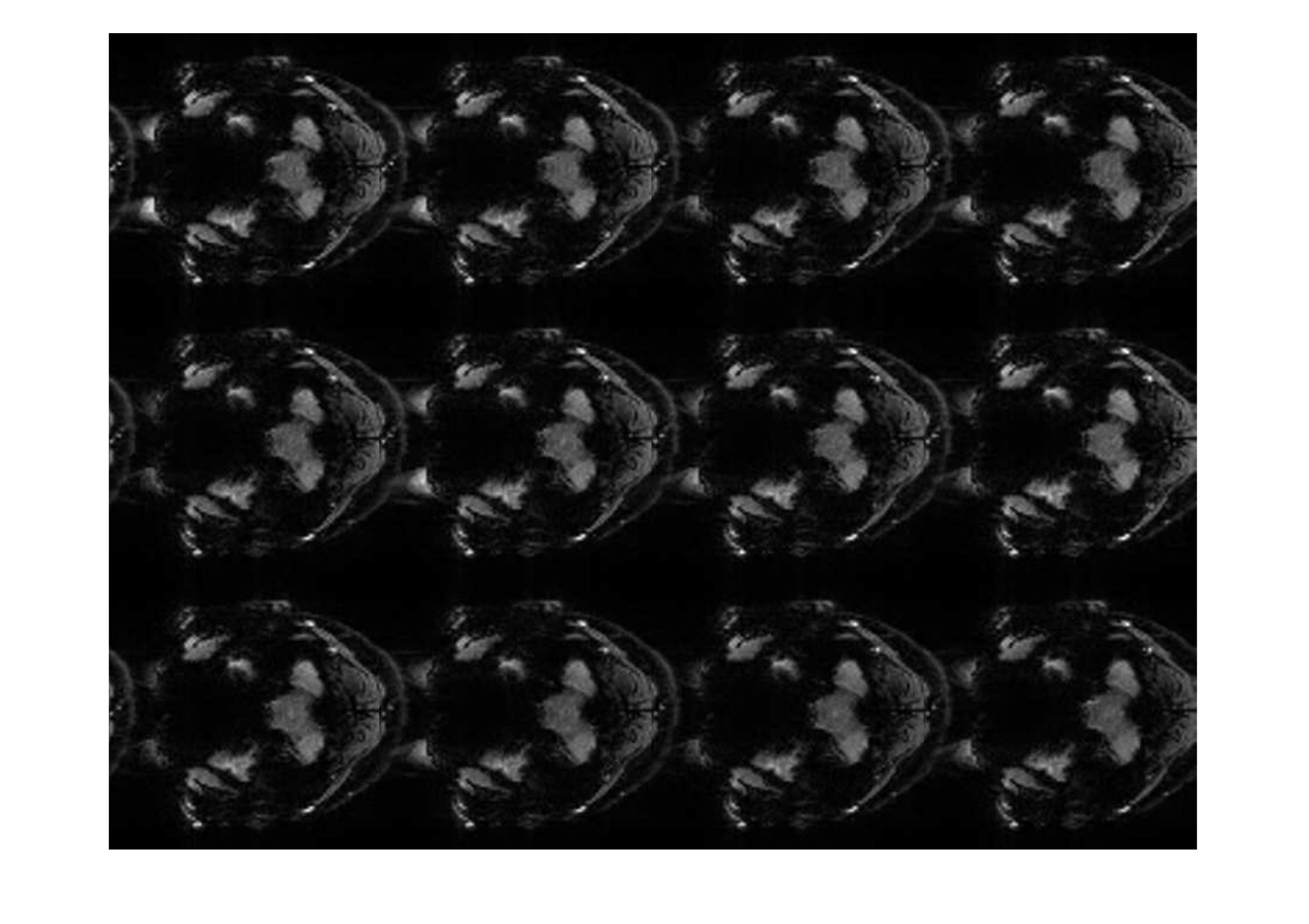}
		\caption{fully sampled scan}
	\end{subfigure}
	\begin{subfigure}[t]{0.24\linewidth}
		\includegraphics[width=1.0\linewidth]{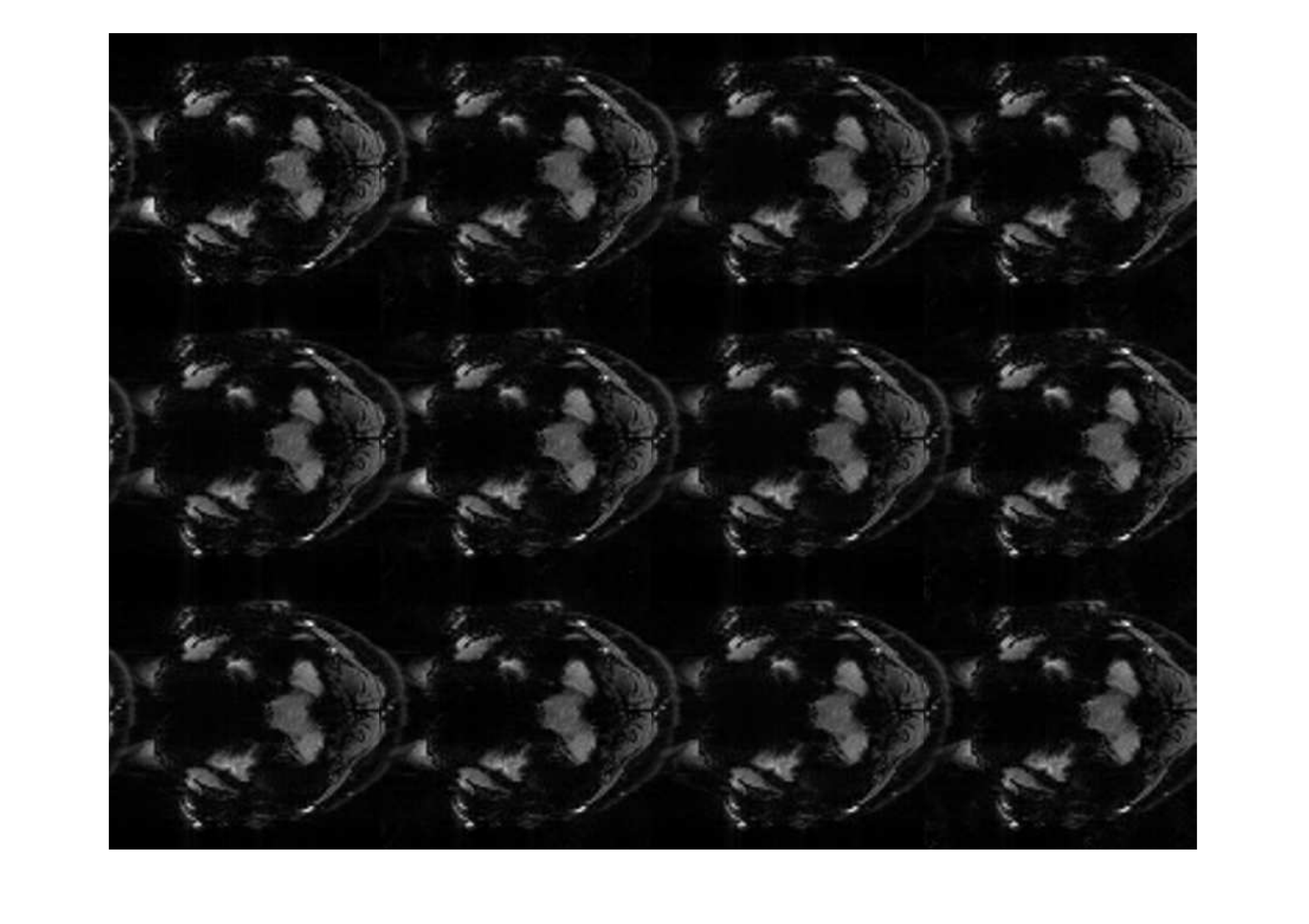}
		\caption{\texttt{RETSINA}}
	\end{subfigure}
	\begin{subfigure}[t]{0.24\linewidth}
		\includegraphics[width=1.0\linewidth]{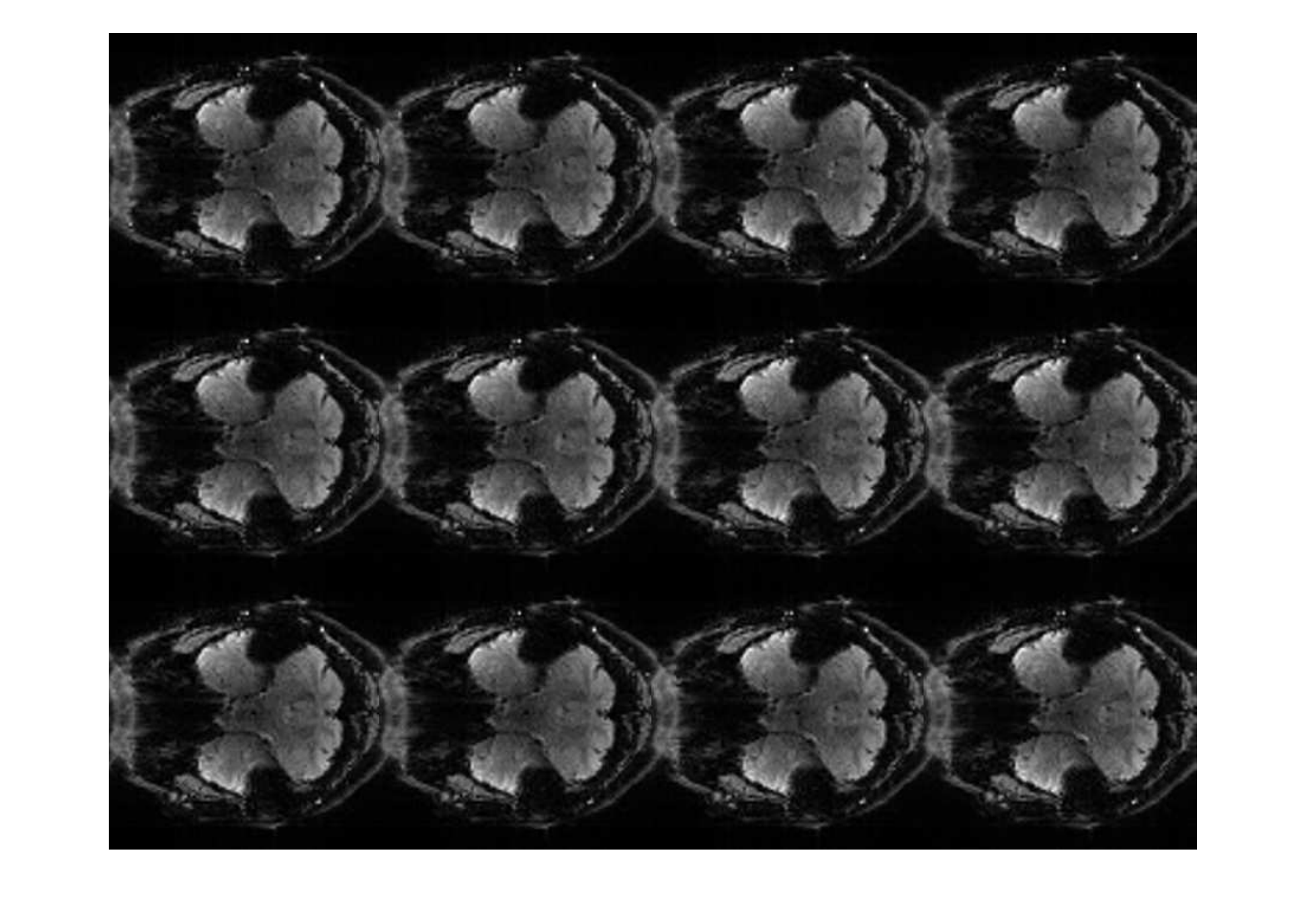}
		\caption{fully sampled scan}
	\end{subfigure}
	\begin{subfigure}[t]{0.24\linewidth}
		\includegraphics[width=1.0\linewidth]{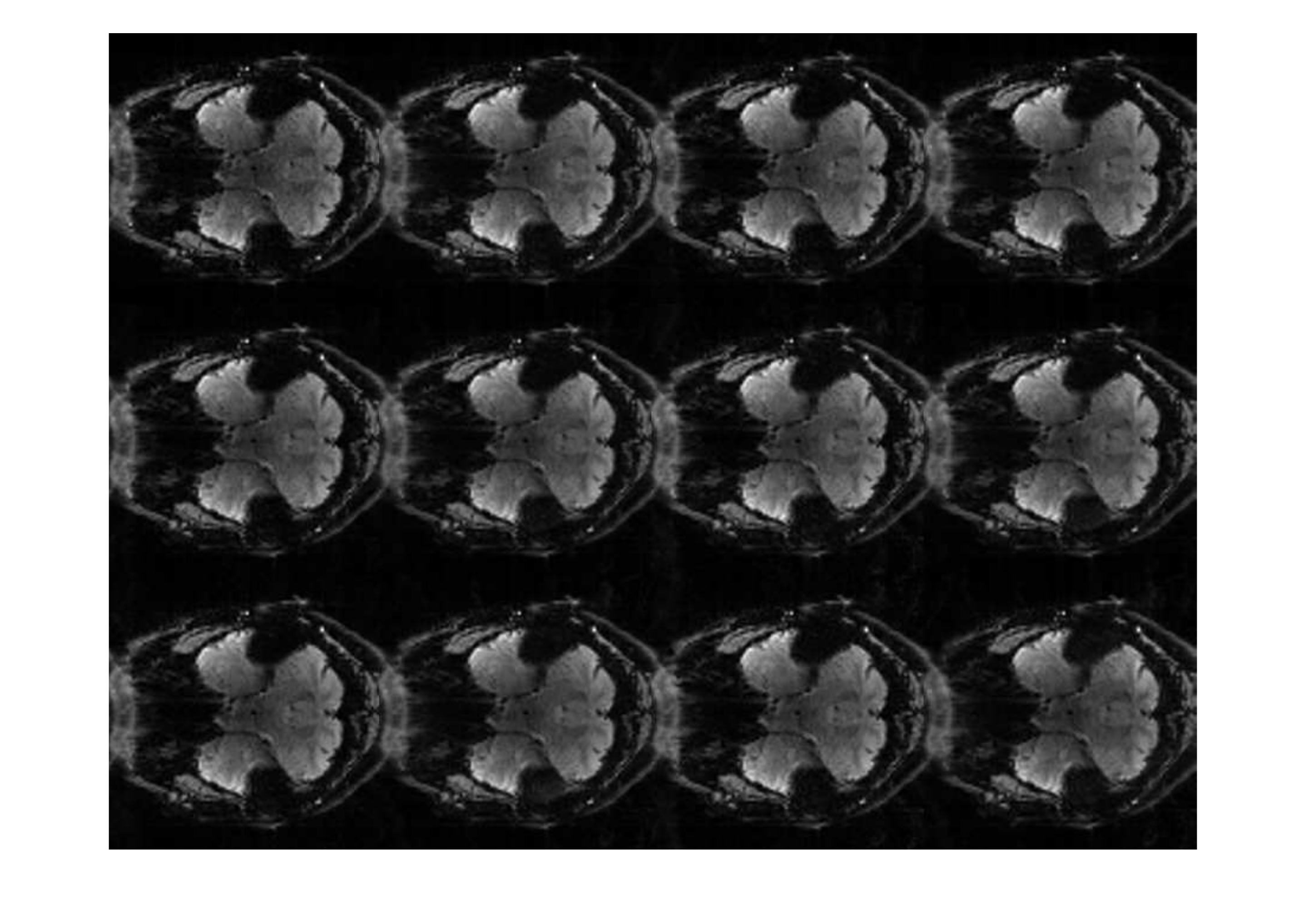}
		\caption{\texttt{RETSINA}}
	\end{subfigure}
	\\
	\begin{subfigure}[t]{0.24\linewidth}
		\includegraphics[width=1.0\linewidth]{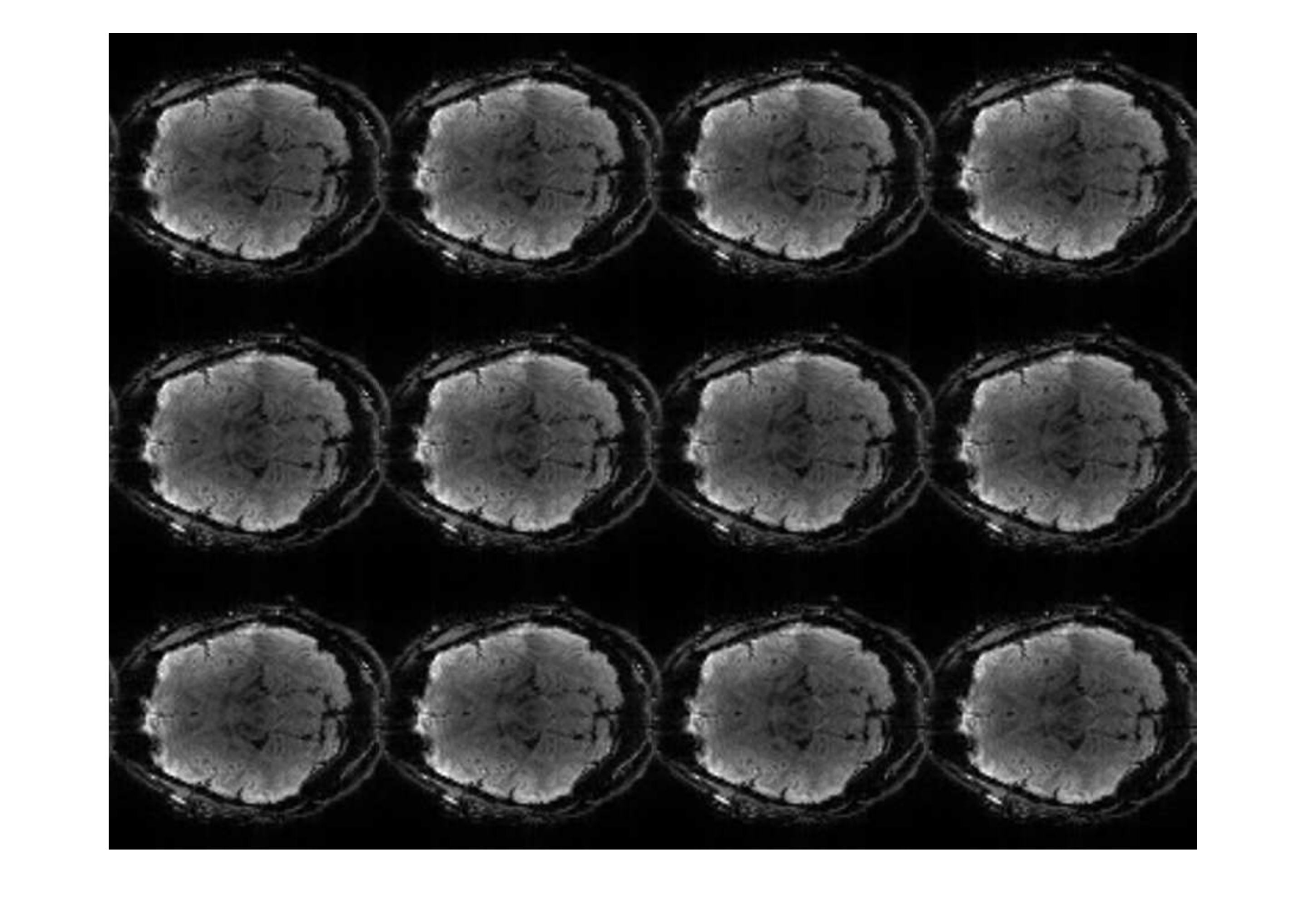}
		\caption{fully sampled scan}
	\end{subfigure}
	\begin{subfigure}[t]{0.24\linewidth}
		\includegraphics[width=1.0\linewidth]{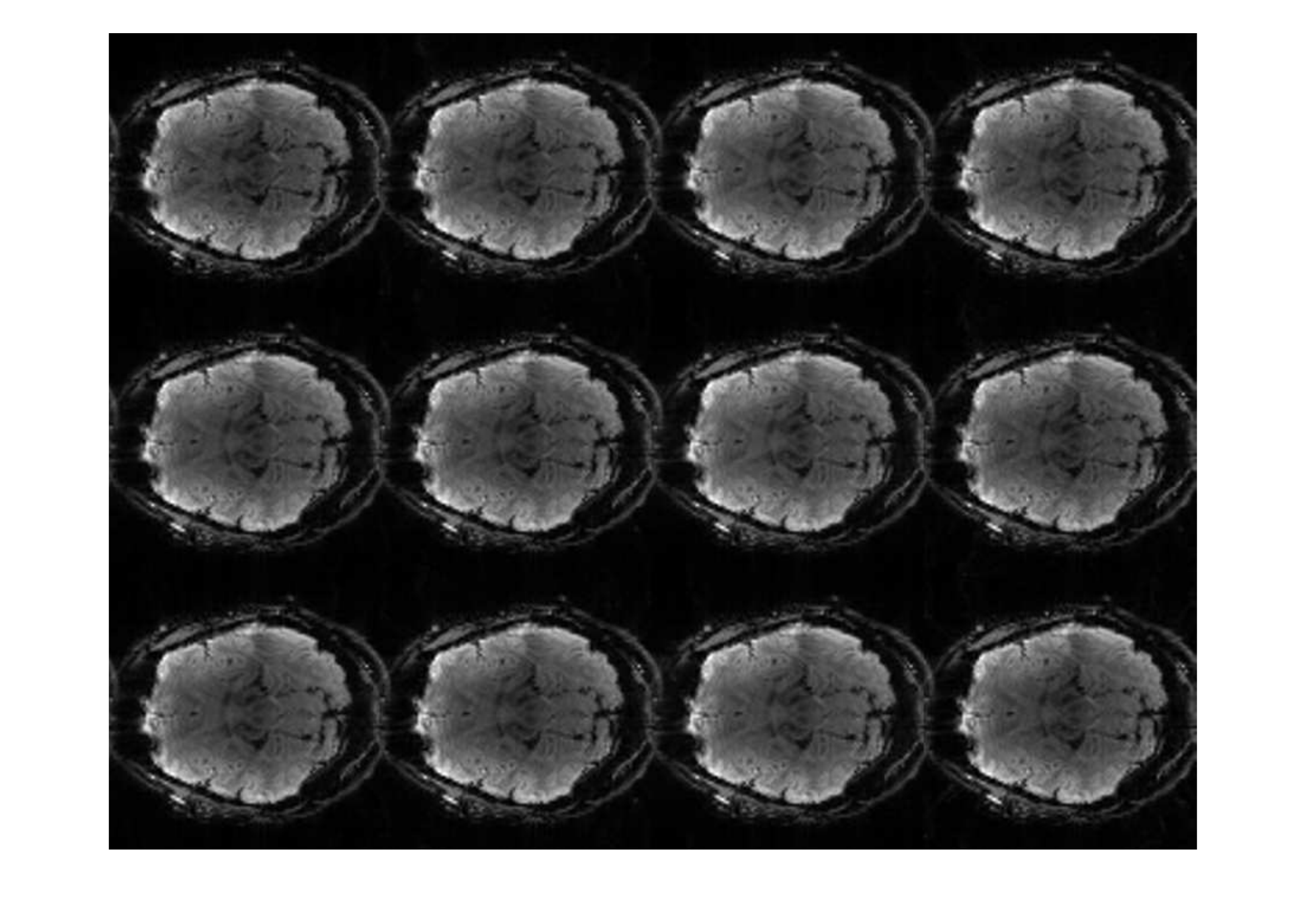}
		\caption{\texttt{RETSINA}}
	\end{subfigure}
	\begin{subfigure}[t]{0.24\linewidth}
		\includegraphics[width=1.0\linewidth]{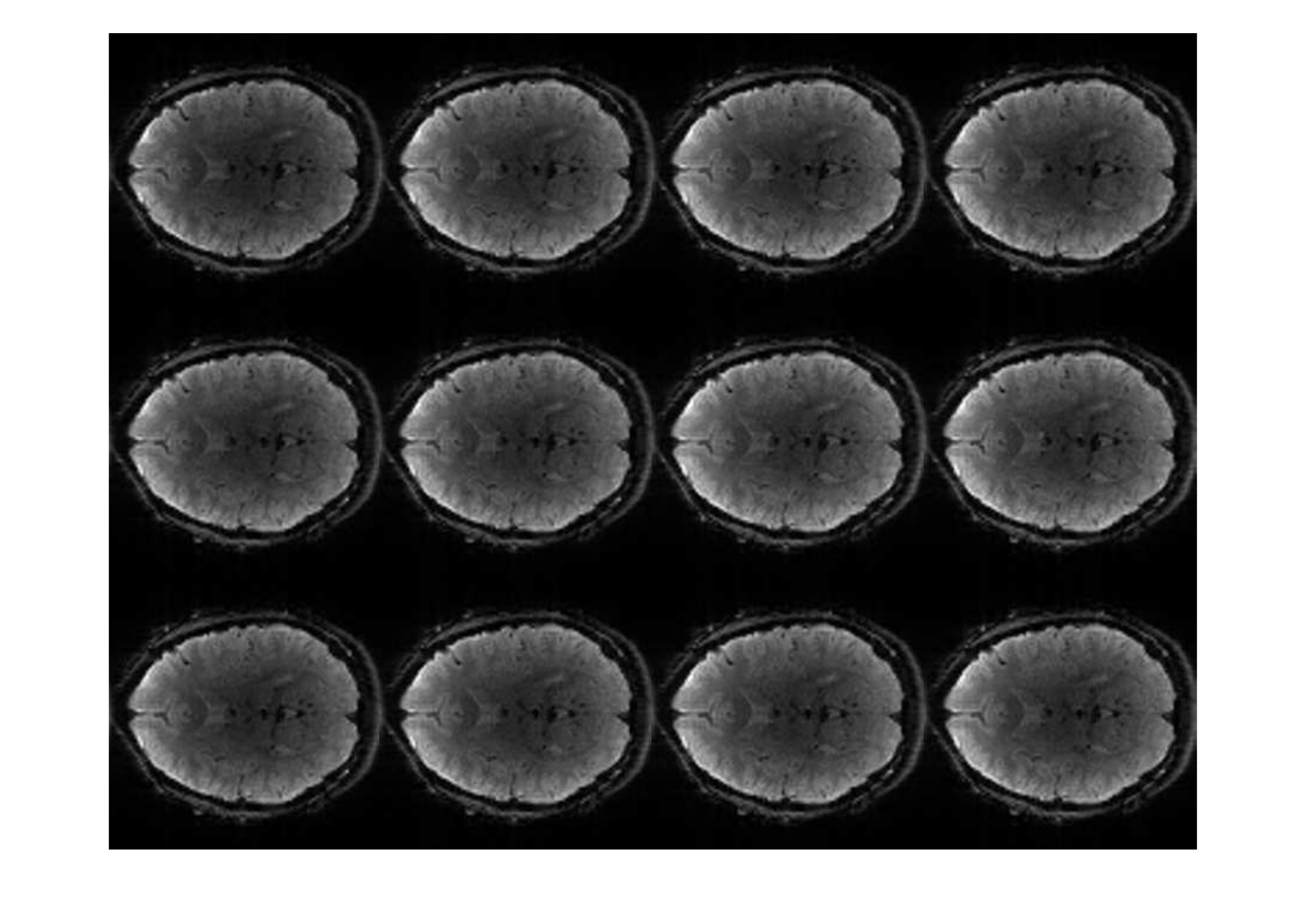}
		\caption{fully sampled scan}
	\end{subfigure}
	\begin{subfigure}[t]{0.24\linewidth}
		\includegraphics[width=1.0\linewidth]{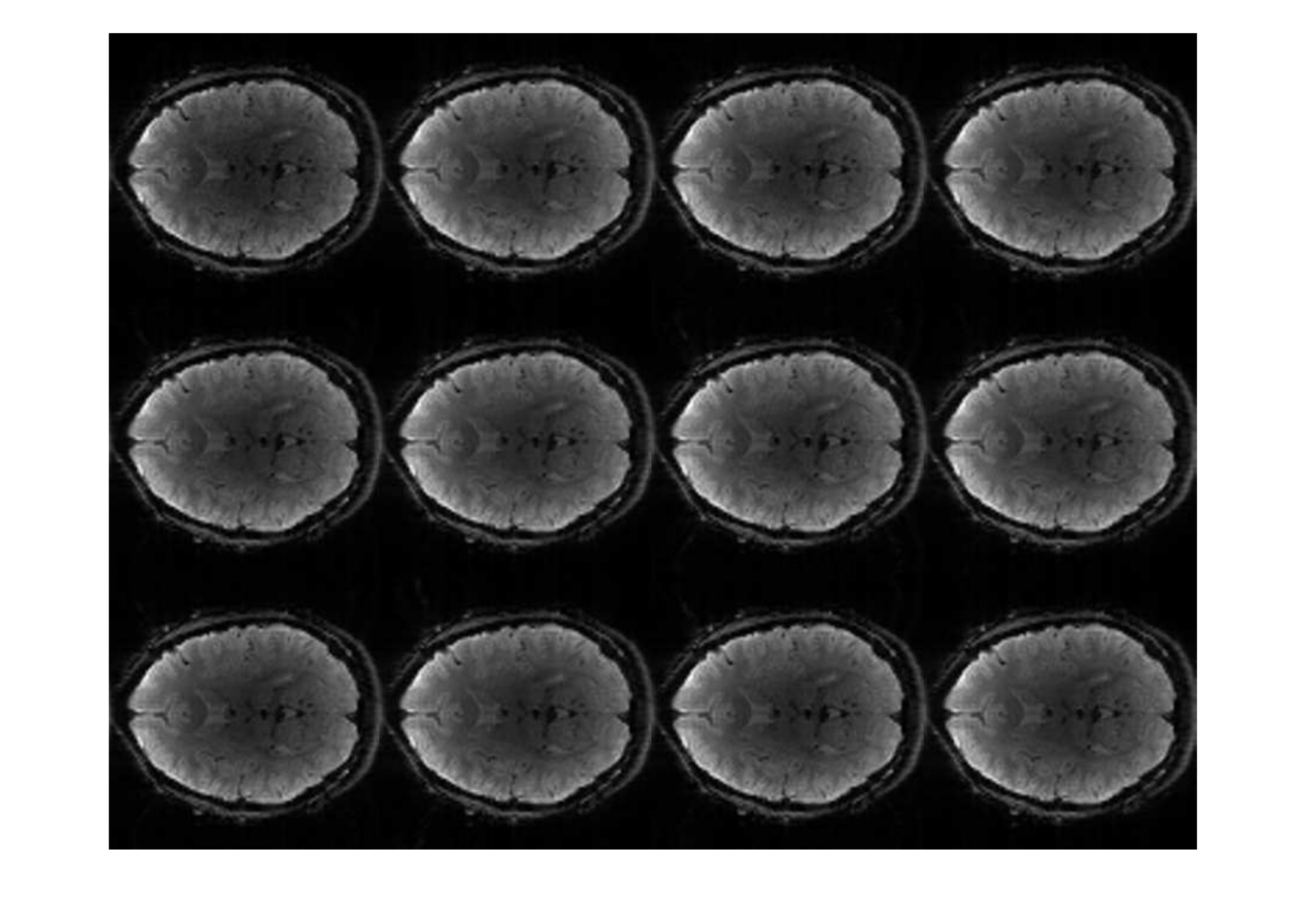}
		\caption{\texttt{RETSINA}}
	\end{subfigure}
	\\
	\begin{subfigure}[t]{0.24\linewidth}
		\includegraphics[width=1.0\linewidth]{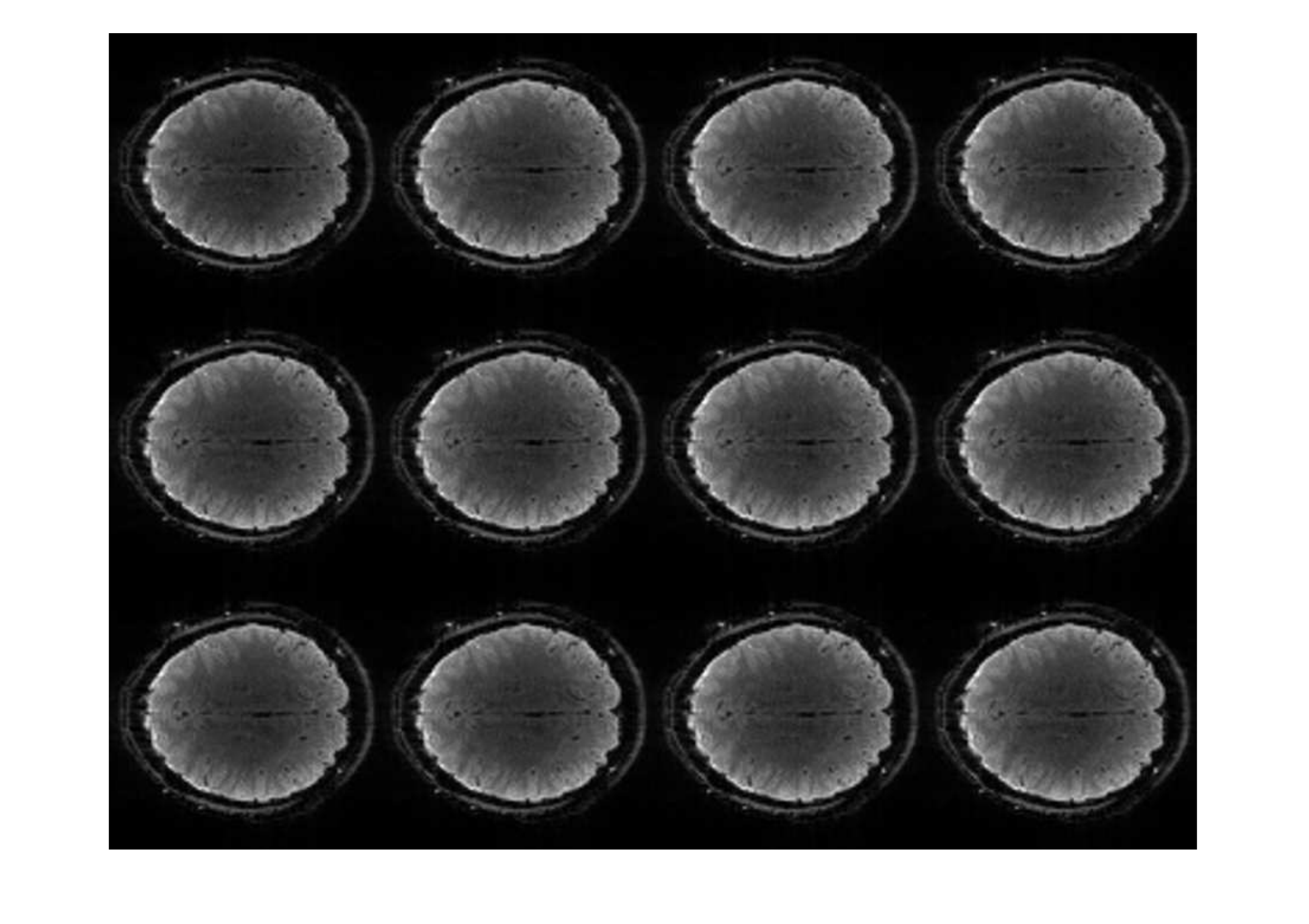}
		\caption{fully sampled scan}
	\end{subfigure}
	\begin{subfigure}[t]{0.24\linewidth}
		\includegraphics[width=1.0\linewidth]{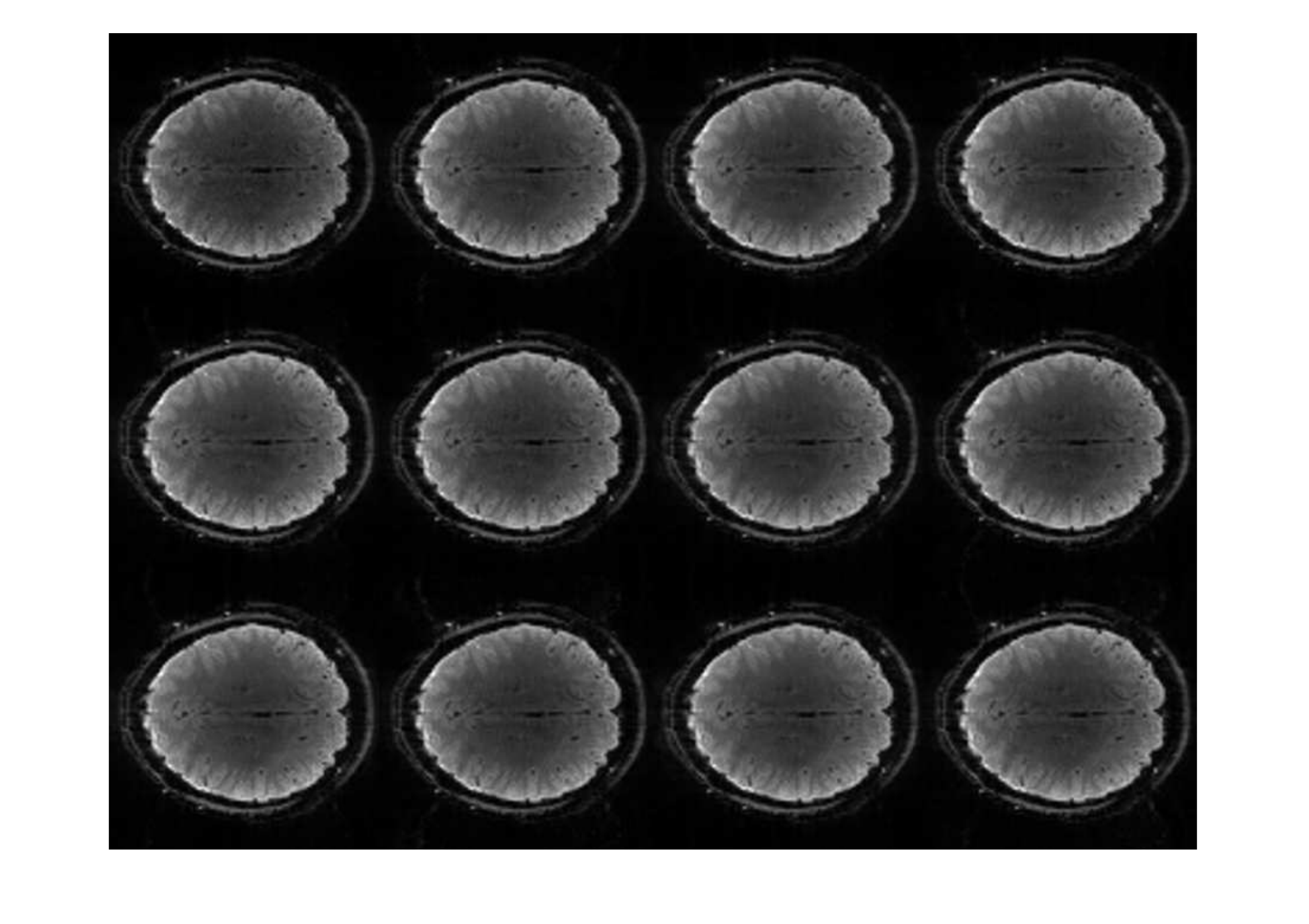}
		\caption{\texttt{RETSINA}}
	\end{subfigure}
	\begin{subfigure}[t]{0.24\linewidth}
		\includegraphics[width=1.0\linewidth]{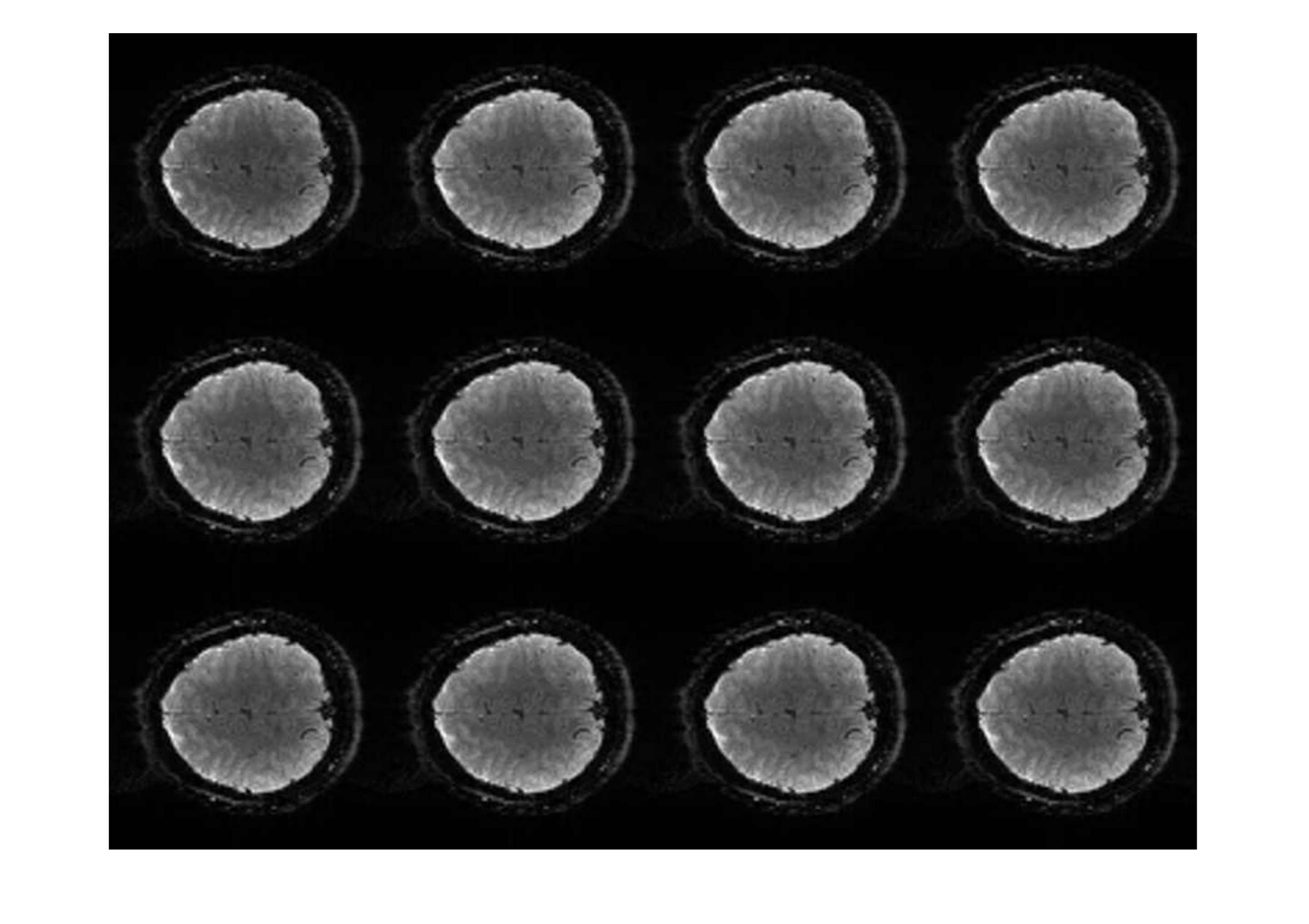}
		\caption{fully sampled scan}
	\end{subfigure}
	\begin{subfigure}[t]{0.24\linewidth}
		\includegraphics[width=1.0\linewidth]{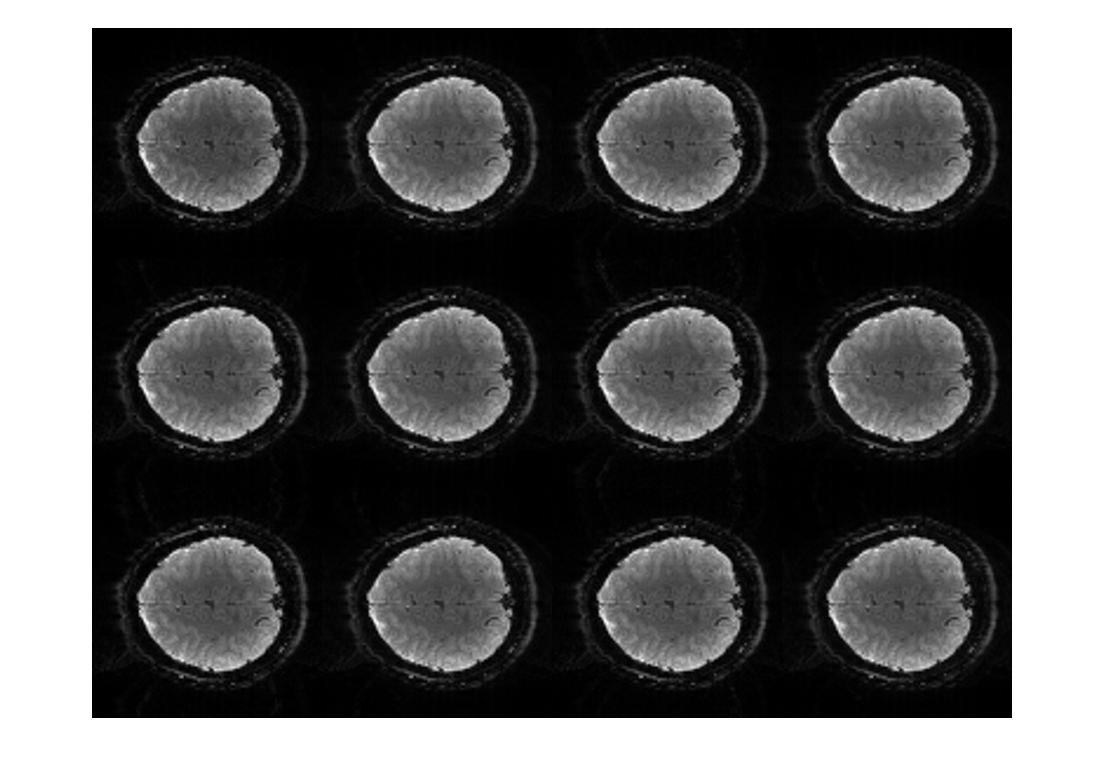}
		\caption{\texttt{RETSINA}}
	\end{subfigure}
	\\
	\begin{subfigure}[t]{0.24\linewidth}
		\includegraphics[width=1.0\linewidth]{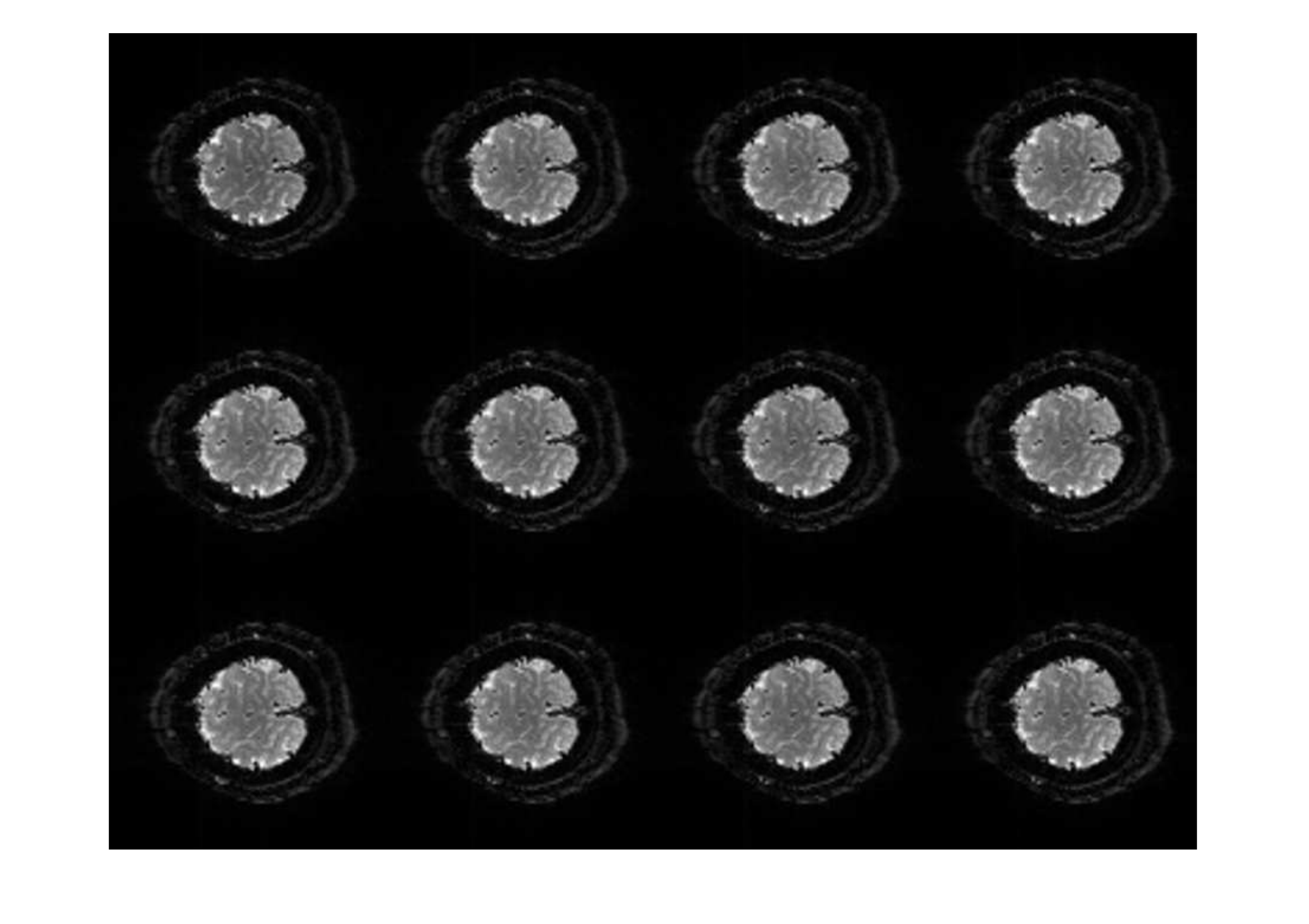}
		\caption{fully sampled scan}
	\end{subfigure}
	\begin{subfigure}[t]{0.24\linewidth}
		\includegraphics[width=1.0\linewidth]{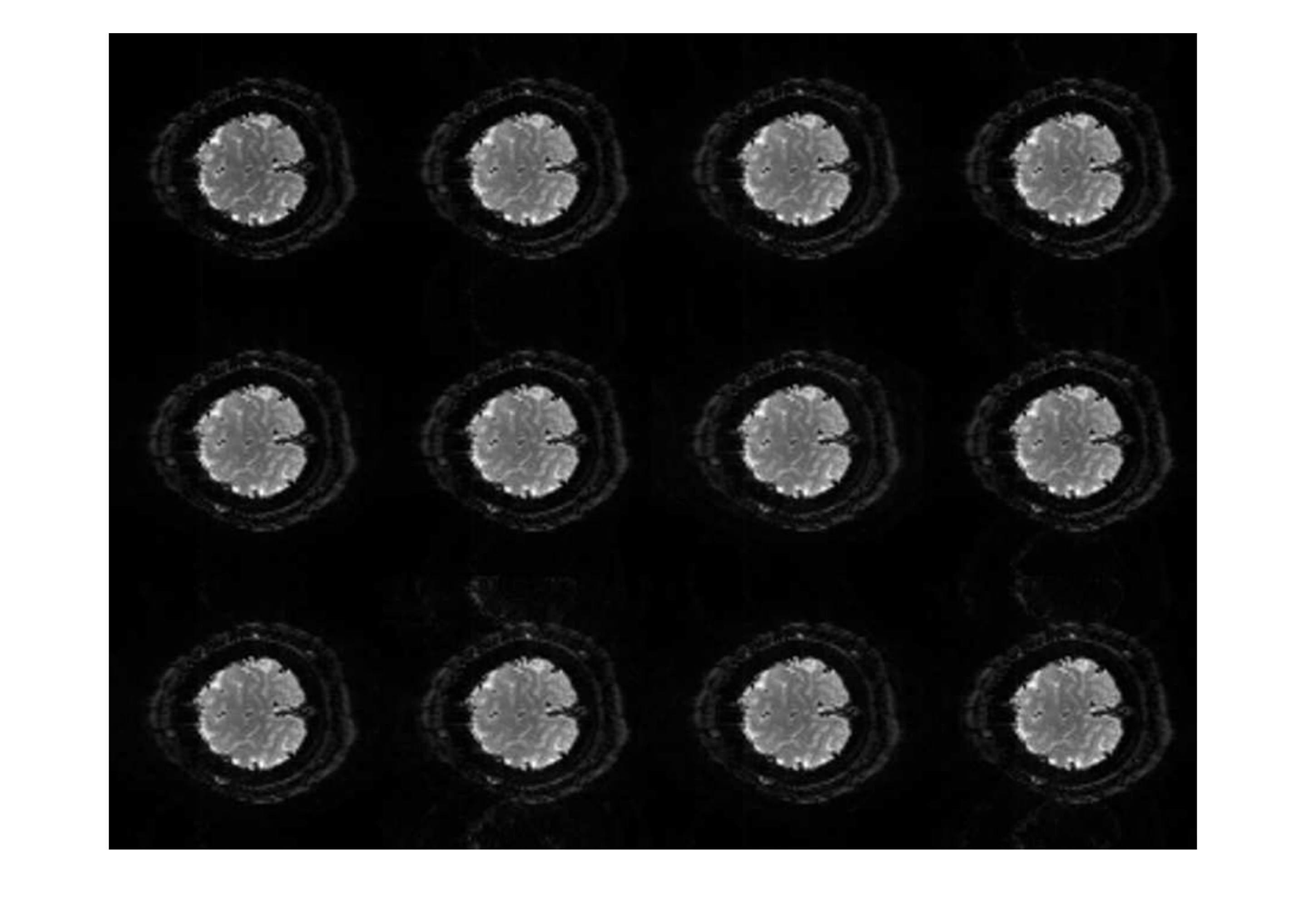}
		\caption{\texttt{RETSINA}}
	\end{subfigure}
	\begin{subfigure}[t]{0.24\linewidth}
		\includegraphics[width=1.0\linewidth]{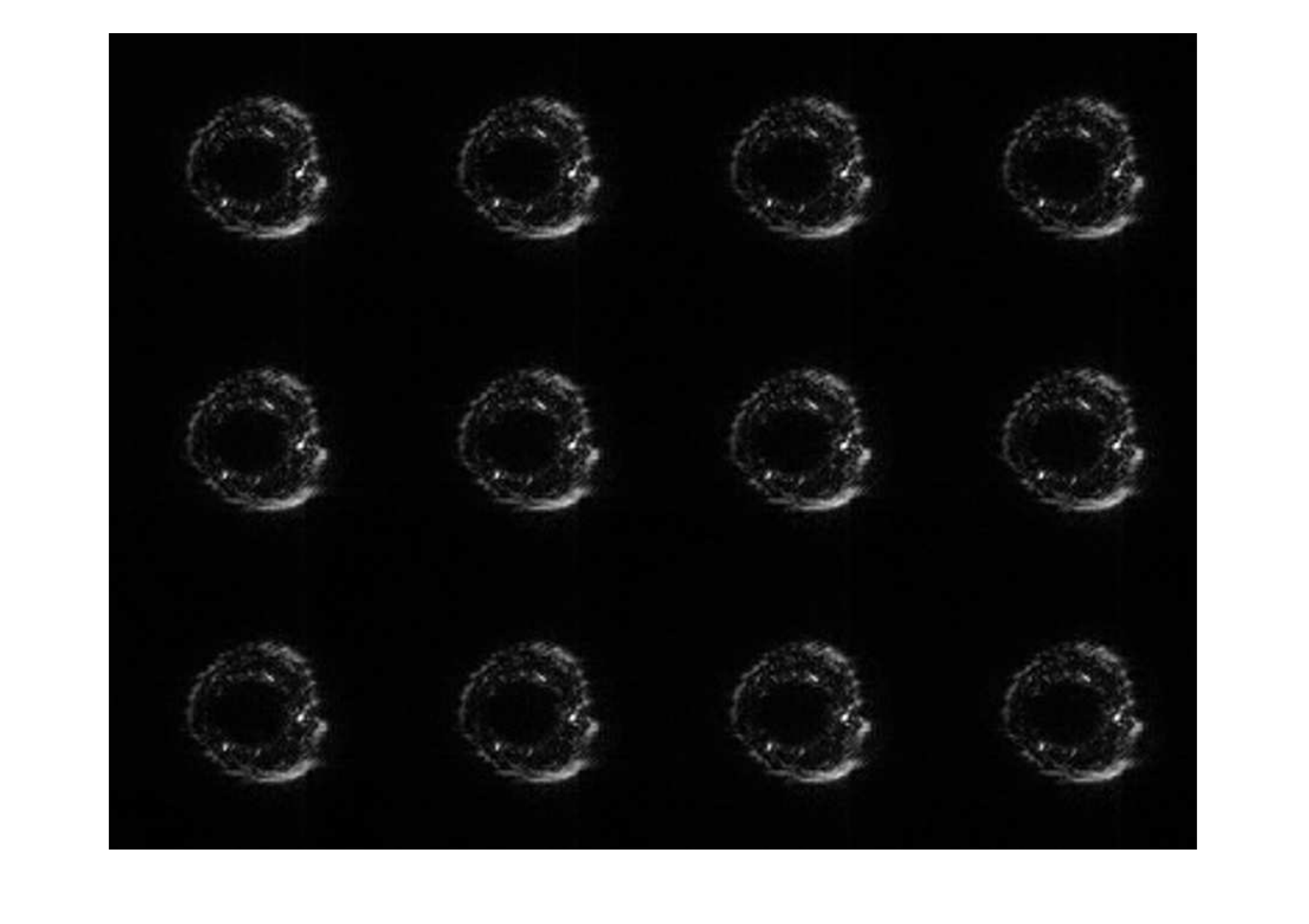}
		\caption{fully sampled scan}
	\end{subfigure}
	\begin{subfigure}[t]{0.24\linewidth}
		\includegraphics[width=1.0\linewidth]{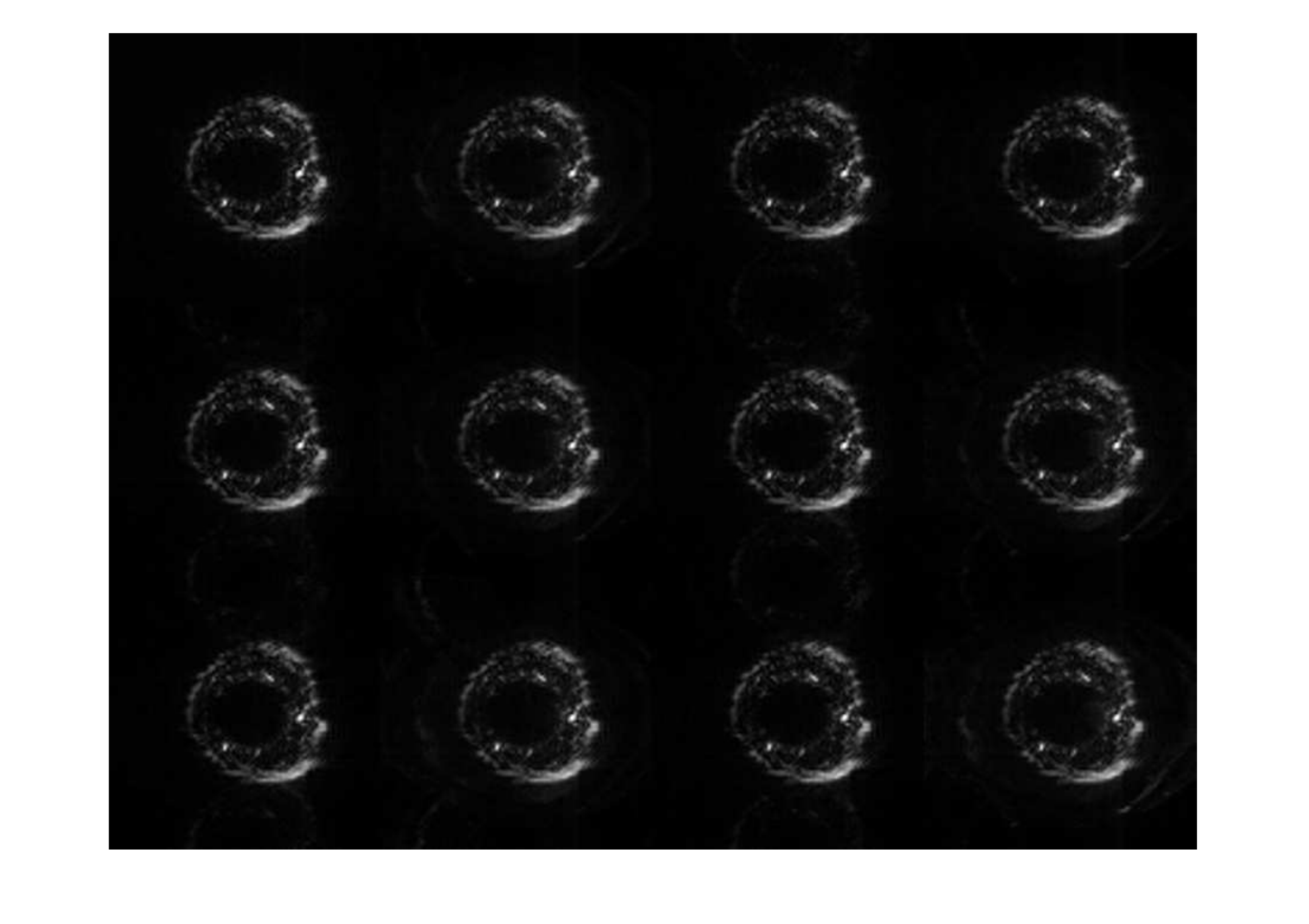}
		\caption{\texttt{RETSINA}}
	\end{subfigure}
	\caption{fMRI reconstruction with 3-fold acceleration}
	\label{fig:rec2}
\end{figure*}
	\vspace{-.1cm}
	\section{Conclusion}
In this paper we studied the sampling and reconstruction of tensors under various schemes. Compared to CS, LRMC, as well as other tensor works, we provide concrete conditions, deterministic and generic, under which exact reconstruction of the tensor was proven to be attainable from regular or even equispaced samples. Furthermore, we cast the fMRI acceleration task as regular tensor sampling process and provided an efficient algorithmic framework to approach the problem. Simulations with synthetic data as well as fMRI scans in the $k$-space show the validity and effectiveness of our approach.			

				\appendices

\section{Proof of theorem \ref{theorem:generic_slab}}\label{appendix:proof_slab}
First, we adjust lemma 1 from \cite{kanatsoulishsrtsp} for complex numbers and selection matrices, which is essential for the proofs.
\vspace{-0.2cm} 
\begin{Lemma}\cite{kanatsoulishsrtsp}\label{lemm:proj}
	Let $\tilde{\Z}=\bm Q\Z$, where the elements of $\Z$ are drawn from an absolutely continuous joint distribution with respect to the Lebesgue measure in $\mathbb{C}^{I F}$ and $\bm \Q\in\mathbb{R}^{I'\times I}$ is a row selection matrix with full row rank. Then the joint distribution of the elements in $\tilde{\Z}$ is absolutely continuous with respect to the Lebesgue measure in $\mathbb{C}^{I'F}$
\end{Lemma}
It follows that $\bm P_1^{(1)}\bm A,\bm P_3^{(2)}{\bm C}$ are drawn from non-singular absolutely continuous distributions. Then Theorem 1 determines the conditions under which the factors of $\underline{\bm Y}_1$ or $\underline{\bm Y}_2$ can be identified. Let's consider the case where $\underline{\bm Y}_1$ is identifiable. Under the conditions of Theorem \ref{theorem:generic_slab}, $\underline{\Y}_1 =  \left\llbracket {\bm A}_1, {\bm B}_1,{\bm C}_1\right\rrbracket$ is essentially unique and from \eqref{compressed1} holds that:
\begin{subequations}
	\begin{align*}
	&{\bm A_1}={\bm P}_1{\bm A}{\bm \Pi}{\bm \Lambda}_1,~{\bm B_1}={\bm B}{\bm \Pi}{\bm \Lambda}_2,~{\bm C_1}={\bm C}{\bm \Pi}{\bm \Lambda}_3,
	\end{align*}     
\end{subequations} 
 where ${\bm \Pi}$ is a permutation matrix and ${\bm \Lambda}_i$ is a full rank diagonal matrix such that ${\bm \Lambda}_1{\bm \Lambda}_2{\bm \Lambda}_3={\bm I}$. Recall that $\underline{\bm Y}_2$ admits a (possibly non-unique) PD $\underline{\bm Y}_2=\left\llbracket\A,{\bm P}_2\B,{\bm C}\right\rrbracket$. Matricizing $\underline{\bm Y}_2$ and plugging in $\B_1$ and $\C_1$ leads to:
\begin{align}
\Y_2^{(1)}&=(\C_1\odot{\bm P}_2\B_1)\A_2^T =(\C\odot{\bm P}_2\B)\bm \Lambda_2\bm \Lambda_3\bm \Pi\A_2^T \label{eq:c_h}
\end{align}
Since $JK_2\geq \lceil F\rceil$, then $\C\odot \bm(P_2\B)$ has full column rank almost surely \cite{jiang2001almost}, and $\A_2=\A\bm \Lambda_1\bm \Pi$ can be identified from \eqref{eq:c_h}. Therefore $\hat{\underline{\Y}} =  \left\llbracket {\bm A}_2, {\bm B}_1,{\bm C}_1\right\rrbracket$ reconstructs signal $\underline{\Y}$.

{The proof shares insights with that of
Theorem 2 \cite{kanatsoulishsrtsp}. The basic difference lies
in the fact that $\bm P_1^{(1)},~\bm P_3^{(2)}$ are full row rank selection matrices
instead of full rank dense matrices.}
\section{Proof of theorems \ref{theorem:generic_fiber},\ref{theorem:generic_entry}}\label{appendix:proof_generic_fiber_entry}

To begin, we use Lemma 1 and observe that $\bm P_1^{(d)}\bm A,$ $\bm P_2^{(d)}{\bm B},\bm P_3^{(d)}{\bm C}$ are drawn from non-singular absolutely continuous distributions. Note that $\bm P_1^{(d)},\bm P_2^{(d)},\bm P_3^{(d)}$ have full row rank by construction. Then we use Theorem 1 to claim identifiability of the factors of each sub-tensor $\underline{\Y}_d$. Under the conditions of Theorems \ref{theorem:generic_fiber},~\ref{theorem:generic_entry}, $\bm P_1^{(d)}\bm A,\bm P_2^{(d)}{\bm B},\bm P_3^{(d)}{\bm C}$ can be identified, which corresponds to identifying all the rows of $\bm A,{\bm B},{\bm C}$, up to column permutation and scaling. The caveat is that the rows of the factors are subject to column permutation and scaling mismatch, since they are obtained by the CPD of independent sub-sampled tensors. For example, let $\underline{\Y}_d =  \left\llbracket {\bm A}_d, {\bm B}_d,{\bm C}_d\right\rrbracket$ and $\underline{\Y}_{d^{\prime}} =  \left\llbracket {\bm A}_{d^{\prime}}, {\bm B}_{d^{\prime}},{\bm C}_{d^{\prime}}\right\rrbracket$. Then, from equations \eqref{eq1}, it becomes clear that in order to obtain $\A,\B,\C$ from $\A_d,\B_d,\C_d,~d=1,\dots,D$ and complete $\underline{\X}$, the permutation and scaling mismatch should be resolved, i.e., ${\bm \Pi}^{(d)}={\bm \Pi}^{({d^{\prime}})}$, ${\bm \Lambda}_i^{(d)}={\bm \Lambda}_i^{({d^{\prime}})}$ for every $d,~d^{\prime}$. To do so the following lemma is being used:
\vspace{-0.2cm}
	 \begin{Lemma}
	 	Assume the entries of $\C\in\mathbb{C}^{K\times F}$ are jointly drawn from an absolutely continuous distribution over $\mathbb{C}^{KF}$. Then $\C(i,f)\neq\C(i^{\prime},f^{\prime})$ almost surely.
	 \end{Lemma}
	 \begin{IEEEproof}
	 	The proof is similar to the proof of Corollary 1 in \cite{jiang2001almost} and uses the fact that $\C(i,f)-\C(i^{\prime},f^{\prime})$ is a non-trivial analytic function of the entries of $\C$ and thus $\C(i,f)-\C(i^{\prime},f^{\prime})\neq0$ almost surely.
	 \end{IEEEproof}

Finally, to resolve the mismatch, we utilize the rules in \eqref{constraints_fiber1}, \eqref{constraints_entry1}. Particularly, when the original tensor is fiber sampled $\C_d=\C_{d^{\prime}}={\bm C},~\forall d,~d^{\prime}$ up to column permutation and scaling. Then, column permutation can be fixed to be the same for all $\underline{\Y}_d$s, since the entries of $\C$ are not equal almost surely. In order to reconcile for scaling mismatch, \eqref{rule:graph}, guarantees that there exist at least one row of $\A$ or $\B$ that is identified (up to permutation and scaling) from 2 different sub-sampled tensors $\underline{\Y}_d$. This is sufficient to resolve the CPD scaling mismatch between every $\underline{\Y}_d$-$\underline{\Y}_{d^{\prime}}$ couple, due to Lemma 1. The entry sampling mechanism, differs to the fiber sampling one, in the fact that $\C$ can only be partially identified from each sub-sampled version $\underline{\Y}_d$. Following same principles as before, permutation and scaling mismatch on the CPD of different $\underline{\Y}_d$s is resolved by \eqref{rule:graph2}, \eqref{rule:perm} along with Lemma 1.

\section{Proof of Theorems \ref{theorem:deterministic_general},\ref{theorem:deterministic_general2}\label{appendix:proof_deterministic}}
The proof is similar to that of Theorem~\ref{theorem:generic_slab},~\ref{theorem:generic_fiber},~\ref{theorem:generic_entry}. The main difference lies in the fact that Theorem 2 is now employed, to establish identifiability on the CPD of each sub-sampled tensor and therefore recoverability of the original tensor. Furthermore, permutation and scaling alignment is performed using the rows of the latent factors which are common among the sub-tensors. This is accomplished, since factors with repeated entries are not allowed.

					\bibliographystyle{IEEEtran}
					\bibliography{mybib2}{}

\begin{thebibliography}{10}
\bibitem{whittaker1915xviii}
E.~T. Whittaker, ``On the functions which are represented by the expansions of
  the interpolation-theory,'' \emph{Proceedings of the Royal Society of
  Edinburgh}, vol.~35, pp. 181--194, 1915.
\end{thebibliography}

				\end{document}